\documentclass[10pt,twocolumn]{revtex4}
\usepackage{epsfig,latexsym}
\usepackage{amsmath,amscd,amssymb}
\usepackage{bm}
\usepackage{graphicx}
\usepackage{color}

\newcommand{\be}{\begin{equation}}
\newcommand{\ee}{\end{equation}}
\newcommand{\bea}{\begin{eqnarray}}
\newcommand{\eea}{\end{eqnarray}}

\newcommand \lan {\langle}
\newcommand \ran {\rangle}

\newcommand{\br}{\mathbf{r}}
\newcommand{\tbr}{\tilde{\mathbf r}}

\newcommand{\tU}{\tilde U}
\newcommand{\tD}{\tilde D}

\newcommand{\bxi}{\mbox{\boldmath${\xi}$}}
\def\rpar{\mathbf{r}_\parallel}

\begin{document}
\title{Probing DNA conformational changes with high temporal resolution by Tethered Particle Motion}

\author{Manoel Manghi,$^{\dagger\ddagger1}$ Catherine Tardin,$^{\S\P1}$ Julien Baglio,$^{\dagger\ddagger\S\P}$ Philippe Rousseau,$^{\parallel**}$ Laurence Salom\'e,$^{\S\P}$ and Nicolas Destainville$^{\dagger\ddagger}$\footnote{Correspondence: \texttt{nicolas.destainville@irsamc.ups-tlse.fr}}
}

\affiliation{$^{\dagger}$Universit\'e de Toulouse; UPS; Laboratoire de Physique Th\'eorique (IRSAMC); F-31062 Toulouse France;
$^{\ddagger}$CNRS; LPT (IRSAMC); F-31062 Toulouse France; $^{\S}$CNRS; IPBS (Institut de Pharmacologie et Biologie Structurale) 205 route de Narbonne, F-31077 Toulouse France; $^{\P}$Universit\'e de Toulouse; UPS; IPBS; F-31077 Toulouse France; $^\parallel$Universit\'e de Toulouse; UPS; Laboratoire de Microbiologie et G\'en\'etique Mol\'eculaires; F-31062 Toulouse France; and $^{**}$CNRS; LMGM; F-31062 Toulouse France.\\
$^{1}$M. Manghi and C. Tardin contributed equally to this work.}
                         
\date{\today}

\begin{abstract}
The Tethered Particle Motion (TPM) technique informs about conformational changes of DNA molecules, e.g. upon looping or interaction with proteins, by tracking the Brownian motion of a particle probe tethered to a surface by a single DNA molecule and detecting changes of its amplitude of movement. We discuss  in this context the time resolution of TPM, which strongly depends on  the particle-DNA complex relaxation time, i.e. the characteristic time it takes to explore its configuration space by diffusion. By comparing theory, simulations and experiments, we propose a calibration of TPM at the dynamical level: we analyze how the relaxation time grows with both DNA contour length (from 401 to 2080 base pairs) and particle radius (from 20 to 150~nm). Notably we demonstrate that, for a particle of radius 20~nm or less, the hydrodynamic friction induced by the particle and the surface does not significantly slow down the DNA. This enables us to determine the optimal time resolution of TPM in distinct experimental contexts which can be as short as 20~ms.
\end{abstract}

\maketitle

\section*{INTRODUCTION}

\begin{figure}[t]
\begin{center}
\includegraphics[height=6cm]{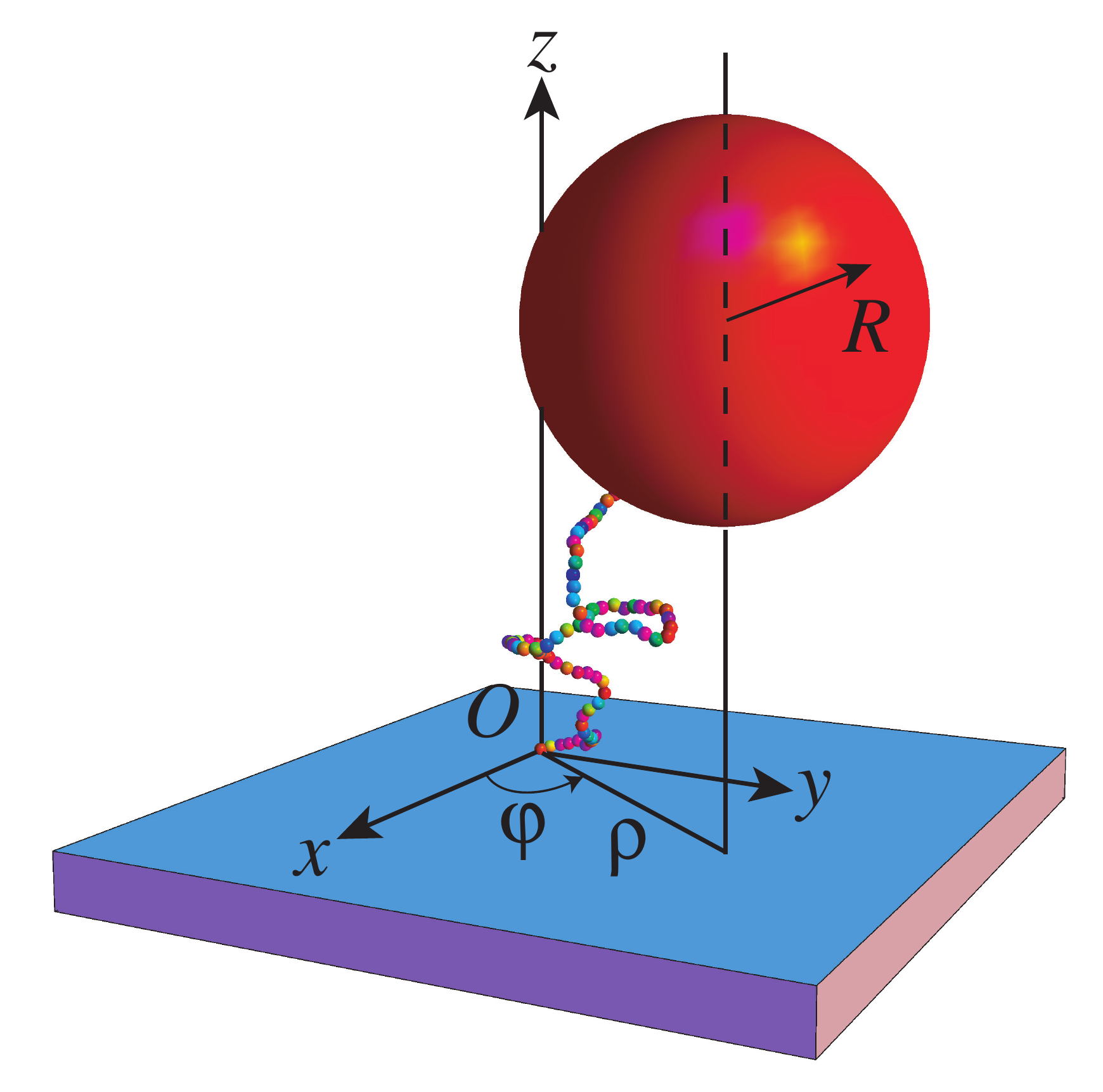}
\caption{\footnotesize Snapshot of TPM Monte-Carlo simulation: The tethered DNA molecule is modeled as a coarse-grained chain of $N$ connected spheres (various hues), fixed to the glass surface at one end and to the tracked particle at the other end. Here, $N=80$, the DNA contour length is $L=2080$ base pairs ($\simeq 700$~nm) and the particle radius is $R=150$~nm. 
%Only the polar coordinates $\rho$ and $\varphi$  in the $(xy)$ plane are accessible by conventional TPM. 
\label{TPMsketch}}
\end{center}
\end{figure}

Biophysical techniques at the single molecule level have become an integral part of the available toolbox to investigate biomolecular machineries. The rapid development of experimental techniques for the exploration of conformations and dynamics of single DNA molecules emphasizes the need for suitable theoretical tools to interpret the large amount of data collected. Among the many techniques used in biology or biophysics laboratories, Tethered  Particle Motion (TPM) \cite{Schafer91, finzi95,Pouget04,Grigoriev,Pouget06,Nelson06,Zurla07,Han09,Seol07,Towles09,Brinkers09,Rutkauskas,wuite} is very promising because it explores the {\it equilibrium} statistical mechanics of the biopolymer in the absence of external force, by contrast to optical or magnetic tweezers experiments~\cite{Bustamante03}.
One end of a DNA molecule is immobilized on a glass surface, and the other end is attached to a particle, the diameter of which ranges from a few tens to several hundreds of nanometers (figure~\ref{TPMsketch}).  By measuring DNA end-to-end distance (or ``effective length''), tracking by video-microscopy the particle trajectory informs about DNA conformations  in real time. Hence TPM also gives access to {\it dynamical} properties.
Getting insights into the dynamics of biomolecular events is of great significance. Should its time resolution be sufficient, TPM has the capability to get access to the dynamics of DNA conformational changes, such as looping/unlooping~\cite{Pouget06,Zurla07,Rutkauskas,wuite}, curvature variations due to protein binding/unbinding~\cite{dixit,Pouget06}, hybridization/dehybridization~\cite{zocchi}, or changes induced by enzyme processing~\cite{Schafer91,Grigoriev}.
Such conformational changes are detected through variations of the particle amplitude of movement, which is calculated on sliding time intervals of a given duration,  $T_{\rm av}$. On the one hand,  $T_{\rm av}$ must be long enough to have a good estimate of the amplitude of movement, and thus to discriminate different amplitudes associated with different DNA conformations. On the other hand, $T_{\rm av}$ sets the TPM time resolution because events shorter than $T_{\rm av}$ are smeared out and thus cannot be detected. Therefore $T_{\rm av}$ must be optimally chosen. We shall see below that its optimal value is proportional to the relaxation time of the particle-DNA complex, i.e. the characteristic time the complex takes to explore its configuration space by diffusion. Thus the knowledge of the relaxation time is a prerequisite to estimate the TPM time resolution. The experimental conditions, especially the proximity of the surface and the attachment of the particle, are likely to perturb the polymer and to increase the relaxation time. Therefore 
knowing the relaxation time in function of both the DNA contour length and the particle radius is of primary importance to anticipate TPM time resolution capabilities.

To our knowledge, dynamical consequences of the setup geometry, in particular of the attached particle, have not been quantified extensively yet. Calibration of TPM experiments, at the dynamical level, remains to be performed.  We focus on the DNA relaxation time, technically defined as the characteristic time of the slowest elastic mode only. We analyze how it scales with particle radius and  DNA length, by using small particles with radii down to 20~nm, for DNA molecules of length $L$ ranging between 401 and 2080 base-pairs (bp; 1~bp = 0.34~nm). The DNA is semi-flexible since $L$ is on the order of the persistence length, $\ell_{\rm p} \simeq 147$~bp. To this end, we combine TPM experiments and dynamical Monte Carlo simulations that take into account hydrodynamic effects in the vicinity of the surface. Comparing both data sets, we demonstrate that the surface and the particle do not affect significantly the polymer dynamics provided that the particle radius $R$ remains small as compared to the DNA contour length. Quantitatively, for the polymers considered in this work, this amounts to $R<L/6$. A radius $R\lesssim20$~nm satisfies this condition for all DNA lengths usually studied by TPM. In addition we address rigorously a critical instrumental issue: the detectors used in TPM experiments always have a finite exposure time, ranging from milli-seconds to a fraction of second~\cite{Allemand,Savin05,Nelson06,BJ06,wong,Towles09}. We show that experimental studies must take into account the finiteness of this exposure time in order to extract valuable relaxation times. Finally, we discuss the intrinsic time resolution of TPM experiments when monitoring DNA conformational changes and give an illustrative numerical example in the case of DNA looping/unlooping.

\section*{MATERIALS AND METHODS}

\subsection*{Tethered Particle Motion (TPM) Experiments}

{\footnotesize DNA substrates were obtained by PCR amplification from plasmid templates with a 21-digoxigenin-modified forward primer and a 21-biotin-labelled reversed primer (Eurogentec) as described in Ref.~\cite{Pouget04}. The DNA substrates DNAR401, DNAR798, DNAR1500, DNARL2080 were produced using pAPT72 as a template (positions: 1460-1861, 1063-1861, 361-1861, 4625-1861, respectively). Their lengths are $L=401$, 798, 1500 and 2080~bp respectively.

Experiments with fluorescent latex particles of radii 20 and 100~nm (Fluospheres Neutravidin, Molecular Probes) were performed using a protocol similar to the one described in Ref.~\cite{Pouget04}. A coverslip flow chamber (30 $\mu$L volume) was incubated with the anti-digoxygenin antibody (20~mg/L; Roche) in phosphate-buffered saline (PBS) for 20 min at room temperature. After washing, the chamber was incubated with casein (1~g/L) in PBS at 4$^{\circ}$C for 4 hours. 
Experiments with the $R=150$~nm particles (Anti-dig fluorescent particles, Indicia) were performed using chambers whose surfaces were derivatized with a mixture of polyethylene glycol and biotinylated polyethylene glycol (Sigma, Nanocs) based on the protocol of Ref.~\cite{Cuvelier}.  The chamber was incubated with neutravidin (20~mg/L, Molecular Probes) in PBS for 20~min at room temperature. 
The chamber was subsequently rinsed and incubated with the mix of DNA and neutravidin (20~nm, 100~nm) or anti-dig (150~nm) coated particles (ratio 1:1, 1~pM) that had been prepared 1 hour before in PBS with 0.1~g/L BSA. Due to the lack of small particles coated with antiDig, we could not use the PEG passivated coverslips, known to reduce unspecific binding of particle-DNA complex on the glass, in all the conditions.

The tethered particles of 150~nm (resp. 100~nm and 20~nm) were visualized by fluorescence video-microscopy,  with a magnification of 63x4x, on a CCD camera Coolsnap (resp. Cascade, RoperScientific) at a recording time of 25 frames/s (resp. 74 frames/s and 112 frame/s), which corresponds to an acquisition period $T_{\rm ac}$ equal to 40~ms (resp. 13.5~ms and 8.9~ms). In all cases, we could restrict the exposure time $T_{\rm ex}$  to 5~ms by the use of an AOTF, characterized by an extinction ratio exceeding 7400 (or by entering such a command in the acquisition parameters of the Cascade CCD). 

The two-dimensional projection $\rpar$ of the particle position was determined on successive images as the spot centroid using a home built image analysis program (Labview). 
The accuracy of the position detection (pointing error), calculated as the standard deviation of the positions of immobilized particles accumulated during 30~s, is equal to 36, 21, and 8~nm for the particles of radius 20, 100, and 150~nm, respectively. Trajectories have an average duration of 90~s. Along a trajectory, the DNA anchoring point at time $t$ is determined by averaging the particle position over an interval of duration $T_{\rm av}=2$~s centered at $t$, and then subtracted from $\rpar$. This sets $\langle \rpar \rangle =0$ and subtracts the instrumental drift. 
Since $T_{\rm av}$ is much larger than the diffusion relaxation time (see below), this anchoring point is determined with a good accuracy.
The trajectories exhibiting asymmetry were discarded following~\cite{meiners}.
The amplitude of movement of the particle is defined as the variance of $\rpar$ averaged on the trajectory as
\be
\Delta \rpar^2=\langle \rpar^2\rangle.
\label{deltar}
\ee
A minority of trajectories appeared to have an amplitude of movement $\Delta\br_\parallel^2$ significantly shorter than the majority ones, which yields a bimodal distribution of $\Delta\br_\parallel^2$ with a small population of low amplitudes. These trajectories, that might be related to multi-DNA-particle complexes, have been discarded. The final number of trajectories for each condition ($L$, $R$, $T_{\rm ex}$, $T_{\rm ac}$) ranges from 19 to 60. Details are given in the Supplementary Data.

\subsection*{DNA coarse-grained model}

The labeled DNA polymer is modeled as a chain of $N$ connected small spheres of radius $a$, whose positions are denoted by $\br_i(t)$ where $i=0,\ldots,N-1$, and a larger final particle of radius $R\geq a$, of position $\br(t)$  (see figures~\ref{TPMsketch} and~S2). The DNA contour length is $L=2a (N-1)$. In this work, $a$ ranges between 1.4 and 7~nm, and $R$ between 0 (i.e. no particle) and 200~nm.
The internal structure of the double-stranded DNA is not considered at this level of modeling. Denaturation bubbles are too scarce at room temperature to have an effect on the global chain conformation~\cite{PRL,BJ}. The persistence length value $\ell_{\rm p}=147$~bp is averaged over the nucleotide sequence. Torsional degrees of freedom are omitted as a first hint into the full problem. The polymer is grafted on a surface which sets $\br_0=0$. 
%Due to the geometry of the experiment, it is convenient to 
We use cylindrical coordinates $(\rho,\varphi,z)$: $\br = \rho {\bf e}_{\rho}+z{\bf e}_z =  \br_\parallel +z{\bf e}_z$. Since the polymer motion is limited to the upper half-plane, we impose the following ``hard wall'' boundary conditions: $z_i>0$ for monomer spheres and $z>R-a$ for the particle. We treat a freely rotative joint to the glass coverslip~\cite{Seol07}, by fixing the first sphere center at a height $a$ above the surface (see also the Supplementary Data). 

% -> SM However, it might happen that, for special types of experimental constructions, the joint is better modeled as a rigid anchor (clamped polymer). In this case, the first two spheres are fixed perpendicularly to the surface.

All spheres interact \textit{via} stretching and bending forces: the potential $U$ is the discrete version of the {\it extensible} worm-like chain potential and depends on the sphere positions $\br_i$ and the particle one $\br$ as $U =  \sum_{i=0}^{N-2} \left[ \frac{\gamma}{4a} \, (|\br_i -\br_{i+1}|-2a)^2 + \frac{\varepsilon}{2a} \,(1-\cos \theta_i) \right] +\frac{\gamma}{4a} (|\br_{N-1} -\br|-R-1)^2 + \frac{\varepsilon}{2a} (1-\cos \theta_{N-1})$ where $\theta_i$ is the angle between neighbouring bonds of sphere~$i$. The first term ensures the polymer connectivity and the second term is the bending energy with zero spontaneous curvature. The last two terms are dedicated to the particle. The parameters $\gamma$ and $\varepsilon$ are the stretching and bending moduli~\cite{Mano06}. The persistence length is given by $\ell_{\mathrm{p}}=\varepsilon/(k_{\rm B}T)$. We choose  $\gamma a^2= 4\varepsilon$, which is exact for an isotropic elastic cylinder with radius $a$. Therefore the key parameters in the simulation are $L/\ell_{\rm p}$, $R$ and $N$. Mutual penetration of monomers is prevented by an excluded volume interaction.

A fully realistic description would be to fix the sphere radius, $a$, equal to the DNA half-width, i.e. about 1~nm. Hence one sphere would model roughly 6~bp and 25 spheres would correspond to the typical DNA persistence length of 147~bp at physiological temperature and salt concentration. Time limitation in the simulations led us to concentrate on $N=50$ (or 25) for a DNA of 400 to 2000~bp and thus to choose $a>1$~nm. As far as equilibrium properties are concerned, the DNA statistical mechanics are insensitive to the choice of $N$, provided that $a$ remains small as compared to $\ell_{\rm p}$, so that a sphere represents a DNA segment that is actually rigid.

\subsection*{Dynamical Monte Carlo simulations}

Out-of-equilibrium dynamics can be tackled numerically by Dynamical Monte Carlo (DMC) simulations that become equivalent to Brownian dynamics when the variation of energy at each time step, $\Delta U$, satisfies $\Delta U \ll k_{\rm B}T$~\cite{Newman}. We have also performed Brownian dynamics simulations of the system which yield very similar results on equilibrium and dynamical properties (see Supplementary Data).
At each Monte Carlo Step (MCStep) of physical duration $\delta t$, a bead is chosen uniformly at random among the $N+1$ possible ones (monomer spheres and labeling particle). Then a random move $\delta \mathbf{r}$ is attempted for this bead, uniformly in a ball of center 0 and radius $R_b$, thus  $\langle \delta \mathbf{r}^2 \rangle = 4\pi\int_0^{R_b} r^4 \mathrm{d}r / 4\pi\int_0^{R_b} r^2 \mathrm{d}r=(3/5) R_b^2$. This quantity must be equal to $6 D_0 \delta t$, where $D_0$ is the diffusion coefficient of the spherical bead, depending on its diameter. In practice, for monomer spheres, $R_b=a/5$ (unless stated differently), where $a$ is their radius. Then $\delta t$ is set subsequently through $\delta t=R_b^2/(10D_0)$, and,  for the particle, $R'_b=\sqrt{10D'_0\delta t}$ is fixed with $D'_0=D_0a/R$. Interactions between adjacent beads are treated {\em via} the interaction potential energy $U$, whereas interactions between non-adjacent beads are of hard core nature, like surface-bead interactions: whenever a move would lead to the penetration of a bead into an other one or the surface, it is rejected.
A Monte Carlo Sweep (MCS) is a sequence of $N+1$ MCSteps. The physical time is incremented of $\delta t$ following each MCS~\cite{Newman}. Typically, a simulation lasts between $10^9$ and $10^{10}$ MCS, which leads to satisfactory error bars, estimated by usual techniques~\cite{Newman} [see also equation~(\ref{rel:err})]. In the simulations, $N=50$ or 25, depending on the slowness of the physical process. Since $a \ll \ell_{\rm p}$ in all cases, we expect that dynamics do not depend on $N$ either. A simulation snapshot is shown in figure~\ref{TPMsketch}. Finally, note that in our DMC simulations, moves are local (one bead at once).  This choice is more time-consuming than the global Monte Carlo sampling of Refs.~\cite{Segall,Nelson06}, but has the advantage to give access to dynamical properties, which is the main goal of the present work.

\subsection*{Diffusion near the surface: hydrodynamic effects}

The motion of a spherical particle is slowed down near a flat surface due to the no-slip condition for the solvent velocity flow at the wall. The induced hydrodynamic interactions cause a variation, with the distance to the surface, of the diffusion coefficient as compared to its bulk value $D_0$. This new diffusion tensor can be split into a component parallel to the wall, $D_{\parallel}$, and a perpendicular component, $D_{\perp}$.
For a sphere of radius $b$ (here $b=a$ or $R$), the center of which is at a distance $h$ from the wall, parallel and perpendicular diffusion coefficients are derived from Fax\'en's law~\cite{Happel}. At order 3 in $b/h$, 
\begin{eqnarray}\label{DperpDpar}
D_{\perp} &=& D_0 \left(1-\frac98 \frac{b}{h} +  \frac12 \frac{b^3}{h^3}  \right)\\
D_{\parallel} &=& D_0 \left(1-\frac9{16} \frac{b}{h} +  \frac18 \frac{b^3}{h^3}  \right).\nonumber
\end{eqnarray}
Monte Carlo simulations must be modified to take into account these spatially varying diffusion coefficients. First of all, random moves $\delta \mathbf{r}$ are now randomly chosen in an ellipsoid to account for anisotropy.
In addition, careful attention must be paid to the discretization of the equations of motion in this case. A vertical drift term $({\rm d} D_\perp/ {\rm d}h)\mathbf{e}_z$ must be added to compensate the variation of $D_\perp$ with $h$ and to restore the detailed balance condition~\cite{Ermak78}. As a consequence, this improvement of dynamics does not affect equilibrium properties such as chain statistics.

\subsection*{Extracting relaxation times from experimental and numerical data}

% -> A key observable in polymer dynamics is the relaxation or correlation time. It corresponds to the time taken by the chain to explore its whole configuration space by diffusion, and can also be viewed as the time after which the memory of the initial condition is lost. We focus here on t
The relaxation time $\tau_{\parallel}$ associated with $\rpar$ can be defined through the two-time correlation function averaged over a trajectory
\begin{equation}
C(t)=\langle\rpar(s+t)\rpar(s)\rangle_s \approx \langle \rpar^2\rangle \exp(-t/\tau_{\parallel}),
\label{Cdet}
\end{equation}
if one assumes without loss of generality that $\langle \rpar\rangle=0$, or through the 2D Mean Square Deviation (MSD):
\begin{eqnarray}
{\rm MSD}(t) &=& \langle(\rpar(s+t)-\rpar(s))^2\rangle_s = 2 \langle \rpar^2\rangle - 2C(t)\nonumber\\ 
&\approx& 2 \langle \rpar^2\rangle [1-\exp(-t/\tau_{\parallel})].
\label{correl}
\end{eqnarray}

At short times $t \ll \tau_{\parallel}$, one expects to recover the 2D diffusion law ${\rm MSD}(t)=4Dt$, with $D$ the apparent particle 2D diffusion coefficient. Thus the correlation time for this particle in a 2D trap with variance $\Delta \rpar^2$ is taken to be~\cite{BJ06}
\begin{equation}
\tau_{\parallel} = \frac{\Delta \rpar^2}{2D}.
\label{tau_gaussian}
\end{equation}

For simulated trajectories, the relaxation times are fitted from $C(t)$ and MSD$(t)$, using equations~(\ref{Cdet},\ref{correl}), leading to two relaxation times, $\tau_C$ and $\tau_{\rm MSD}$. Below, we report the mean values $\tau_{\rm m}=(\tau_C+\tau_{\rm MSD})/2$, with error bars taken as  $|\tau_C-\tau_{\rm MSD}|/2$. In practice, fits are performed on an interval $t \in [0,t_{\rm sup}]$. Since numerical error bars on $C(t)$ and MSD$(t)$ are larger and larger as $t$ grows, the smallest possible value of $t_{\rm sup}$ must be used. On the other hand, for MSD$(t)$, one must have $t_{\rm sup}$ larger than a few $\tau_{\rm MSD}$ in order to fit properly the exponential decay. We have chosen $t_{\rm sup}=4\tau_{\rm MSD}$, which is a good compromise between both constraints. The fitting procedure, consisting in fitting $\tau_{\rm MSD}$ on $[0,t_{\rm sup}]$ and then adjusting $t_{\rm sup}=4\tau_{\rm MSD}$, is iterated a few times until $\tau_{\rm MSD}$ is converged. Similarly, $\tau_C$ is obtained by measuring the slope of $\ln[C(t)/C(0)] \simeq -t/\tau_C$, on the interval $[0,\tau_C]$, with the same iterative procedure. However, in some instances ($R\leq20$~nm),  $\ln[C(t)/C(0)]$ appears to display a short transient, equal to a small fraction of $\tau_C$, because of slow diffusion modes. In this case, linear regressions are performed on a suitably chosen interval $[t_{\rm inf},\tau_C]$, where they appear to be very good (correlation coefficients $|r|>0.9995$). 

For experimental data, $\tau_{\parallel}$ is fitted from $C(t)$ as follows. The raw correlation function  averaged over all available trajectories is denoted by $C_{\rm raw}(t)$. The systematic pointing error is taken into account: the detected position, $\mathbf{r}_{\parallel,{\rm raw}}$, is the sum of the actual position, $\mathbf{r}_{\parallel}$ and the pointing error, $\mathbf{r}_{\rm e}$, two independent random variables. Thus $\Delta \mathbf{r}_{\parallel,{\rm raw}}^2 = \lan \mathbf{r}_{\parallel,{\rm raw}}^2\ran=\lan \mathbf{r}_{\parallel}^2\ran+\lan \mathbf{r}_{\rm e}^2 \ran$. The second contribution is systematically subtracted from measured values $\Delta \mathbf{r}_{\parallel,{\rm raw}}^2 \equiv C_{\rm raw}(0)$, using the pointing error values as given above.
This modified correlation function is denoted by $C_{\rm m}(t)$. In addition, the subtraction of drift induces systematic anti-correlations at short times leading to the following fitting form
\begin{equation}
\frac{C_{\rm m}(t)}{C_{\rm m}(0)}=\left(1+2\frac{\tau_{\rm m}}{T_{\rm av}}\right) e^{-t/\tau_{\rm m}} -2\frac{\tau_{\rm m}}{T_{\rm av}},
\end{equation}
with a {\em single} fitting parameter, $\tau_{\rm m}$. Examples are given in figure~S1. The prescription is the same as above: the fitting interval $[0,t_{\rm sup}]$ is chosen so that $t_{\rm sup}=4\tau_{\rm m}$.
Error bars on $\tau_{\rm m}$ correspond to the standard deviation of the measurements on individual trajectories.

\subsection*{Correction of detector time-averaging effects}

Finally, one has to correct time-averaging effects  in experimental results. In Refs.~\cite{Allemand,Savin05,Nelson06,BJ06,wong,Towles09}, the time-averaging (or blurring) effect due to the finite exposure time of detectors in single molecule (or particle) tracking experiments was investigated. When tracked molecules or particles diffuse in confined regions, diffusion constants can be significantly under-estimated, as well as sizes of confining domains. 
This effect was quantified by exact analytical arguments, in the contexts of diffusion in membrane domains~\cite{BJ06} and optical traps~\cite{wong}, leading to the same correction to $\Delta \rpar $.
In the present work, the situation is similar because the motion of the particle is restricted by the DNA tether, thus all the analytical derivation of Refs.~\cite{BJ06,wong} remains valid: the real relaxation time $\tau_\parallel$, domain size (or root-mean-square excursion from the attachment point) $\Delta \rpar$ and diffusion coefficient $D$ can be related to their measured counterparts, $\tau_{\rm m}$, $\Delta \mathbf{r}_{\parallel,\mathrm{m}}\equiv \sqrt{C_{\rm m}(0)}$ and $D_{\rm m}$. Whenever $\tau_{\rm m} \geq 2\;T_{\rm ex}/3$~\cite{BJ06}
\begin{equation}
\tau_{\parallel} \simeq  \tau_{\rm m} - T_{\rm ex}/3,
\label{zerel}
\end{equation}
where $T_{\rm ex}$ is the detector exposure time, and
\begin{equation}
\Delta \rpar = \Delta \mathbf{r}_{\parallel,\mathrm{m}} \left[2 \frac{\tau_{\parallel}}{T_{\rm ex}} - 2 \left(\frac{\tau_{\parallel}}{T_{\rm ex}}\right)^2 \left(1 - e^{-\frac{T_{\rm ex}}{\tau_{\parallel}}} \right)\right]^{-1/2}.\label{ell:ellm0}
\end{equation}
If $\tau_{\rm m} < 2\;T_{\rm ex}/3$, no correction can be applied~\cite{BJ06} and one ought to switch to a faster acquisition device.
}

\section*{RESULTS AND DISCUSSION}

\subsection*{Equilibrium distributions of particle positions}

Before considering the dynamics of the particle-DNA complex, we first briefly analyze the particle positions at equilibrium through the histograms of projected particle excursions, $\rho=| \rpar |$, i.e. the probability distributions $p(\rho)$, and through the standard deviations $\Delta \mathbf{r}_{\parallel}=\langle \mathbf{r}_{\parallel}^2 \rangle^{1/2}$. We compare the latter to the fitting formula suggested by Towles \textit{et al.}~\cite{Towles09}, in order to test both our simulation and experimental results.

We simulate DNA molecules of lengths $L=401,798,1500$ and $2080$~bp, corresponding to the semi-flexible case ($2<L/\ell_p<14$) for which no analytical expression for $p(\rho)$ is known. The theoretical study of the limiting cases, rigid rod ($L \ll \ell_{\rm p}$) and flexible chain ($L \gg \ell_{\rm p}$), is reported in the Supplementary Data. In addition to the experimental particle radii $R=20$, 100, and 150~nm, we also examined numerically the radii $R=40$, 80, and 200~nm, as well as $R=0$ (no particle). The numerical distributions $p(\rho)$ and their deviation from a Gaussian are discussed in the Supplementary Data (figures~S4--S6).

 \begin{figure}[t]
\begin{center}
\includegraphics[width=8cm]{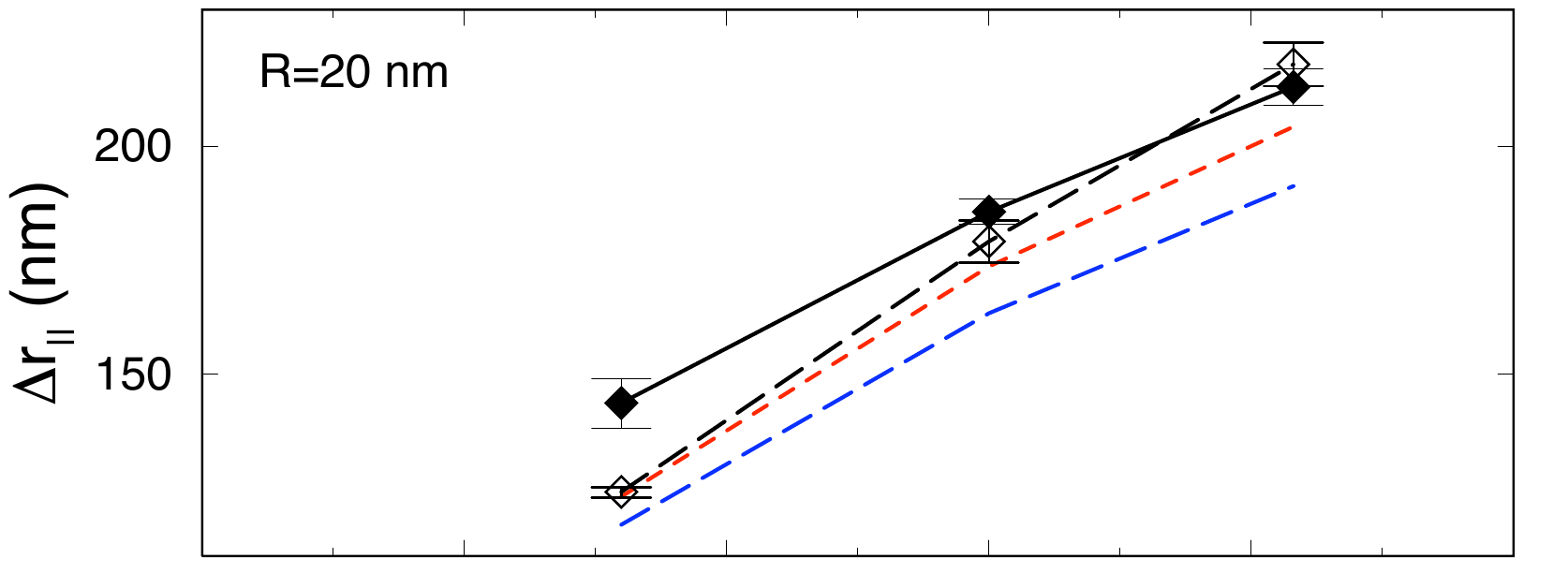}
\includegraphics[width=8cm]{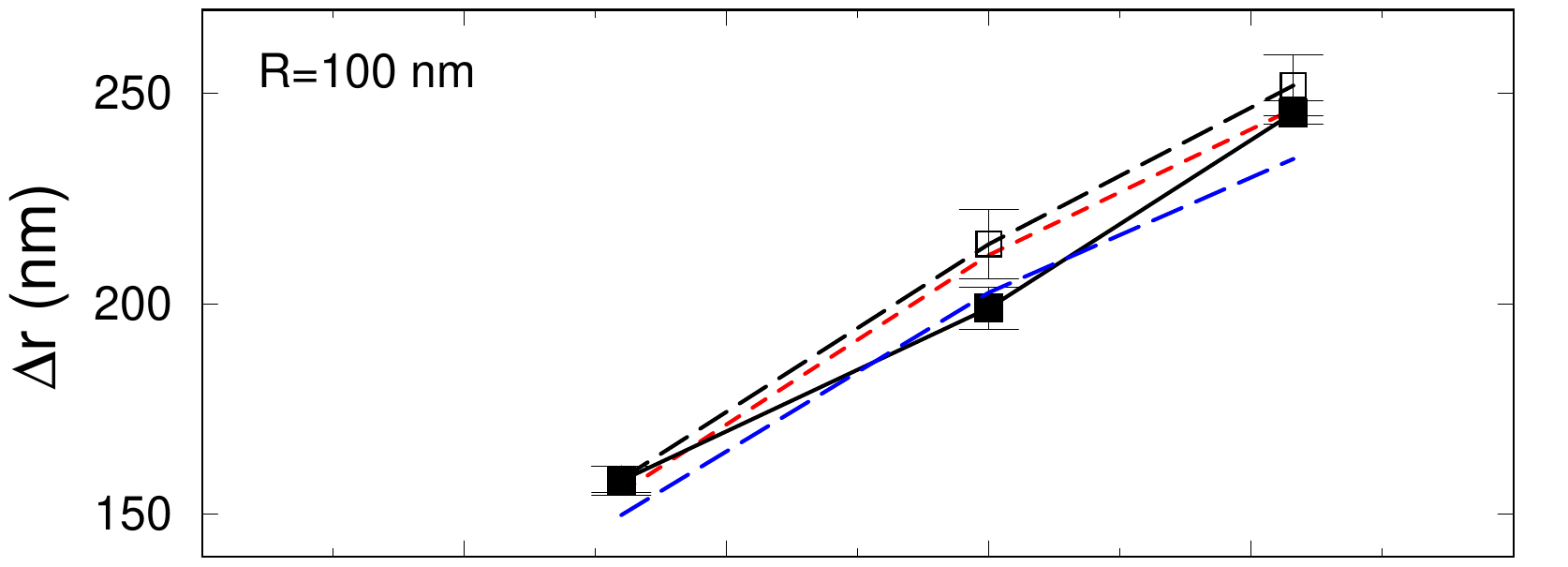}
\includegraphics[width=8cm]{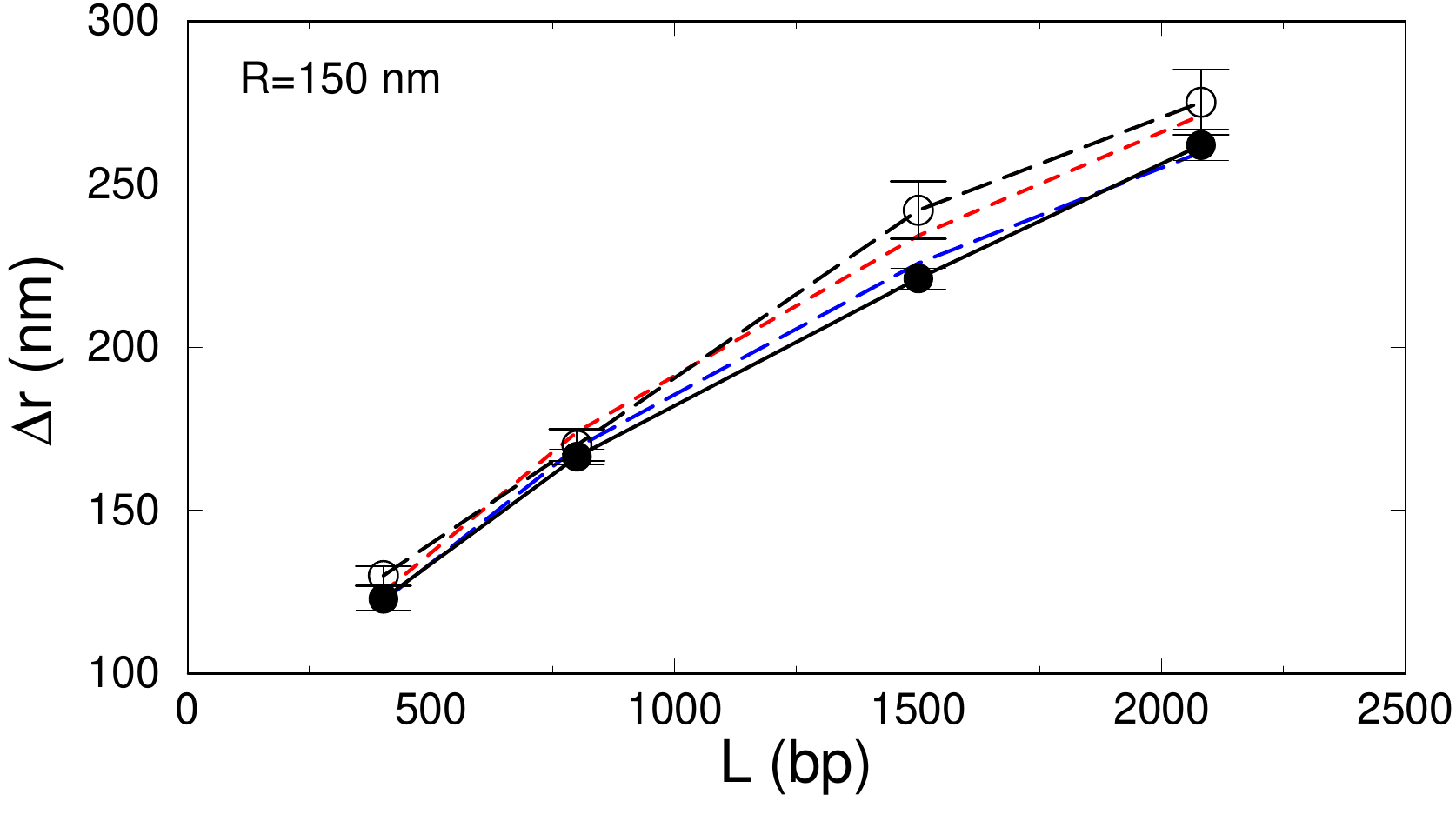}
\caption{\footnotesize Comparison between experimental (solid symbols) and numerical (open symbols) standard deviations $\Delta \rpar=\langle \mathbf{r}_{\parallel}^2 \rangle^{1/2}$. Experimental values have been corrected for systematic errors. Error bars are 95~\% confidence intervals. Red and red dashed lines are predictions from Ref.~\cite{Nelson06} for $\ell_{\rm p}=147$ and 128~bp, respectively.
\label{fig:DeltaR}}
\end{center}
\end{figure}

The two distributions $p(\rho)$ extracted from both TPM experiments and simulations are compared in the Supplementary Data (see figures~S5,~S6). The agreement is systematically good, even though some small discrepancies appear. 
In order to quantify them, we plot in figure~\ref{fig:DeltaR} the values of $\Delta \mathbf{r}_{\parallel}$ as a function of $L$. Interpolating functions evaluated from Monte Carlo simulations have been proposed in Ref.~\cite{Nelson06} that give $\Delta \rpar$ in function of $L$, $R$ and $\ell_{\rm p}$.
We have checked that our numerical results are in agreement with these interpolations when $R\geq100$~nm, even though they were calibrated for $R>190$~nm. In figure~\ref{fig:DeltaR}, one observes that experimental $\Delta \mathbf{r}_{\parallel}$ are shorter than simulated ones. Nelson \textit{et al.}~\cite{Nelson06} have already noticed that assuming the persistence length $\ell_{\rm p}=147$~bp (50~nm) in data analysis leads to larger $\Delta \mathbf{r}_{\parallel}$ than observed. To circumvent this issue, they pointed out that setting $\ell_{\rm p}=128$~bp (43.4~nm) yields better agreement, attributing this low value to the buffer. We also observe that setting $\ell_{\rm p}=128$~bp leads to values of $\Delta \rpar$ closer to our experimental observations for $R=150$~nm.

However, for $R=20$~nm, the interpolating functions proposed in Ref.~\cite{Nelson06} are less reliable (see figure~\ref{fig:DeltaR}). In particular, decreasing the persistence length value to $\ell_{\rm p}=128$~bp is no longer adequate and goes into the wrong direction (147~bp is a better choice). 

Note that the actual value of $\ell_{\rm p}$ is certainly not the unique source of discrepancy between theory and experiments. For example, the interaction between the functionalized surface on the one hand, and the particle (and DNA) on the other hand, is more complex than a simple excluded volume one. Therefore inferring a precise value of $\ell_{\rm p}$ from such experiments is a challenging task~\cite{Brinkers09,chen}. Given the approximations used in the physical modeling, we conclude that the agreement with experiments (discrepancies are less than 15\%) is satisfactory.

\subsection*{Relaxation times of the particle-DNA complex}

In this section, we compare the relaxation times $\tau_\parallel$ extracted from both experimental and numerical trajectories (see Methods).
We first present theoretical considerations on dynamical properties of a DNA fluctuating freely in solution and then give basic insights into the influence of the attached particle. We finally address the issue of the whole particle-DNA-surface system, both experimentally and numerically. 

\medskip\textit{Theoretical aspects on relaxation times and diffusion coefficients}~-- For a free chain of polymerization index $N$ and radius of gyration $R_G$, the correlation time is~\cite{Doi}
\be
\tau_{\parallel} \simeq \frac{R_G^2}{D_{\rm chain}}= \frac{N R_G^2}{D_0}.
\label{rouse}
\ee
In the second equality, we have used the Rouse model ignoring hydrodynamic interactions and valid for flexible chains ($L/\ell_{\rm p}\gg1$), which sets that $D_{\rm chain}=D_0/N$ where $D_0 = k_{\rm B}T/(6 \pi \eta a)$ is the Stokes' formula for the diffusion coefficient of a monomer sphere of radius $a$  in a liquid of viscosity $\eta$. For a Gaussian chain, $R_G^2 \propto N$ and we get $\tau_{\parallel} \sim N^2/D_0$ at large $N$. For a rigid rod, $R_G^2 \propto N^2$ and ${\tau_{\parallel}}\sim N^3/D_0$.

Note that at the level of modeling chosen in the present work where the real polymer is modeled by $N$ connected monomer spheres,  dynamics (and thus $\tau_\parallel$) is not affected by the choice of $N$ as soon as $a \ll \ell_{\rm p}$. Indeed, $R_G$ being insensitive to the choice of $a$ (or $N$), the only dependence in $N$ comes from the ratio $N/D_0$ in equation~(\ref{rouse}). But with the definition of $D_0$, $N/D_0 \propto Na \propto L$, independently of $a$.

How is this chain relaxation time modified when a particle is grafted at one end? 
The time evolution of the DNA polymer and the particle is governed by the over-damped Langevin equation~\cite{Doi,Mano06}
\begin{equation}
\zeta_i \dot{\br}_i(t) = - \nabla_{\br_i} U(\br_0,\ldots,\br_{N-1},\br_{\rm part})+ {\bm \xi}_i(t),
\label{langevin0}
\end{equation}
where $\br_i$ are the position of spheres ($i=0,\ldots,N-1$) and the particle ($i={\rm part}$) (we neglect hydrodynamic interactions as a first hint into the full problem). This equation relates the linear response of each object $i$ to forces applied to it: on the one hand, the derivative of the potential $U$ (sum of pairwise potentials) between the $N$ objects; on the other hand, the stochastic forces ${\bm \xi}_i(t)$ mimic the action of a thermal heat bath and obeys the fluctuation-dissipation relation: $\left\langle{\bm \xi}_i(t)\cdot{\bm \xi}_j(t')\right\rangle=6k_{\rm B}T\zeta_i\delta(t-t')\delta_{ij}$, where the friction coefficient $\zeta_i$ ($=\zeta_0$ for monomer spheres and $\zeta_0R/a$ for the particle) is related to the diffusion coefficient $D_i$ through $D_i=k_{\rm B}T/\zeta_i$.

We relate the diffusion coefficient of the particle-DNA complex, $D_{\rm c}$, to those of the particle alone, $D_{\rm part}$, and the DNA polymer alone, $D_{\rm chain}$. 
To get rid of the interaction forces, we consider the barycenter of the particle-DNA complex
$\br_{\rm c}=(\zeta_{\rm chain} \br_{\rm chain} + \zeta_{\rm part} \br_{\rm part})/\zeta_{\rm c}$ where $\br_{\rm chain}=\sum \br_i/N$ is the DNA center of mass.
Then a linear combination of Langevin equations~(\ref{langevin0}) yields $\zeta_{\rm c}\dot{\br}_{\rm c}(t) =  {\bm \xi}_{\rm c}(t)$, where ${\bm \xi}_{\rm c}={\bm \xi}_{\rm chain} + {\bm \xi}_{\rm part}$. The correlation function $\langle {\bm \xi}_{\rm c}(t)\cdot{\bm \xi}_{\rm c}(t')\rangle=6k_{\rm B}T\zeta_{\rm c}\delta(t-t')$ sets the diffusion coefficient of the particle-DNA complex, $D_{\rm c}$:
\begin{equation}
D_{\rm c}^{-1} = D_{\rm part}^{-1} + D_{\rm chain}^{-1}.
\label{D:complex}
\end{equation}
We emphasize that $D_{\rm c}$ should not be confused with the measured effective particle diffusion coefficient, $D$, because $\br_{\rm c} \neq \br_{\rm part}$. The only case when they are equal is for $D_{\rm part} \ll D_{\rm chain}$. 
By contrast, the relaxation time $\tau_{\parallel}$, obtained by tracking the particle only, is a feature of the dynamics of the whole complex. This is the reason why we focus on this observable in the following.

\begin{figure}[t]
\begin{center}
\includegraphics[height=6cm]{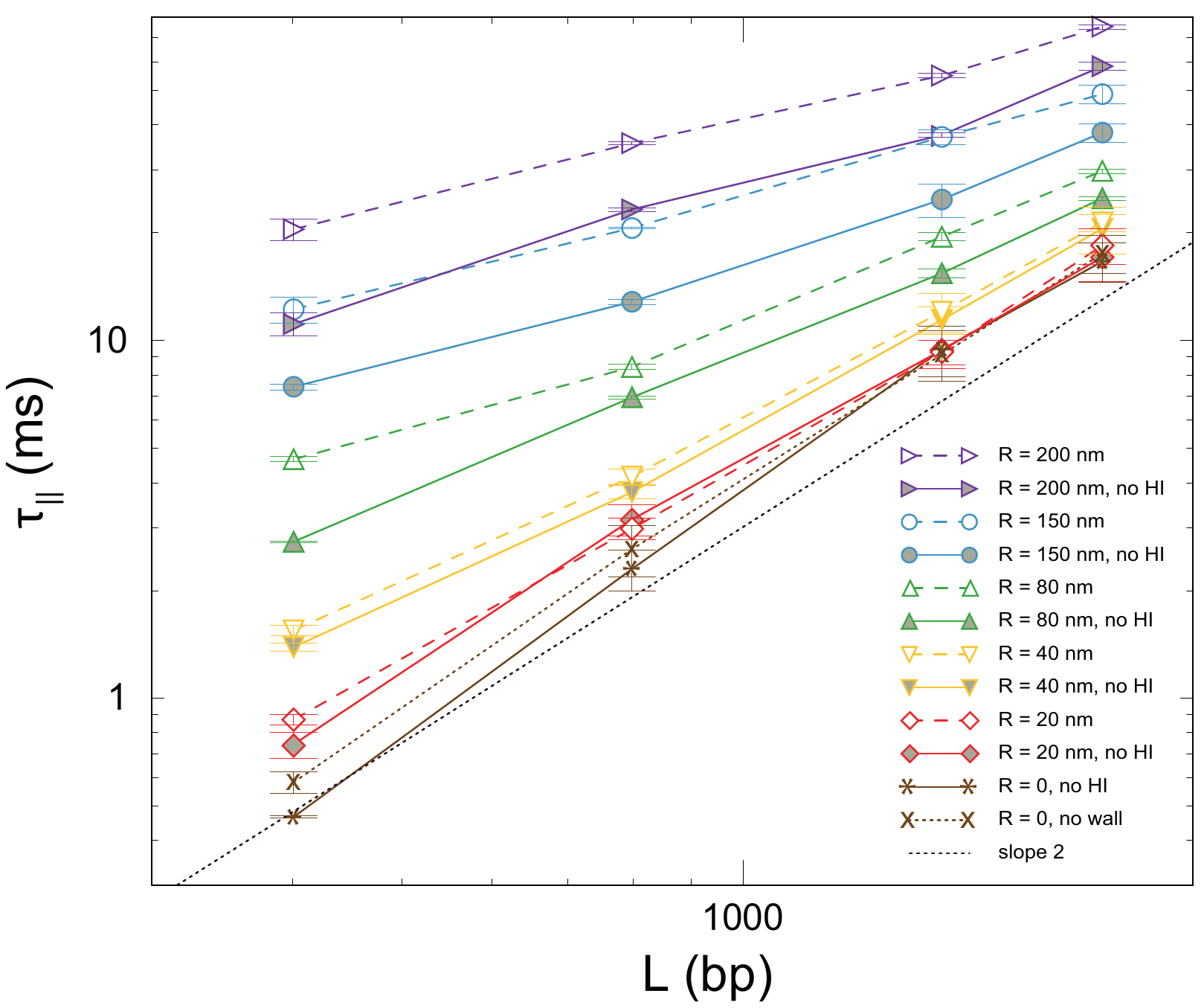}
\caption{\footnotesize Numerical relaxation time $\tau_{\parallel,0}^{\rm sim}$ (solid lines: without wall hydrodynamic friction (no HI); $N=50$, except for $L=401$~bp, $R=150$~nm and $L\leq798$~bp, $R=80$ and 200~nm, where $N=25$) and $\tau^{\rm sim}_\parallel$ (dashed lines: with wall hydrodynamic friction;  $N=25$ for $L=401$ and 798~bp, $N=50$ for $L=1500$ and 2080~bp) \emph{versus} the polymer length $L$ for various particle sizes $R$, in log-log coordinates. In the case $R=0$, $\tau_{\parallel}$ is also plotted when there is no hard wall. The dotted line of slope 2 shows the expected scaling for long DNAs (Rouse model for a random walk). \label{figMC}}
\end{center}
\end{figure}

\medskip\textit{Simulation results without hydrodynamic interactions}~-- We first display in figure~\ref{figMC} the relaxation times fitted from numerical trajectories simulated without hydrodynamic interactions with the surface. They are denoted by $\tau_{\parallel,0}^{\rm sim}$. We checked that finite-$N$ effects appear to be negligible for $a \ll \ell_{\rm p}$ (data not shown). We also explored the reference case $R=0$, where no particle is attached to DNA. We have seen above that the Rouse model predicts $\tau_{\parallel} \sim L^2$ at large $L$. Figure~\ref{figMC} shows that this regime is also a good approximation even for the finite sizes considered here.
We also plot in figure~\ref{figMC} the relaxation times in the case where $R=0$ in the absence of hard wall at $z=0$. One can see that the wall does not hinder DNA dynamics significantly.

When $R>0$, the particle slows down dynamics because viscous drag increases with the particle size. However our simulations predict that DNAs are not significantly affected by a small particle ($R=20$~nm). A medium particle, $R=40$~nm (resp. 80~nm) does not perturb DNAs of lengths $L\geq 800$~bp (resp. 1500~bp). In the other cases, increasing $R$ from 0 to 150~nm at fixed $L$ monotonically increases $\tau_{\parallel,0}^{\rm sim}$, by a factor ranging from 2 ($L=2080$~bp) to 10 ($L=401$~bp). 

In the Supplementary Data (see figure S8), we compare our numerical values of figure~\ref{figMC} without wall hydrodynamic friction to simple calculations where the DNA molecule is modeled by an ideal spring. In this case, the relaxation time would be equal to the ratio of the 
particle (or particle-DNA complex) friction coefficient to the DNA spring constant~\cite{Brinkers09}. This approach appears to be insufficient because the interplay between the polymer, the bead and the wall is more complex than this simple image.

\medskip\textit{Simulation results with hydrodynamic interactions with the surface}~-- Hydrodynamic interactions between the particle-DNA complex and the surface are implemented in the numerical code using  equation~(\ref{DperpDpar}). Figure~\ref{figMC} shows that the diffusion is slowed down by hydrodynamic corrections for large particles ($R\geq80$~nm) but is weakly affected for small ones ($R\leq40$~nm). Indeed, small particles are far away from the surface, in that sense that $R \ll z$ and $b/h=R/(z+a)$ is generically small in equation~(\ref{DperpDpar}). Only the first DNA spheres, for which $b/h=a/(z_i+a)$ is of order 1, are slowed down by the proximity of the surface but this is not sufficient to affect the whole dynamics. 
This is related to the observation that the exact modeling of the DNA-surface joint is not a relevant issue (see Supplementary Data).
By contrast, when $R$ is large, $R/(z+a)$ is close to 1 for the particle itself, and hydrodynamic corrections play a significant role. In the following, we use these numerical results with hydrodynamic interactions.

\begin{figure}[t]
\begin{center}
\includegraphics[height=6cm]{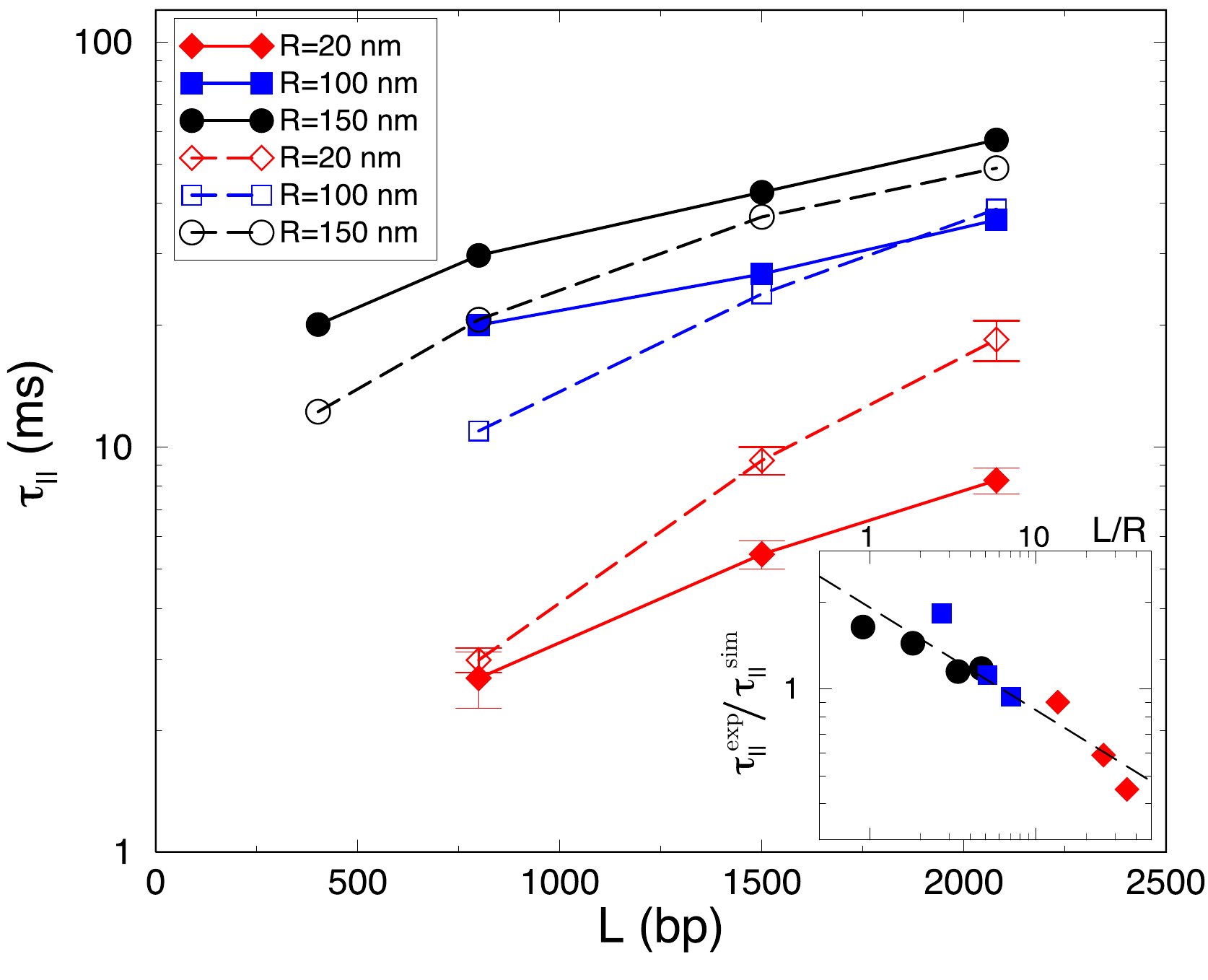}
\caption{\footnotesize Experimental ($\tau_{\parallel}^{\rm exp}$, solid symbols) and numerical ($\tau_{\parallel}^{\rm sim}$, using $z$-dependent diffusion coefficients, open symbols) relaxation times for different DNA lengths, $L$, and particle radius, $R$, in linear-log coordinates. Displayed error bars are explained in the text for numerical data and are 95~\% confidence intervals for experimental data. Other error bars are smaller than symbol sizes. \emph{Inset:} Ratio $\tau_{\parallel}^{\rm exp}/\tau_{\parallel}^{\rm sim}$ \emph{versus} $L/R$ with the same symbols as above, in log-log coordinates. The dashed line shows the best linear regression, with slope $-0.355 \simeq -1/3$. }
\label{tau:compare}
\end{center}
\end{figure}

\medskip\textit{Comparison between experimental and numerical relaxation times}~-- Figure~\ref{tau:compare} displays our experimental [corrected from detector averaging effects using equation~(\ref{zerel})] and numerical values of the relaxation times $\tau_{\parallel}$ extracted from functions $C(t)$ or MSD$(t)$ (see Materials and Methods). Experimental and numerical values are found in good agreement, with ratios of experimental to numerical values varying from 0.5 to 2. These ratios are plotted in the Inset of figure~\ref{tau:compare}. They appear to be well correlated to the ratios $L/R$ and to follow the approximate power-law: $\tau_{\parallel}^{\rm exp}/\tau_{\parallel}^{\rm sim} \propto (R/L)^{1/3}$.
This suggests that the observed discrepancies are not due to statistical errors (as supported by the small error-bars), but have a physical origin.
When $R \ll L$, $\tau_{\parallel}^{\rm exp}<\tau_{\parallel}^{\rm sim}$, i.e. experiments are faster than simulations. If $R \geq L/5$, then $\tau_{\parallel}^{\rm exp}\geq\tau_{\parallel}^{\rm sim}$. 
Indeed, if one neglects hydrodynamics,  the measured diffusion coefficient of the particle-DNA complex, $D_{\rm c}$, is related to the particle one, $D_{\rm part}$, and to the polymer one, $D_{\rm chain}$, through equation~(\ref{D:complex}):
$D_{\rm c}^{-1} = D_{\rm part}^{-1} +D_{\rm chain}^{-1}$.
Thus the particle does not slow down the complex provided that $D_{\rm part} \gg D_{\rm chain}$. Now $D_{\rm part} = K/R$ with $K=k_{\rm B}T/(6 \pi \eta)$, and $D_{\rm chain}=K/(Na)=2K/L$ in the Rouse approximation. If $2R \ll L$, then $D_{\rm part} \gg D_{\rm chain}$ and the DNA dominates the particle-DNA dynamics, as if the particle were absent. But in this case, hydrodynamic interactions between DNA segments should be taken into account. Since such interactions are known to accelerate polymer dynamics~\cite{Doi,Mano06}, one expects $\tau_{\parallel}$ to be shorter than in simulations, as indeed observed in experiments.
Conversely, if $2R \gg L$, then the particle dominates the particle-DNA dynamics, as if the role of the DNA polymer was limited to its spring properties. We have simplified the complex hydrodynamics of this particle in the vicinity of a wall through the expansions in equation~(\ref{DperpDpar}). But these expansions are valid in the small $b/h$ limit only and we cannot expect them to be correct when $b/h\simeq 1$, which is the case when $2R \gg L$. In Ref.~\cite{Happel} Section~7-4, it is proved that equation~(\ref{DperpDpar}) overestimates $D_\perp$ and $D_\parallel$ in this case. Consequently, $\tau_\parallel$ is underestimated as observed. Describing correctly the particle dynamics would require a full integration of all hydrodynamic interactions, between the monomer spheres, the particle and the surface. This is beyond the scope of this work.

\subsection*{Choice of TPM parameters to infer DNA dynamics}

The preceding analysis leads us to suggest the following recommendations to extract the most valuable information on DNA dynamics from TPM experiments.

\medskip\textit{Small particles of radius $R \lesssim 20$~nm do not slow down DNAs as short as 400~bp}~-- Comparing simulation data for $R=0$ and $R>0$ (figure~\ref{figMC}) shows  that  the presence of the particle does not increase the relaxation times for small particles or long DNAs. Indeed, provided that  $R \ll L/2$, dynamics are dominated by the polymer. Accordingly, figure~\ref{figMC} shows that if $R < L/6$ then $\tau_{\parallel}$ is less than twice the relaxation time of DNA without particle, which demonstrates that slowing-down is weak or moderate in this case. In particular, slowing down is very weak when $R=20$~nm for $L=401$ to 2080~bp.

They are three contributions to hydrodynamics: i) slowing down of the chain and the particle, due to the no-slip condition at the surface, ii) hydrodynamic interactions between the chain and the particle, and iii) interactions between monomer spheres. For creeping flows, these three contributions are additive~\cite{Happel}. We have shown above that, for small $R$, (i) affects only very weakly the relaxation times. Contributions (ii) and (iii), which are not taken into account in simulations, moderately accelerate the dynamics of the particle-DNA complex, as corroborated by experimental measurements (diamonds in figure~\ref{tau:compare}). By linearity, we thus expect that neither the wall nor the small particle slow down DNA dynamics as compared to a DNA chain fluctuating freely in solution.

\medskip\textit{Taking into account exposure times and acquisition periods}~-- In the Materials and Methods section, we have explained how the finiteness of the exposure time, $T_{\rm ex}$, modifies the root-mean-square excursion of the particle, $\Delta \mathbf{r}_{\parallel}$, the particle-DNA relaxation time, $\tau_\parallel$, and the effective particle diffusion coefficient, $D$  (when $T_{\rm ex}\leq 3 \tau_{\parallel}$). We have prescribed how to recover their real values rigorously, contrary to the rough approximation of equation~(\ref{ell:ellm0}) proposed by Ref.~\cite{Towles09}. Indeed, the formula given in this anterior work largely overestimates the averaging effect, thus requiring the introduction of a phenomenological time scale $\tau^*$.

To this respect, we have analyzed two particle-DNA complexes ($R=150$~nm and $L=2080$ and 401~bp) with two different exposure times, $T_{\rm ex}=5$ and $40$~ms and the same acquisition period $T_{\rm ac}=40$~ms in both cases. The measured values before  and after applying equations~(\ref{zerel},\ref{ell:ellm0}) are given in Table~\ref{tab:corr}. We observe that the corrected values are in excellent agreement for both DNA lengths.
These values validate our method to correct detector averaging effects the experimental data.
\begin{table}[t]
\begin{tabular}{cccccc}
\hline
L(bp) & $T_{\rm ex}$(ms) & $\tau_{\rm m}$(ms) & $\Delta \br_{\parallel,{\rm m}}$(nm)& $\tau_\parallel$(ms) & $\Delta \br_{\parallel}$(nm) \\ 
\hline
2080 & 40 & 69.5 & 223 & 56.1 & 250 \\
     & 5 &  59.1 & 258 & 57.4 & 261 \\ 
\hline
401  & 40 & 32.8 & 91  & 19.5 & 122 \\
     &  5 & 21.7 & 118 & 20.0 & 123 \\
\hline
\end{tabular}
\caption{\footnotesize Relaxation times and movement amplitudes before ($\tau_{\rm m}$ and
$\Delta \br_{\parallel,{\rm m}}$) and after correction ($\tau_\parallel$ and $\Delta \br_{\parallel}$) of detector time averaging effects, using equations~(\ref{zerel},\ref{ell:ellm0}), for two DNA lengths $L$ and two exposure times $T_{\rm ex}$. Here the particle radius is $R=150$~nm.
\label{tab:corr}}
\end{table}

The values of $\tau_{\parallel}$ given in figure~\ref{tau:compare} will help experimentalists to anticipate the appropriate required values of the exposure time $T_{\rm ex}$ in future works. Owing to the complexity of the problem, no simple formula can be given that expresses $\tau_{\parallel}$ as a function of $R$ and $L$. In practice, a linear interpolation of the values of figure~\ref{tau:compare} will provide a good estimate of $\tau_{\parallel}$ for other values of $R$ and $L$, in the range studied.

However, reducing $T_{\rm ex}$ may remain insufficient when one is interested in measuring $\tau_{\parallel}$. The acquisition period $T_{\rm ac}$ being imposed by the camera, the only points available to fit MSD$(t)\propto 1-\exp(-t/\tau_{\parallel})$ or $C(t)\propto\exp(-t/\tau_{\parallel})$ plots are measured at discrete values $t=0,T_{\rm ac},2T_{\rm ac}$ and so forth. So if $T_{\rm ac}$ is large as compared to $\tau_{\parallel}$, MSD($T_{\rm ac}$) and $C(T_{\rm ac})$ have already reached their asymptotic values, respectively $2\Delta \mathbf{r}_{\parallel}^2$  and 0, making any fitting procedure hopeless. If one is interested in dynamical properties, the true technical limit obviously remains the camera acquisition period, which must be of the same order of magnitude as $\tau_{\rm m}$ or smaller, 
typically $T_{\rm ac} \lesssim 2\tau_{\rm m}$ (as detailed in the Supplementary Data).
  
\medskip\textit{Measuring variations of $\langle \mathbf{r}_{\parallel}^2 \rangle$ in real time}~-- Beyond the measurement of equilibrium statistics or relaxation rates, TPM experiments can also be used to monitor DNA conformational changes. To what extent can those conformational changes be monitored in real time? In TPM, such a change is characterized by a variation of the movement amplitude, $\Delta\br^2_\parallel=\langle \mathbf{r}_{\parallel}^2 \rangle$, due to a given molecular event such as protein binding or DNA looping, leading to a variation of the apparent DNA length. But measuring the mean value $\langle \mathbf{r}_{\parallel}^2 \rangle$ supposes to average $\mathbf{r}_{\parallel}^2$ on a sufficiently long time interval, of duration again denoted by $T_{\rm av}$~\cite{Pouget06}. By definition of the relaxation time $\tau_{\parallel}$, the average will be accurate provided that $T_{\rm av} \gg \tau_{\parallel}$. Once $\tau_{\parallel}$ is known, what averaging time $T_{\rm av}$ should be  chosen in practice? In other words, what is the minimum time scale of conformational changes that TPM can reach? Formally speaking, $\langle \br_{\parallel}^2 \rangle \simeq \int_{t_0}^{t_0+T_{\rm av}} \mathbf{r}_{\parallel}^2(t) \; {\rm d}t/T_{\rm av}$. The relaxation time of $\br_{\parallel}^2 $, $\tau_{\parallel,2}$, is again defined by $\langle \br_{\parallel}^2 (s+t)\br_{\parallel}^2 (s)\rangle_s - \langle \br_{\parallel}^2  \rangle^2 \approx \sigma^2_{\br^2_\parallel} \exp(-t/\tau_{\parallel,2})$ (we use the notation $\sigma_{\br^2_\parallel}$ for the r.m.s. of $\br^2_\parallel$, to avoid confusion with $\Delta \br_{\parallel}^2$, the variance of $\br_\parallel$). Then the statistical error on the measurement of $\langle \br_{\parallel}^2  \rangle$ (68\% confidence interval), denoted by ${\rm Err}(\br_{\parallel}^2)$, can be quantified as follows~\cite{Newman}: if $\tau_{\parallel,2} \gtrsim T_{\rm ac}$, then ${\rm Err}(\br_{\parallel}^2)= \sqrt{2\tau_{\parallel,2}/T_{\rm av}}\sigma_{\br^2_\parallel}$. This relation takes into account statistical correlations between successive frames, measured by $\tau_{\parallel,2}$. If $\tau_{\parallel,2} \ll T_{\rm ac}$, then ${\rm Err}(\br_{\parallel}^2)= \sqrt{T_{\rm ac}/T_{\rm av}}\sigma_{\br^2_\parallel}$, because all frames are statistically independent. 

Now, what is the relationship between $\tau_{\parallel,2}$ and $\tau_{\parallel}$? Since $\br_{\parallel}^2 =x^2+y^2$, $\tau_{\parallel,2}=\tau_{x^2}=\tau_{y^2}$, the (identical) relaxation times of the observables $x^2$ and $y^2$. If $x$ and $y$ were Gaussian random variables, then $\tau_{x^2} = \tau_{x}/2 = \tau_{\parallel}/2$. But $x$ is not exactly Gaussian. In practice, we have measured numerically in our case that $\tau_{\parallel,2} \simeq \tau_{\parallel}/c$ with $c$ ranging from 2.5 to 3.5, depending on $L$ and $R$. Thus $\tau_{\parallel,2} \leq \tau_{\parallel}/2$. Furthermore, we have also measured numerically that $\sigma_{\br^2_\parallel}< \langle \br_{\parallel}^2 \rangle$ for the values of $L$ and $R$ studied here. Were $\mathbf{r}_{\parallel}$ be exactly Gaussian, a simple calculation shows that $\sigma_{\br^2_\parallel}= \langle \br_{\parallel}^2 \rangle$ (because $\mathbf{r}_{\parallel}$ is two-dimensional). Thus in the regime where $\tau_{\parallel,2} \gtrsim T_{\rm ac}$, 
\begin{equation}
\lambda\equiv\frac{{\rm Err}(\br_{\parallel}^2)}{\langle \br_{\parallel}^2  \rangle}\leq \sqrt{\frac{\tau_{\parallel}}{T_{\rm av}}}.
\label{rel:err}
\end{equation}
The equality defines the limiting value $T_{\rm av}^* =\tau_\parallel/\lambda^2$. 
For example, if $T_{\rm ac}=10$~ms and $\tau_{\parallel} \approx 30$~ms, $T_{\rm av} \geq T_{\rm av}^*= 100 \; \tau_{\parallel} \approx 3$~s yields a relative statistical error $\lambda\leq10\%$.  If the conformational changes lead to variations of $\lan \br_{\parallel}^2\ran$ larger than 10\% (the statistical noise), then a thresholding method~\cite{Pouget06,Zurla07,wuite,dixit} can detect them in real time (provided that their time scale is higher than 3~s). Using shorter DNAs, smaller particles, and a faster acquisition device improves the time resolution. Concrete illustrative examples are given in figures~S9, S10 and thereafter.

\medskip {\textit{Numerical example of conformational changes detection: DNA looping/unlooping}~-- To further illustrate TPM capabilities in terms of detection of conformational changes, we propose the following example. It is related to TPM experiments where DNA loops are promoted by proteins binding to
two specific DNA loci, and DNA conformation alternates between looped and unlooped
states~\cite{Pouget06,Rutkauskas,wuite}. In the looped state, the effective DNA contour length is
shorter, resulting in a smaller $\Delta\rpar^2$. Switching between looped and unlooped states can be detected in
real time provided that their lifetime is long enough. One of the
purposes of the present example is to show how the minimum detectable
lifetime can be determined {\em a priori} from the knowledge of $\tau_\parallel$.}
\begin{figure}[h]
\begin{center}
\includegraphics[width=0.9\linewidth]{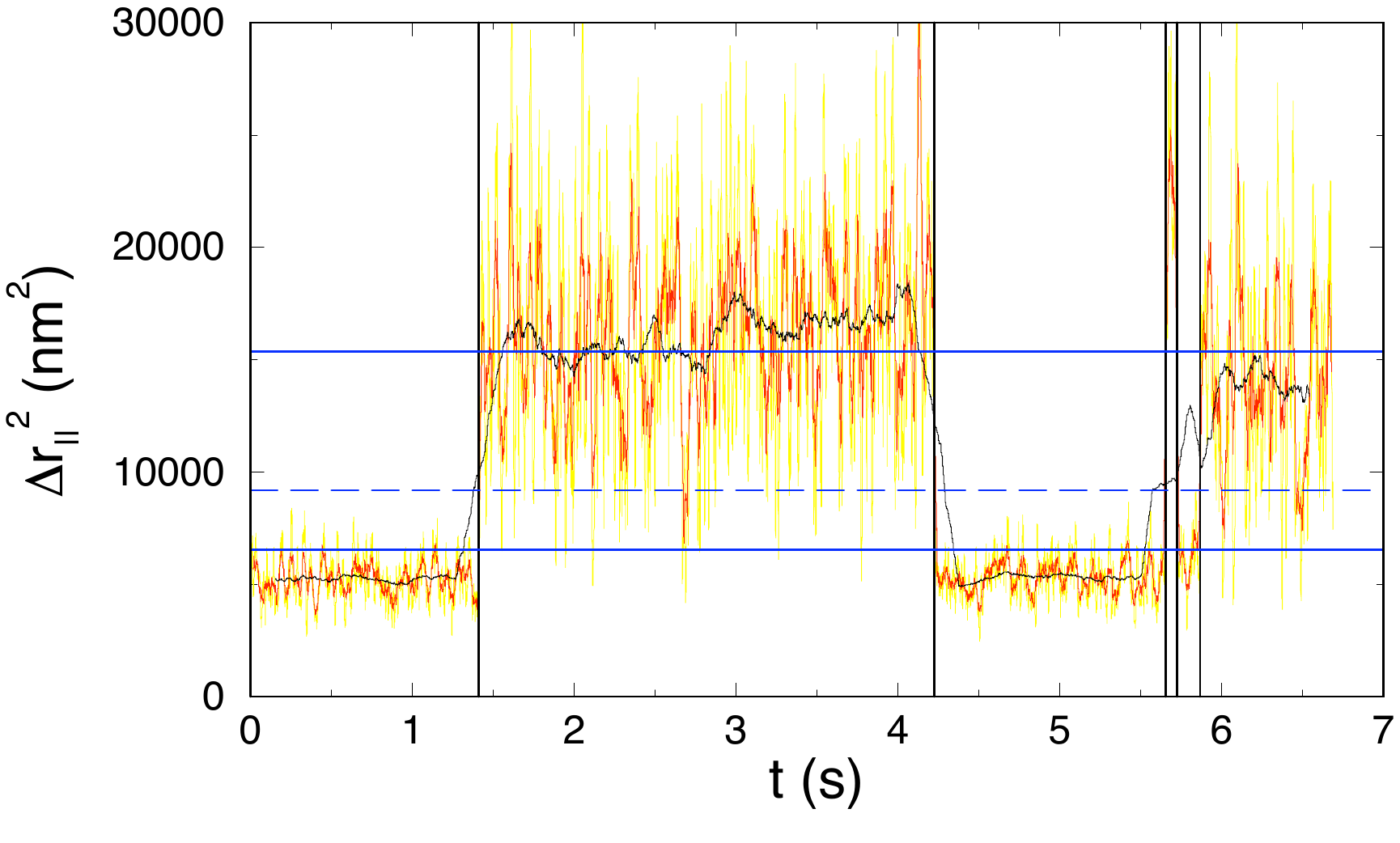}\\
\includegraphics[width=0.9\linewidth]{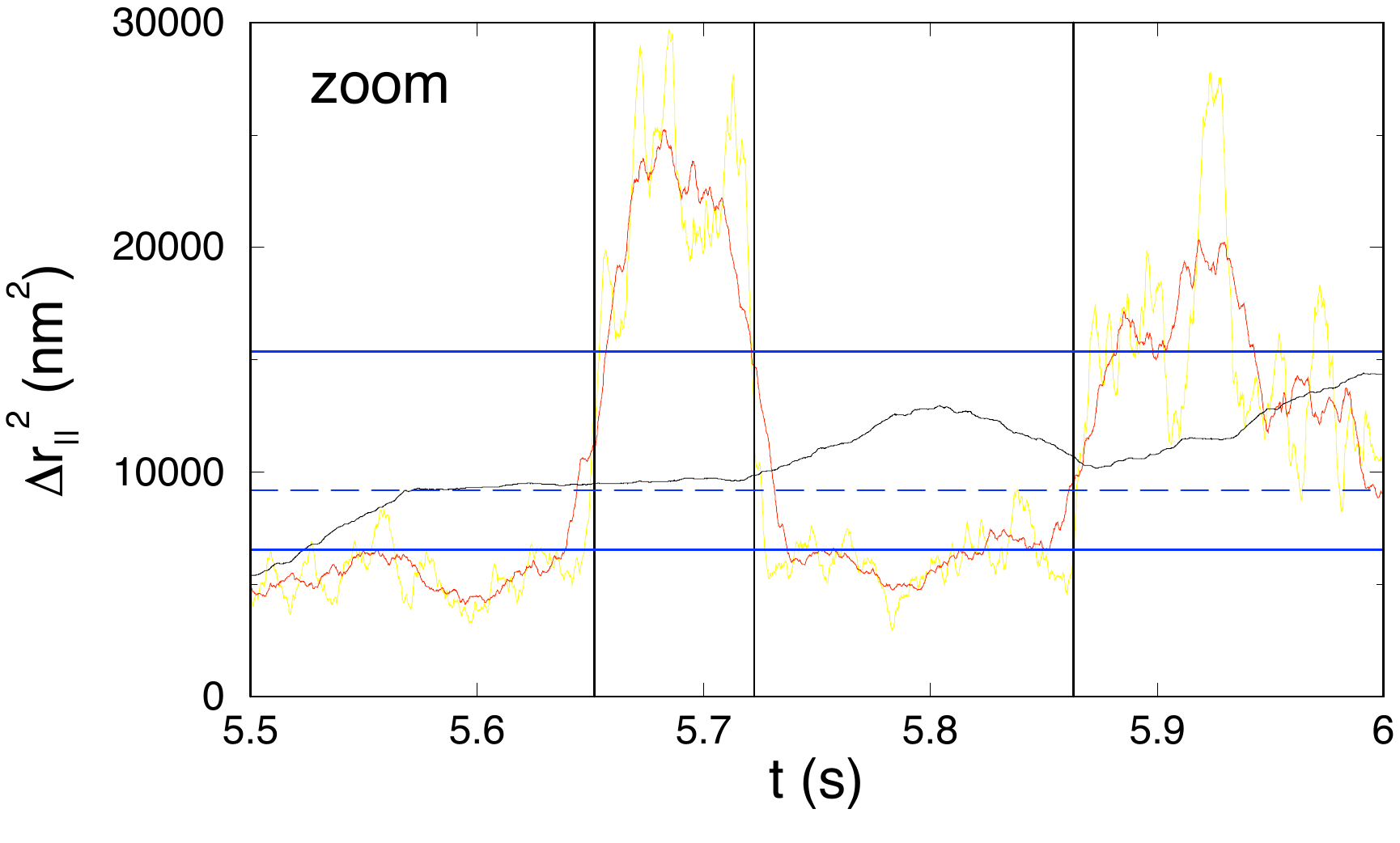}
\caption{Simulated amplitude of movement $\Delta
  \mathbf{r}_\parallel^2(t)$ for $L=798$~bp, $R=20$~nm and three averaging times
  $T_{\rm av}=3$ (yellow), 30 (red) and 300~ms (black). The
  vertical lines indicate the looping and unlooping events in the
  simulation. The horizontal solid lines show the expected values
  of $\Delta \mathbf{r}_\parallel^2$ in the looped (bottom) and
  unlooped (top) states. The dashed line indicates the threshold separating
  these two states for detection purposes. The bottom
  plot is a zoom of the top one on a time interval where
  conformational changes are very close. 
\label{looping}}
\end{center}
\end{figure}

We focus on $L=798$~bp and $R=20$~nm with $N=25$ beads (denoted by $B_0,\ldots,B_{24}$). Hydrodynamic
interactions with the wall are taken into account. The numerical relaxation time is $\tau_{\parallel}=2.98\simeq 3$~ms (see figure~\ref{figMC}), and $\Delta \mathbf{r}_\parallel=124$~nm. The two loci where protein
binding occurs are situated on beads $B_6$ and $B_{18}$. Thus the
looped DNA is roughly equivalent to a 400~bp long DNA, with
$\Delta \mathbf{r}_\parallel=81$~nm. 
In the simulation, looping is promoted by an
attractive quadratic potential between beads $B_6$ and $B_{18}$:
$V=2 k_{\rm B} T(\mathbf{r}_{18}-\mathbf{r}_{6})^2$.
When we turn on (resp. off) the potential, the DNA molecule switches into the looped (unlooped)
state. In figure~\ref{looping}, $\Delta \mathbf{r}_\parallel^2$ is plotted in
function of time for different averaging-interval lengths $T_{\rm av}$. To
set the threshold separating looped and unlooped states, we choose the
condition $(1+\lambda)\Delta \mathbf{r}_{\parallel,\rm looped}^2=
(1-\lambda)\Delta \mathbf{r}_{\parallel,\rm unlooped}^2$, leading to
$\lambda=0.4$. This means that a maximum relative error of 40\% is allowed if one wants to detect unambiguously
looped and unlooped states. Using equation~(\ref{rel:err}), this leads to the limiting
value $T_{\rm av}^* =\tau_\parallel/\lambda^2  =  19$~ms (provided that 
$T_{\rm ac} \lesssim \tau_{\parallel,2} \approx 1$~ms). As illustrated in
figure~\ref{looping}, choosing $ T_{\rm av}<T_{\rm av}^*$ leads to very numerous false
detections of looping/unlooping. On the contrary, increasing $T_{\rm av}$ above the threshold $T_{\rm av}^*$
makes false detections unlikely.
However, as is shown in figure~\ref{looping}, increasing $T_{\rm av}$ increases the error on the determination of the moment when the \textit{transition} between looped and unlooped states occurs. Hence the looped/unlooped state duration measure is less precise (see the zoom). This precision is on the order of magnitude of $T_{\rm  av}$. Furthermore, because of averaging, only durations larger than $T_{\rm av}$ can be accurately detected. For instance, in figure~\ref{looping} (zoom), the looped state between $t=5.72$~s and $t=5.86$~s is not detected because its duration is shorter than $T_{\rm av}$ for $T_{\rm av}=300$~ms. Therefore $T_{\rm av}$ must be large enough to distinguish between looped and unlooped
states, but it must be as short as possible to detect the shortest
lifetimes. Equation~(\ref{rel:err}) provides the optimal choice. Of course, the
shortest $\tau_\parallel$, and therefore  particles of radius $R<L/6$, will provide the best detection capabilities. Another illustrative example is given in figure~S8 in the case of DNA bending induced, for instance, by protein binding.
% -> SM ??? In this case, the bent/non-bent conformations cannot be distinguished by looking at the distribution $p(\rho)$, because it is not bimodal. But they can be distinguished by following $\Delta \mathbf{r}_{\parallel}^2(t)$, provided that one knows in advance what averaging time $T_{\rm av}$ to use.}

\section*{CONCLUSION AND OUTLOOK}

This work focuses on the influence on DNA dynamics of the TPM setup geometry, namely the attachment of the DNA tether ends to the surface and the labeling particle. To what extent DNA characteristic times can be inferred from TPM observations? The answer to this complex issue cannot be summarized in simple formulas because many parameters are involved and interrelated, in particular: the particle size $R$, the DNA contour length $L$, the DNA persistence length $\ell_p$, the exposure time $T_{\rm ex}$, the acquisition period $T_{\rm ac}$, and the averaging time $T_{\rm av}$. Here, we used numerical simulations, the results of which were compared to experimental measurements.  A good quantitative agreement (see figures~\ref{fig:DeltaR} and~\ref{tau:compare}) allows us to draw the following two major conclusions, summarizing the survey of experimental constraints on dynamical TPM examined in the last section of our Results and Discussion.

\noindent 1) For DNA lengths $L$ usually used in TPM experiments, i.e., from hundreds to thousands base pairs, the DNA dynamics is weakly perturbed by particles of radius $R<L/6$. As long as $L$ is larger than 400~bp, the influence of the particle is negligible if $R\leq20$~nm. Indeed, in this case, hydrodynamic friction induced by the particle and the surface are weak and the DNA characteristic relaxation time in the plane parallel to the surface, $\tau_{\parallel}$, is almost identical to the case where the particle and the surface are absent.

\noindent 2) In the TPM analysis of DNA conformational changes, the use of particles of radius smaller than 20~nm allows to detect molecular events with a time resolution as short as 20~ms in favorable cases. We believe that such a substantial improvement will provide a more detailed view on DNA-protein interaction processes involving DNA bending~\cite{Pouget06,dixit}, looping~\cite{Pouget06,Rutkauskas,wuite} or migration of a Holliday junction~\cite{Grigoriev} in the future. We have shown in the simulation of looping/unlooping events that the knowledge of the relaxation time $\tau_\parallel$ is a prerequisite in order to extract the most relevant information from TPM experiments exploring DNA conformational changes: if $\tau_\parallel$ is not known, it is not possible to decide whether a detected variation of $\Delta \mathbf{r}_\parallel^2$ is a real conformational change or a spurious statistical fluctuation.

On the theoretical side, the scaling law $\tau_\parallel^{\rm exp}\propto(R/L)^{1/3}\tau_\parallel^{\rm sim}$  in figure~\ref{tau:compare} provides a clue that hydrodynamic interactions play a crucial role. Brownian dynamics simulations (see Supplementary Data) will be the next step in order to elucidate this issue, at least for the fastest complexes (short chains and small beads). They ought to include hydrodynamic interactions between beads~\cite{Mano06,Petrov06}, usually taken into account through the Rotne-Prager tensor, and  hydrodynamic images induced by the presence of the particle and the surface to ensure the no-slip condition on these surfaces~\cite{Mano06}.

\smallskip 
\noindent {\footnotesize {\bf Acknowledgments}~-- We thank Denis Gotta for his participation to this work as part of his Science Degree project, Vincent Croquette and Rob Philipps for stimulating discussions.}

\setcounter{figure}{0}
\setcounter{equation}{0}

\makeatletter
\renewcommand\theequation{S\arabic{equation}}
\@addtoreset{equation}{section}
\makeatother

\makeatletter
\renewcommand\thefigure{S\arabic{figure}}
\@addtoreset{equation}{section}
\makeatother

\begin{center}
{\large \bf
Supporting material to the paper\\
Probing DNA conformational changes with high temporal resolution by Tethered Particle Motion}

\bigskip

Manoel Manghi, Catherine Tardin, Julien Baglio, Philippe Rousseau, Laurence Salom\'e, and Nicolas Destainville
\end{center}

\subsection*{Tethered Particle Motion (TPM) Experiments}

The number of experimental particle trajectories followed in TPM experiments for various DNA lengths $L$, exposure times $T_{\rm ex}$ and particle radii $R$ are given in the following table (distances in nm, times in ms):

%\begin{table}[h]
\begin{center}
\begin{tabular}{|c||c|c|c|c|}
\hline
DNA length & $T_{\rm ex}=40$  & \multicolumn{3}{c|}{$T_{\rm ex}=5$}\\ \cline{3-5}
 $L$ (bp) & $R=150$ & $R=20$ & $R=100$ & $R=150$\\
\hline
401& 46 &  &  & 32 \\
\hline
798 & & 34 & 24 & 32 \\
\hline
1500 & & 38 & 19 & 40 \\
\hline
2080 & 33 & 60 & 41 & 29  \\
\hline
\end{tabular}
\end{center}

An additional set of 35 trajectories were followed for $L=2080$~bp, $R=150$~nm, $T_{\rm ex}=5$~ms and $T_{\rm ac}=12.8$~ms.

\subsection*{Fitting procedure}

\begin{figure}[t]
\begin{center}
\includegraphics[height=6cm]{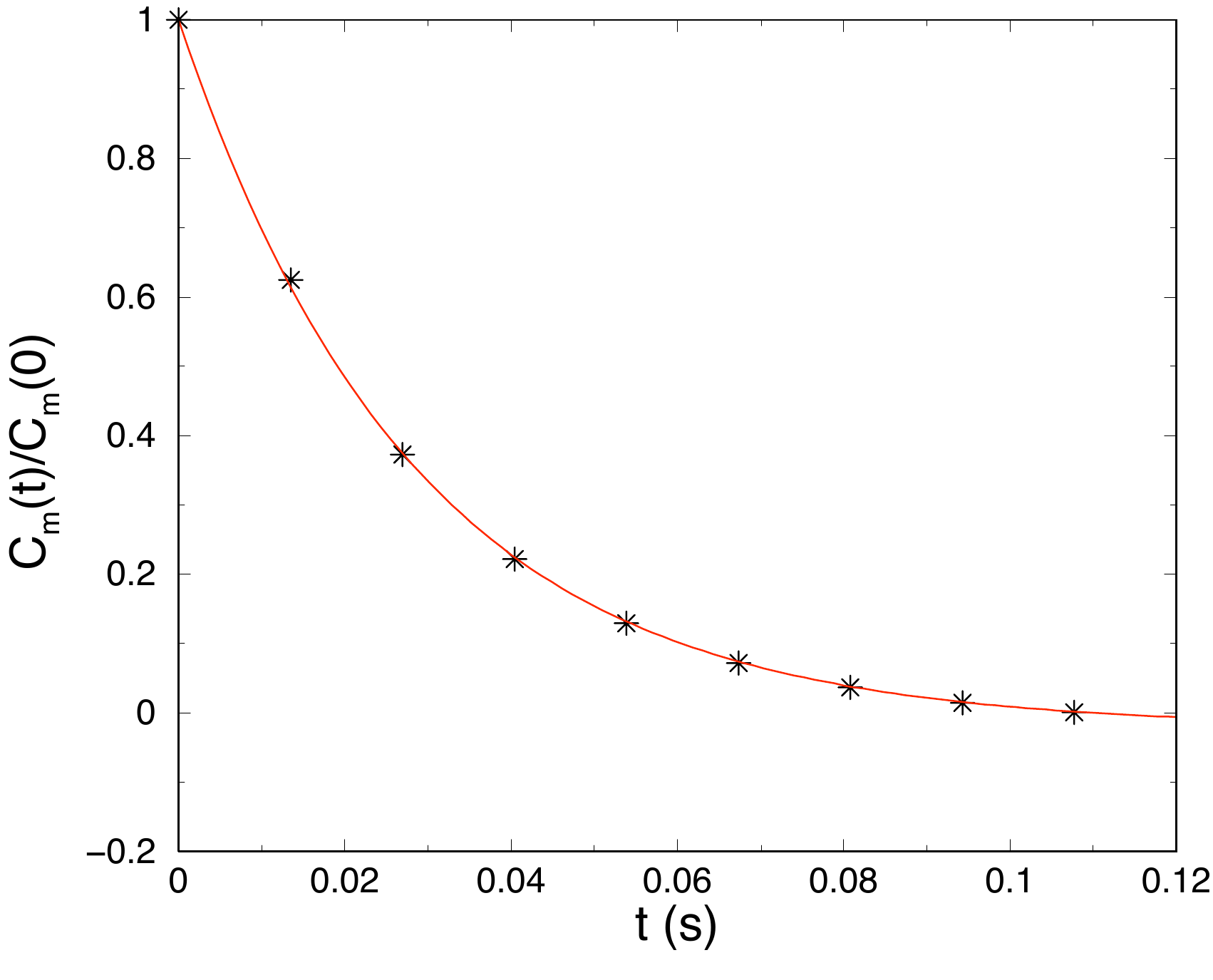} \\
\includegraphics[height=6cm]{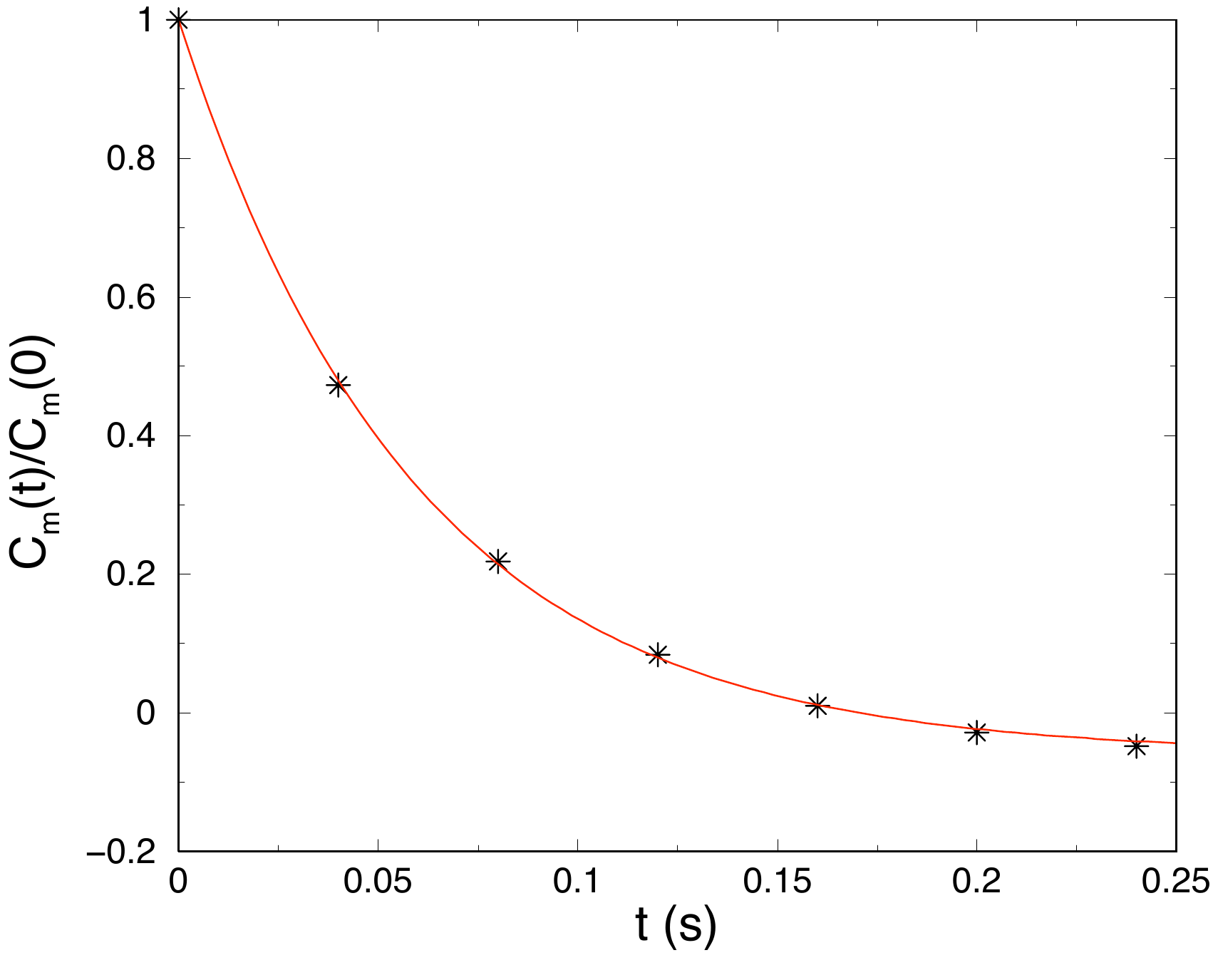}
\caption{\footnotesize Two fits of experimental correlation functions  $C_{\rm m}(t)/C_{\rm m}(0)$ according to equation~(6). Stars are experimental points and red curves are fits. Left: $L=1500$~bp, $R=100$~nm, $T_{\rm ac}=13.5$~ms and $T_{\rm ex}=5$~ms. The fit gives $\tau_{\rm m}=25.6$~ms. Right: $L=2080$~bp, $R=150$~nm, $T_{\rm ac}=40$~ms and $T_{\rm ex}=5$~ms. The fit gives $\tau_{\rm m}=59.0$~ms.
\label{fits:C:t}}
\end{center}
\end{figure}

The fitting procedure to extract the relaxation time $\tau_\parallel$ from experimental data is presented in the Material and Methods section of the paper. The fitting form is given in equation~(6), 
{which we first demonstrate. The modified correlation function $C_{\rm m}(t)$ is obtained after subtraction of noise, that we thus ignore here, and of instrumental drift. To subtract  instrumental drift, the measured position $\rpar$ is first averaged on a time interval of duration $T_{\rm av}$ centered at $t$: 
\begin{equation} 
\mathbf{\bar r}_\parallel (t) \equiv \frac1{T_{\rm av}} \int_{t-T_{\rm av}/2}^{t+T_{\rm av}/2}
 \rpar(\tilde t) {\rm d}\tilde t,
\label{rbar}
\end{equation}
and then subtracted from $\rpar(t)$. Thus
\begin{equation}
C_{\rm m}(t) = \langle [\rpar(s+t)-\mathbf{\bar r}_\parallel (s+t)]\cdot[\rpar(s)-\mathbf{\bar r}_\parallel (s)] \rangle_s.
\end{equation}
We assume that $\tau_{\rm m} \ll T_{\rm av}$ and $t\ll T_{\rm av}$, which is satisfied in practice in this work. We set $b=2\tau_{\rm m}/T_{\rm av} \ll 1$. The measured position $\rpar(t)$ is the sum of the real position, denoted by 
$\rpar^{(0)}(t)$, and of the drift $\mathbf{d}(t)$: $\mathbf{r}_\parallel (t)=\mathbf{r}_\parallel^{(0)} (t)+\mathbf{d}(t)$. The drift is a slowly varying function of time, so that one also has $\mathbf{\bar r}_\parallel (t)=\mathbf{\bar r}_\parallel^{(0)} (t)+\mathbf{d}(t)$. Thus
\begin{eqnarray}
C_{\rm m}(t) &=& \langle [\rpar^{(0)}(s+t)-\mathbf{\bar r}_\parallel^{(0)} (s+t)]\cdot[\rpar^{(0)}(s)-\mathbf{\bar r}_\parallel^{(0)} (s)] \rangle \nonumber \\
 &=&  \langle \rpar^{(0)}(s+t) \cdot \rpar^{(0)}(s) \rangle 
	+ \langle \mathbf{\bar r}_\parallel^{(0)} (s+t)\cdot \mathbf{\bar r}_\parallel^{(0)} (s) \rangle \nonumber \\
 & - & \langle \rpar^{(0)} (s+t) \cdot \mathbf{\bar r}_\parallel^{(0)} (s) \rangle
	- \langle \mathbf{\bar r}_\parallel^{(0)} (s+t)\cdot \rpar^{(0)} (s) \rangle.
\label{CmSI}
\end{eqnarray}	
If, without loss of generality, we assume that $\langle \rpar^{(0)}  \rangle=0$ on each trajectory, then the first average in equation~(\ref{CmSI}) is equal to $\langle [\rpar^{(0)}]^2 \rangle e^{-t/\tau_{\rm m}}$. Using equation~(\ref{rbar}), $\langle \rpar^{(0)}(s) \cdot \rpar^{(0)}(\tilde t) \rangle=\langle [\rpar^{(0)}]^2\rangle e^{-|\tilde t -s|/\tau_{\rm m}}$ and
 $T_{\rm av} \gg t,\tau_{\rm m}$, one gets that the three last averages are equal to $b\langle [\rpar^{(0)}]^2\rangle$. Therefore $C_{\rm m}(t) = \langle [\rpar^{(0)}]^2 \rangle (e^{-t/\tau_{\rm m}} -b)$ and 
$C_{\rm m}(0) = \langle [\rpar^{(0)}]^2 \rangle (1 -b)$. It follows that 
$C_{\rm m}(t)/C_{\rm m}(0) = (1+b)e^{-t/\tau_{\rm m}} -b$ at first order in $b$, as claimed above.}
Fig.~\ref{fits:C:t} provides two generic examples of such fits of modified experimental correlation functions $C_{\rm m}(t)$. These figures illustrate the quality of the fits.

An alternative way to circumvent systematic errors on $C_{\rm raw}(0)$ consists of excluding $C_{\rm raw}(0)$ in the fitting procedure and using the fitting formula
\begin{equation}
C_{\rm raw}(t)=\Delta \mathbf{r}_{\mathrm{m},\parallel}^2\left[\left(1+2\frac{\tau_{\rm m}}{T_{\rm av}}\right) e^{-t/\tau_{\rm m}} -2\frac{\tau_{\rm m}}{T_{\rm av}}\right],
\end{equation}
with two free parameters, $\Delta \mathbf{r}_{\mathrm{m},\parallel}^2$ and $\tau_{\rm m}$. For all data sets, the so-obtained values of $\tau_{\rm m}$ are equal, within a few percents, to the ones coming from equation~(6) of the main text, thus confirming the validity of our fitting procedure.

\subsection*{Brownian Dynamics simulations}

For the numerical iterations, the evolution of each sphere position $\tbr_i=\br_i/a$ is governed by an iterative Langevin equation with a discrete time step $\delta t$. In terms of the discrete time variable $n=t/\delta t$,
\be
\tbr_i(n+1)=\tbr_i(n)- \tD_0 \, \nabla_{\tbr_i}\tilde U(\{\tilde\br_k\}) + \sqrt{2 \tD_0} \, \tilde{\bxi}_i(n),
\ee
where $\tU=U/k_BT$ and the rescaled random displacement $\tilde{\bxi}_i(n)$ has variance unity $\langle\tilde{\bxi}_i(n)\cdot\tilde{\bxi}_j(m)\rangle=3\,\delta_{ij}\delta_{nm}$ [see also equation~(10) and below]. The rescaled bare diffusion coefficient $\tD_0 = D_0\delta t/a^2$ is the diffusion constant in an unbounded space in units of the particle radius $a$ and time step $\delta t$: a sphere typically takes $1/\tD_0$ iterations to diffuse on a distance $a$.  For sufficient numerical accuracy we choose time steps in the range $\tD_0 = 10^{-3}$--$10^{-5}$ to avoid unphysical large displacements and forces. Output values are calculated every $10^3$--$10^4$ steps, total simulation times are on the order of $10^8$--$10^9$ steps, giving reasonable error bars.
Finally, since the polymer motion is limited to the upper half-plane $z>0$, we use the following reflection boundary condition: if a sphere intersects the substrate, its height $z_i$ is replaced by $z_i^{\rm refl}$:
\bea
z_i<a &\Rightarrow& z_i^{\rm refl}=2a-z_i \quad \mathrm{for}\quad i<N\\
z_N<R-a &\Rightarrow& z_N^{\rm refl}=2(R-a)-z_N
\eea
The excluded volume interaction is modeled by a repulsive, truncated Lennard-Jones potential $\tU_{LJ}=\sum_{i<j} [(b/|\tbr_i-\tbr_j|)^{12} - 2(b/|\tbr_i-\tbr_j|)^6+1]$ valid for separation $|\tbr_i-\tbr_j|<2$ and $b=R/a+1$ for $j=N$ and $b=2$ otherwise.

\subsection*{Comparison between Brownian Dynamics and dynamical Monte Carlo simulations}

We have simulated our particle-DNA complex for different choices of $L$ and $R$, using both Brownian Dynamics and Monte Carlo simulations.

The relevant test of the dynamical Monte Carlo (DMC) algorithm reliability is the acceptance ratio $\Upsilon$, i.e. the number of accepted Monte Carlo moves as compared to the number of attempts, which must be close to 1. Indeed, in Monte Carlo algorithms, a move is always accepted if $\Delta U\leq0$ and is accepted with probability $\exp(-\Delta U/k_{\rm B}T)$ if  $\Delta U> 0$. Thus
$\Upsilon \simeq 1/2+\langle  \exp(-\Delta U/k_{\rm B}T) \rangle_{\Delta U> 0}/2 \simeq 1-
\langle  \Delta U \rangle_{\Delta U> 0}/(2k_{\rm B}T)$ and the condition $\Delta U \ll k_{\rm B}T$ is equivalent to $\Upsilon$ close to 1.

We have observed that, up to error bars, statistical and dynamical properties are similar. In terms of acceptance ratio $\Upsilon$, we have measured that our Monte Carlo parameter choices lead to $68 \%<\Upsilon<88 \%$ for all the situations studied in this paper, and $\Upsilon > 80 \%$ for simulations with hydrodynamic interaction with the wall (those compared to experiments). To this end, $R_b$ (defined in the Materials and Methods section) was set to $R_b=a/5$ for all data sets, with the few following exceptions: $R_b=a/7.5$ for $L=798$~bp and $R=0$;  $R_b=a/10$ for $L=401$~bp and $R=0$ or 20~nm. 
 
Thus the criterion $\Delta U \ll k_BT$ is not satisfied {\em stricto sensu}, but the fact that both types of simulations give comparable results even at the dynamical level proves that this value of $\Upsilon$ is sufficient in our case. 
The choice of excluded volume interaction in the DMC code (whenever a move would lead to the penetration of a bead into an other one or the substrate, it is rejected) saves computational time as compared to the calculation of truncated Lennard-Jones potentials used in Brownian Dynamics. Since DMC is favorable in terms of computational time (about 30 times faster in our case because it saves the computation of Lennard-Jones potentials and allows for a larger time step), we restrict our numerical work in the paper to DMC simulations. Note however that these Monte Carlo simulations remain time-consuming: each simulation for a set of parameters typically requires several days on a standard processor.

\subsection*{Exact equilibrium distributions in the rigid and flexible limits}

\begin{figure}[ht]
\begin{center}
\includegraphics[height=6cm]{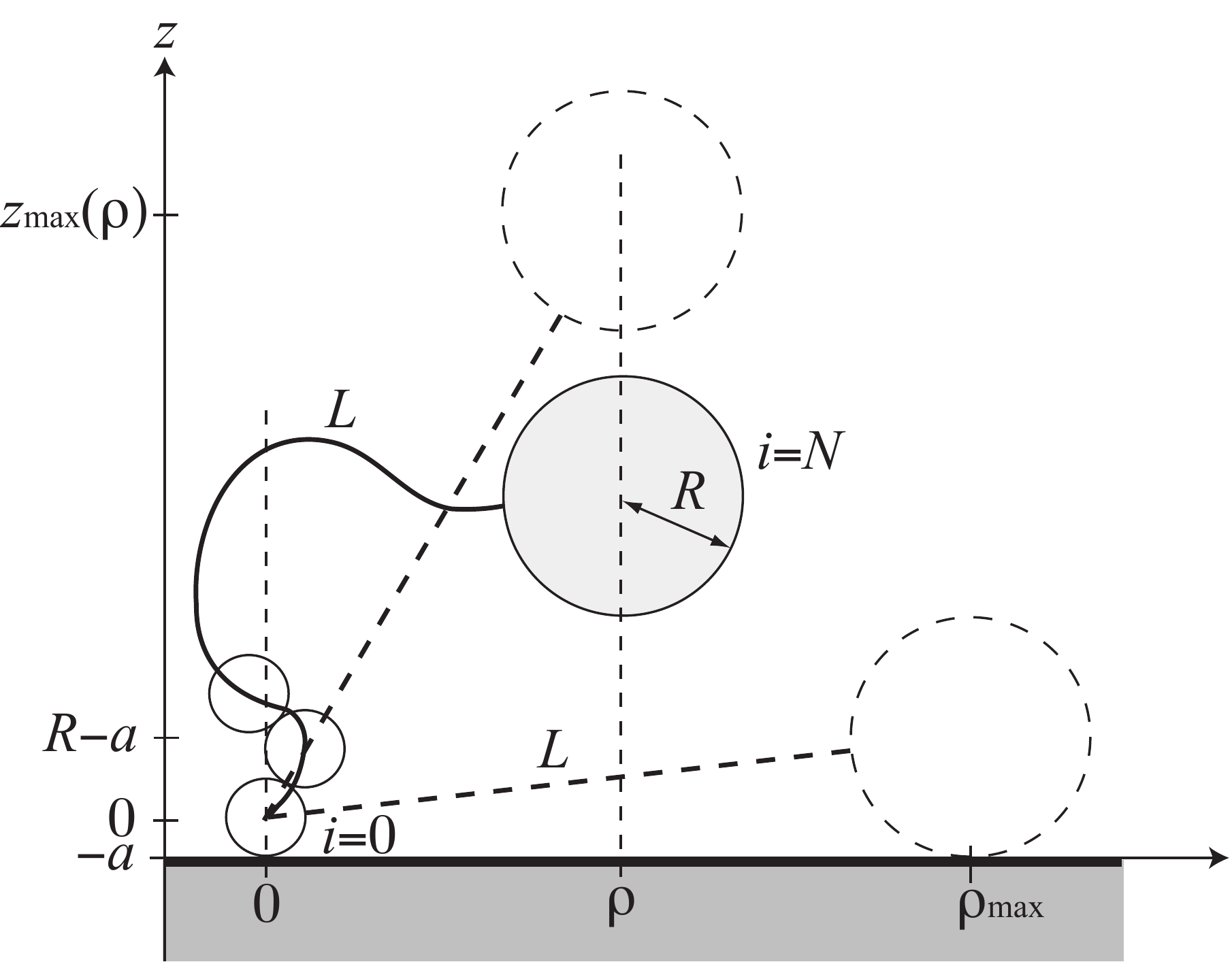}
\caption{TPM geometry. The particle has radius $R$. The polymer, of curvilinear length $L$, is modeled by a sequence of $N$ spheres of radius $a$. The polar coordinates of the particle in the $(xy)$ plane are denoted by $(\rho,\varphi)$, $\rho_{\rm max}$ is the maximum extension and the particle height is denoted by $z$. The whole chain is confined in the upper half-space. For a given $\rho$,  $z \leq z_{\rm max}(\rho)$. Only $\rho$ and $\varphi$ (or $x$ and $y$) are accessible by TPM.
\label{fig1}}
\end{center}
\end{figure}

\begin{figure*}[ht]
\begin{center}
\includegraphics[height=5cm]{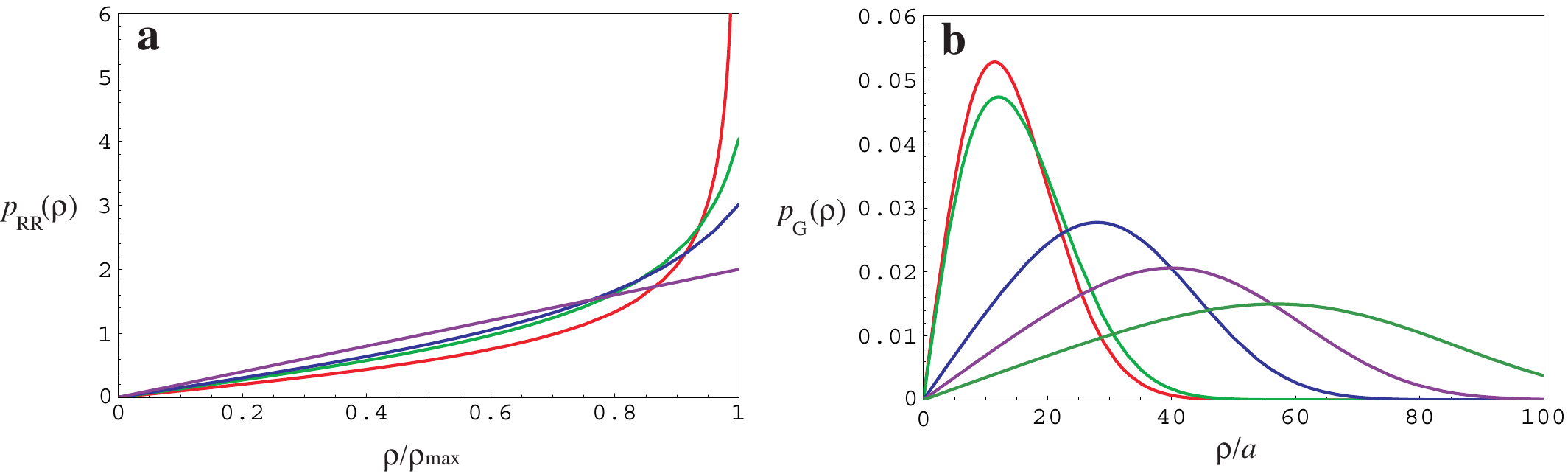}
\caption{Theoretical probability distribution $p(\rho)$ for $N=100$ in the two following limits. (a): Rigid rod for various particle sizes $R$: $a$ (red), $L/2$ (green), $L$ (blue), and  $\infty$ (purple) (from top to bottom at $u=1$). $p_{\rm RR}(u)$ is expressed as a function of the reduced particle coordinate parallel to the substrate, $u\equiv\rho/\rho_{\max}$.(b): Gaussian chain, $p_{\rm G}(\rho)$, $\rho$ in units of $a$ and various particle sizes $R$: 0 [pure Gaussian see equation~(\ref{pGgauss})], $10a$, $50a$, $100a$, and $200a$ from left to right.}
\label{fig2}
\end{center}
\end{figure*}

The particle center distribution $G(\br_0,\br;N)$ is the polymer propagator from $\br_0={\bf 0}$ to the center of the particle $\br$, in $N$ steps (in the $z>0$ half-space).
The experimental observable is not this full distribution  
but the marginal distribution of $\rho$ 
\be
p(\rho)=2\pi \rho\int_{R-a}^{z_{\rm max}(\rho)} G({\bf 0},\br;N)\, \mathrm{d}z,
\label{marg_prob}
\ee 
where $z_{\rm max}(\rho)=\sqrt{(R_0+L)^2-\rho^2}$ is the maximum value for $z$ at fixed $\rho$  and $R_0\equiv R+a$ (Fig.~\ref{fig1}).
This probability distribution can be computed analytically in the two following limits: rigid rod ($L/\ell_{\rm p}\to0$) and phantom Gaussian chain ($L/\ell_{\rm p}\to\infty$).
In the rigid rod limit, the propagator is simply fixed by $| \br | = R_0+L$, in other words $G_{\rm RR}({\bf 0},\br;N)=A\,\delta[\sqrt{\rho^2+z^2}-(R_0+L)]$ where $A =[2\pi(R_0+L)(2a+L)]^{-1}$ is the normalization constant. Using equation~(\ref{marg_prob}), we find 
\be
p_{\rm RR}(\rho)=\frac{\rho}{(2a+L)\sqrt{(R_0+L)^2-\rho^2}}
\label{marg_prob_RR}
\ee
for $0\leq \rho\leq\rho_{\rm max}\equiv\sqrt{(R_0+L)^2-(R-a)^2}$ (Fig.~\ref{fig1}). In Fig.~\ref{fig2}a, we display $p_{\rm RR}(u)$ where $u\equiv \rho/\rho_{\rm max}$. This distribution is linear for small $\rho$. When $R \ll L$, it sharply increases for $\rho \rightarrow  \rho_{\rm max}$ because of projection effects. When $R \gg L$, this distribution becomes $p_{\rm RR}(u) \simeq 2u$, owing to
$p_{\rm RR}(u) = p_{\rm RR}(\rho) \; {\rm d}\rho/ {\rm d}u = \rho_{\rm max}  \; p_{\rm RR}(\rho)$. 

In the flexible or phantom Gaussian regime, the end-to-end distance probability distribution for the chain alone is given by $G_{\rm chain}({\bf 0},\br_{N-1};N-1)\propto \exp\{-3\br_{N-1}^2/[2(N-1)(2a)^2]\}$, where $\br_{N-1}$ is the 3D position of the last sphere. The  distribution of the center of the particle $\br$ is obtained by imposing that 
$|\br-\br_{N-1}|=R_0$: 
\begin{eqnarray}
P (\br)&\equiv &G_G(\br_0,\br;N) \nonumber \\
&=&\int \mathrm{d}^3\br_{N-1}\,G_{\rm chain}({\bf 0},\br_{N-1};N-1) \delta(|\br-\br_{N-1}|-R_0)\nonumber \\
 & = & B \alpha^{3/2}\frac{\sinh(2\alpha R_0|\br|)}{\alpha R_0|\br|}\exp[-\alpha(\br^2+R_0^2)],
\end{eqnarray}
a function of $|\br|$ where $\alpha=\frac3{2(N-1)(2a)^2}$. The constant $B$ is given by the normalization condition $\int_V P (\br) \mathrm{d}^3\br=1$, $V$ being the accessible volume. In order to forbid polymer trajectories intersecting the $z=0$ plane, we substract all the forbidden paths using the mirror reflection argument~\cite{Shmolu} (the mirror image of $z=0$ is in the notation of Fig.~\ref{fig1}, $z=-2a$):
\bea
p_{\rm G}(\rho) &=& 2\pi \rho\left[\int_{R-a}^{z_{\rm max}(\rho)}P(\sqrt{\rho^2+z^2})\, \mathrm{d}z \right. \nonumber  \\
&-&\left.\int_{R-a}^{z_{\rm max}(\rho)}P(\sqrt{\rho^2+(z+2a)^2})\, \mathrm{d}z\right]\\
 &=& 2\pi \rho\left[\int_{R-a}^{z_{\rm max}(\rho)}P(\sqrt{\rho^2+z^2})\, \mathrm{d}z \right. \nonumber  \\
&-&\left.\int_{R+a}^{z_{\rm max}(\rho)+2a}P(\sqrt{\rho^2+z^2})\, \mathrm{d}z\right]\label{pG},
\eea
as plotted in Fig.~\ref{fig2}b. The analytical expression for $p_{\rm G}(\rho)$ is rather complicated but mean values such as $\lan\rho^2\ran\equiv
\langle \rpar^2 \rangle \equiv \Delta \rpar^2$ can,  in principle, be computed without major difficulty.
This probability distribution can be simplified in two limits. For small particle radii, $R\ll \sqrt{L a}$, equation~(\ref{pG}) reads
\bea
p_{\rm G}(\rho) &\approx& 2\pi^{3/2}B \alpha \rho e^{-\alpha (\rho^2+R_0^2)} \nonumber\\
& \times & 
\left\{\mathrm{erf}[\sqrt{\alpha }(R+a)]-\mathrm{erf}[\sqrt{\alpha }(R-a)] \right. \nonumber\\
& - & \left. -\mathrm{erf}[\sqrt{\alpha }(z_{\rm max}(\rho)+2a)]+\mathrm{erf}[\sqrt{\alpha }z_{\rm max}(\rho)]\right\}\nonumber\\
&\approx& 2 \alpha  \rho e^{-\alpha \rho^2},
 \label{pGgauss}
\eea
where $\mathrm{erf}(x)=\frac2{\sqrt{\pi}}\int_0^xe^{-y^2}\mathrm{d}y$ is the error function. As expected, we recover the Gaussian behaviour when $R \rightarrow 0$ and $\langle\rho\rangle=\sqrt\pi\sqrt{aL/3}$ and $\langle\rho^2\rangle=4aL/3$, which corresponds to equation~(10c) of Ref.~\cite{SegallSI} in the limit $R\ll \sqrt{L a}$ (with $\xi=a$).

In the large particle radius limit, $R\gg \sqrt{L a}$, we find $P(\br)\simeq \sqrt{\alpha } \exp[-\alpha (|\br|-R)^2]/(R|\br|)$ and the most probable value for $|\br|$ gets displaced towards non-zero values $\langle|\br|\rangle=R$. Equation~(\ref{pG}) then simplifies to
\be
p_{\rm G}(\rho) \approx 2\sqrt{\frac\alpha\pi} \frac{\rho}{\sqrt{\rho^2+R^2}} \exp[-\alpha (\rho^2+2R^2-2R\sqrt{\rho^2+R^2})].
\label{pGausslargeR}
\ee
When $R$ increases, this probability distribution gets displaced to the right and widens as seen in Fig.~\ref{fig2}b. But due to the projection in 2D, the most probable value of $\rho$ now scales like $\sqrt{R}$ and
\be
\langle\rho^2\rangle=\frac4{\sqrt{3\pi}}R\sqrt{aL}+\frac23 aL,
\ee
which corresponds to equation~(10c) of Ref.~\cite{SegallSI} in the limit $R\gg \sqrt{L a}$.

The intermediate semi-flexible regime, $L/\ell_{\rm p} = \mathcal{O}(1)$, of interest in TPM  experiments, is well described by the worm-like chain model. For instance, in the experiments described in the paper, DNA lengths vary between 400 and 2080~bp, which corresponds to $2<L/\ell_{\rm p}<14$. This model can be tackled analytically~\cite{Samuel02} but it should be kept in mind that real chains are self-avoiding and that the presence of the particle renders the problem even more intricate. 
For instance, the effect of the excluded volume of the particle is to widen and shift to large $\rho$ the Gaussian distributions shown in Fig.~\ref{fig2}b. It then becomes analytically intractable for finite chains, but it can be tackled numerically by Monte Carlo or Brownian dynamics simulations. 

In the limit $L\gg\ell_{\rm p}$ where $\ell_{\rm p}=147$~bp is fixed, as for double stranded DNA, the worm-like chain model tends to the Gaussian limit at a more corse-grained level where the Gaussian unit, the Kuhn length, is no more $2a$ but $2\ell_{\rm p}$. Moreover the number of independent monomers $N$ becomes $L/2\ell_{\rm p}$ and the mean-square end-to-end distance is $\Delta \br^2=\lan\br^2\ran=2L\ell_{\rm p}$. Hence the formula given by equation~(\ref{pGgauss}) is used where the parameter $\alpha$ becomes $\alpha=3/(4L\ell_{\rm p})$. These Gaussian distributions are compared to the numerical ones for $R=0$ in Fig.~\ref{Hist:p:de:rho}a below. When $R>0$, as in Fig.~\ref{Hist:p:de:rho}b, the numerical distributions cannot be compared to the present theoretical ones because of the excluded volume effect mentioned just above.

\subsection*{Equilibrium numerical distributions}

We report the equilibrium distributions computed using Monte Carlo simulations as presented in the Material and Methods.
 
First of all, the criterion $a\ll \ell_{\rm p}$ is satisfied for all DNA lengths if $N=50$. However for small DNA molecules ($L \leq 798$~bp), some calculations were performed choosing $N=25$ to save computational time, and in this case we have checked that the distributions coincide within statistical noise.
In particular, the fact that monomer spheres are in general larger than the actual radius of the DNA molecule ($a>1$~nm) does not affect significantly our physical observables.
This observation is related to the questions of polymer self-avoidance and solvent quality: in our simulations, the short range repulsion, which is essentially of electrostatic origin, is taken into account by the sphere hard-core repulsion, the effective range of which is thus larger than the bare DNA diameter. Moreover, there exist attractive interactions, mediated by the solvent, between two DNA segments which are far away along the chain. They are of hydrophobic nature~\cite{odijk,strey} since water is a bad solvent for DNA. We assume monovalent counter-ions and do not consider situations where DNA condensation occurs~\cite{bloomfield}. These interactions are relevant for long DNAs which are flexible at large length scales and can be neglected in this work, because the average number of sphere contacts is extremely small as $L/\ell_{\rm p}$ remains moderate. Therefore the choice of $a$ appears to be an irrelevant issue as far as chain statistics are concerned, provided that $a\ll \ell_{\rm p}$.
\begin{figure}[ht]
\begin{center}
\includegraphics[height=6cm]{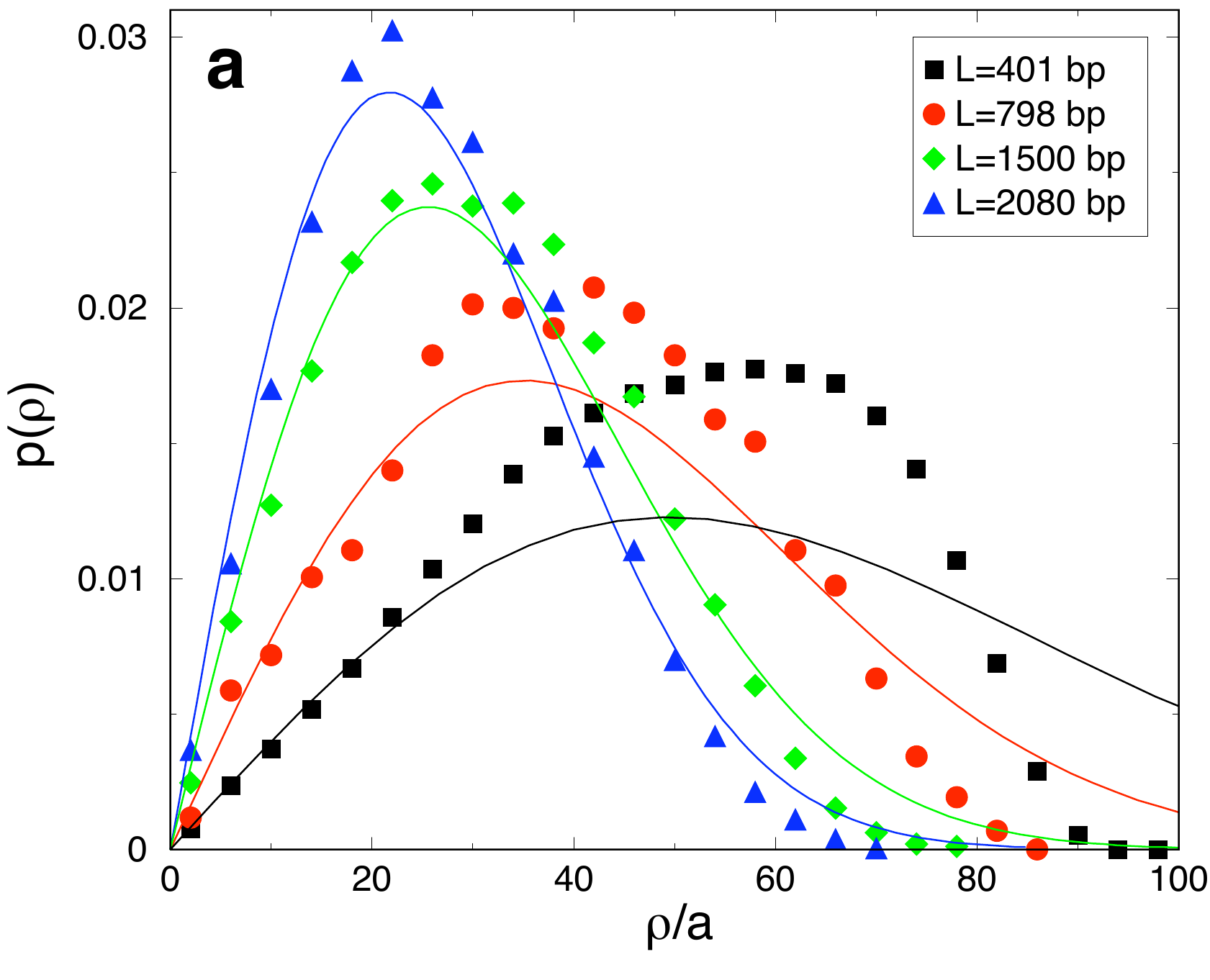}
\includegraphics[height=6cm]{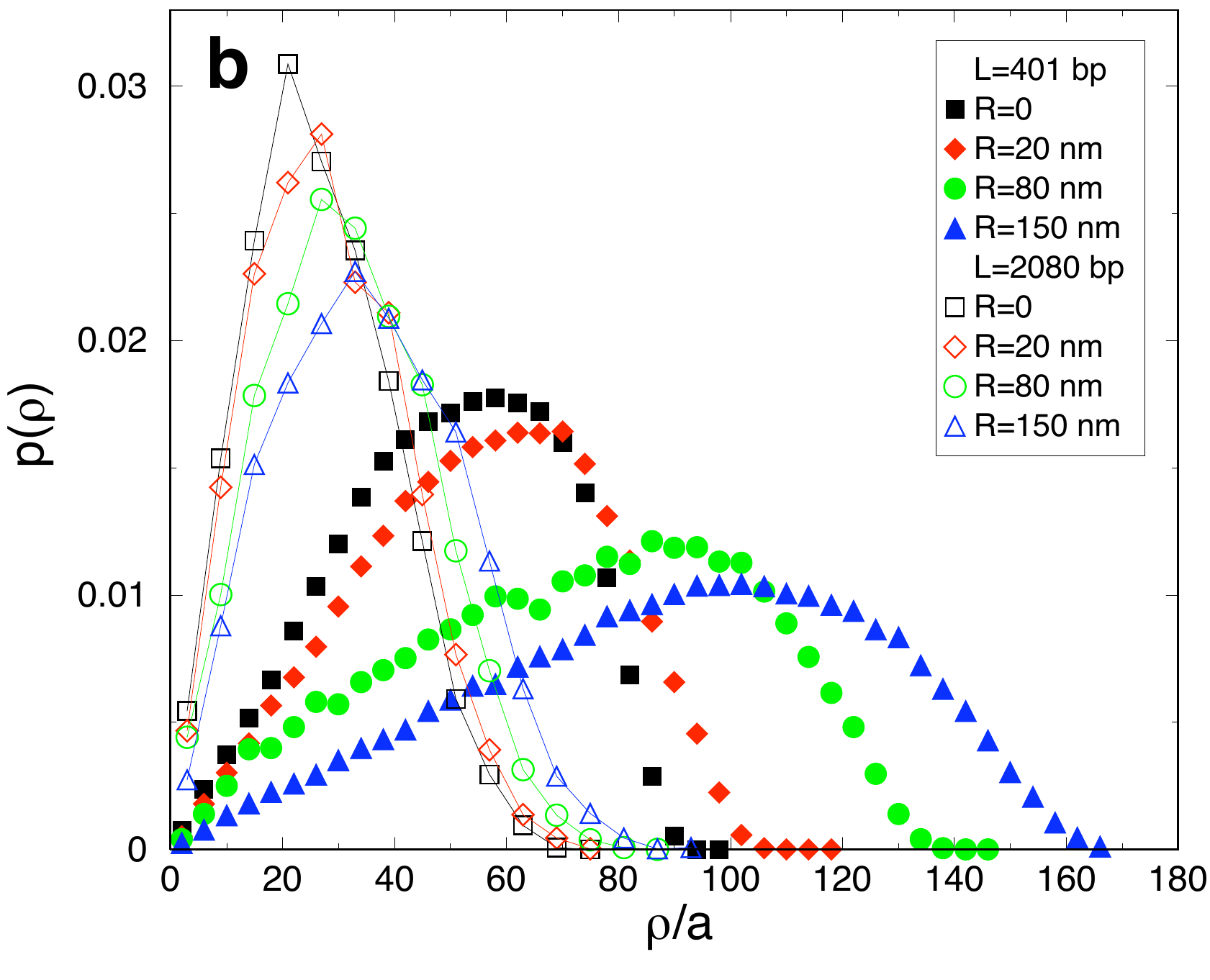}
\caption{\footnotesize Numerical distributions $p(\rho)$, $\rho$ in units of $a$ ($N=50$). \textbf{(a)} Symbols represent the normalized histograms for $R=0$ and $L=401$, 798, 1500 and 2080~bp and continuous lines the theoretical Gaussian distributions for the same values of $L$. \textbf{(b)} Normalized histograms for $L=401$ and 2080~bp and different values of $R$. Here lines are guides for eyes.
\label{Hist:p:de:rho}}
\end{center}
\end{figure}

The numerical distributions $p(\rho)$ are plotted in Fig.~\ref{Hist:p:de:rho}. 
Fig.~\ref{Hist:p:de:rho}a shows $p(\rho)$ when the particle radius $R=0$. Also displayed in the figure are the theoretical Gaussian distributions for $R=0$, as discussed at the end of the previous section. When $L$ is large the numerical distributions and the Gaussian ones are superimposed. This observation supports our previous remarks about the rareness of DNA self-contacts, as self-avoidance is not taken into account in the Gaussian approximation. 

Fig.~\ref{Hist:p:de:rho}b shows the same histograms for $L=401$ and 2080~bp where $R$ varies between 0 and $150$~nm. For $L=2080$~bp, the distributions look like Gaussian for all the values of $R$ studied, but are slightly shifted to larger $\rho$ and widened, as compared to the phantom chain case, due to self-avoidance between the particle and the chain. The Gaussian character is discussed afterwards. When $L$ decreases, numerical distributions deviate from Gaussians: they are skewed to the right with a steep decrease near $\rho_{\rm max}$, a signature of the rigid-rod distribution (see Fig.~\ref{fig2}a). For $L=401$~bp, this decrease is steep for small $R$, but gets smoothed when $R$ increases.

To quantify further the Gaussian character of the numerical distributions, we focus on their two moments $\langle \rho^2 \rangle$ and $\langle \rho^4 \rangle$. For a Gaussian, we expect, in 2D, $\langle \rho^4 \rangle=2 \langle \rho^2 \rangle^2$. Therefore we define
\begin{equation}
A \equiv \frac{\langle \rho^4 \rangle^{1/2}}{\sqrt{2} \langle \rho^2 \rangle}.
\end{equation}
Deviations of $A$ from 1 are indicative of the non-Gaussian character. Some numerical values of $A$ are given in the following table (DNA lengths $L$ in bp, particle radii $R$ in nm). As expected, the closest values to 1 are obtained for long DNAs and small beads, and the Gaussian character is lost for short DNAs and large beads. 

\medskip

\begin{center}
\begin{tabular}{|c||c|c|c|c|}
\hline
 & $L=401$  & $L=798$ & $L=1500$ &  $L=2080$\\
\hline \hline
$R=0$ & 0.85 & 0.89 & 0.92 & 0.94 \\
\hline
$R=20$& 0.85 & 0.89  & 0.93 & 0.93 \\
\hline
$R=80$& 0.85 & 0.88 & 0.90 &  0.92 \\
\hline
$R=200$& 0.84 & 0.87 & 0.87 & 0.89 \\
\hline
\end{tabular}
\end{center}

\medskip

Finally, in order to quantify the effect of the surface, we have also simulated DNAs with no surface, rotating freely in the whole space around their attachment point (data not shown). For the four DNA lengths considered in this work, no effect on the distribution of $\rho$ has been detected (by contrast, $z$ is obviously perturbed by the wall). Note that in both the rigid and Gaussian regimes, this result is shown analytically in equations~(\ref{marg_prob_RR}), (\ref{pGgauss}) and~(\ref{pGausslargeR}). It appears to be also valid in the intermediate semi-flexible regime.

\subsection*{Negligible effects of the substrate-DNA joint}

% -> SM However, it might happen that, for special types of experimental constructions, the joint is better modeled as a rigid anchor (clamped polymer). In this case, the first two spheres are fixed perpendicularly to the surface.

Here we address the question of the modeling of the substrate-DNA joint. As already discussed in Ref.~\cite{Seol07SI} in the different context of optically trapped particles, the elastic properties of the polymer's anchor point cannot be well characterized at the experimental level. For instance, it might happen that, for special types of experimental constructions, the joint is better modeled as a rigid anchor (clamped polymer). 

Therefore we examined two extreme cases: in the paper, we have considered a freely rotating joint (or freely flexible pivot) where all orientations of the chain tangent vector close to the substrate are allowed (except, of course, those intersecting the substrate). We also simulated a clamped joint where this tangent vector is prescribed to be perpendicular to the substrate by preventing the second sphere from moving ($\br_1=2a {\bf e}_z$). {\em A prioiri}, only small DNAs are likely to be affected by this boundary condition, because its ``memory'' along the chain is lost after a few persistence lengths. To clarify this question in the finite $L$ cases considered here, we have also performed DMC simulations with a clamped joint. On the 6 parameter sets studied (namely $L=401$, 798 and 2080~bp, and $R=20$ and 150~nm), no clear tendency can be identified: within error bars, $\tau_{\parallel}$ (and $\langle \mathbf{r}_{\parallel}^2 \rangle$) are equal in both cases (data not shown). Thus the exact way the joint is modeled is not a relevant issue.

The observed fact that for small particles, the replacement of $D_0$ by $D_\perp$ and $D_\parallel$ does not modify $\tau_{\parallel}$, can also be interpreted from this perspective. In this case, the dynamics is dominated by the chain and weakly slowing down the particle does not change this fact. The only effect of hydrodynamic corrections near the wall is to slow down a few monomer spheres close to the substrate. This can be seen as an intermediate case between the two extreme previous cases: these few spheres are fully mobile when the joint is freely rotating, and immobile when it is clamped. Since these two extreme cases lead to the same relaxation times, the intermediate one~-- slowed-down spheres~-- also does.

\subsection*{Comparison between experimental and numerical distributions}

\begin{figure}[ht]
\includegraphics[width=0.8\linewidth]{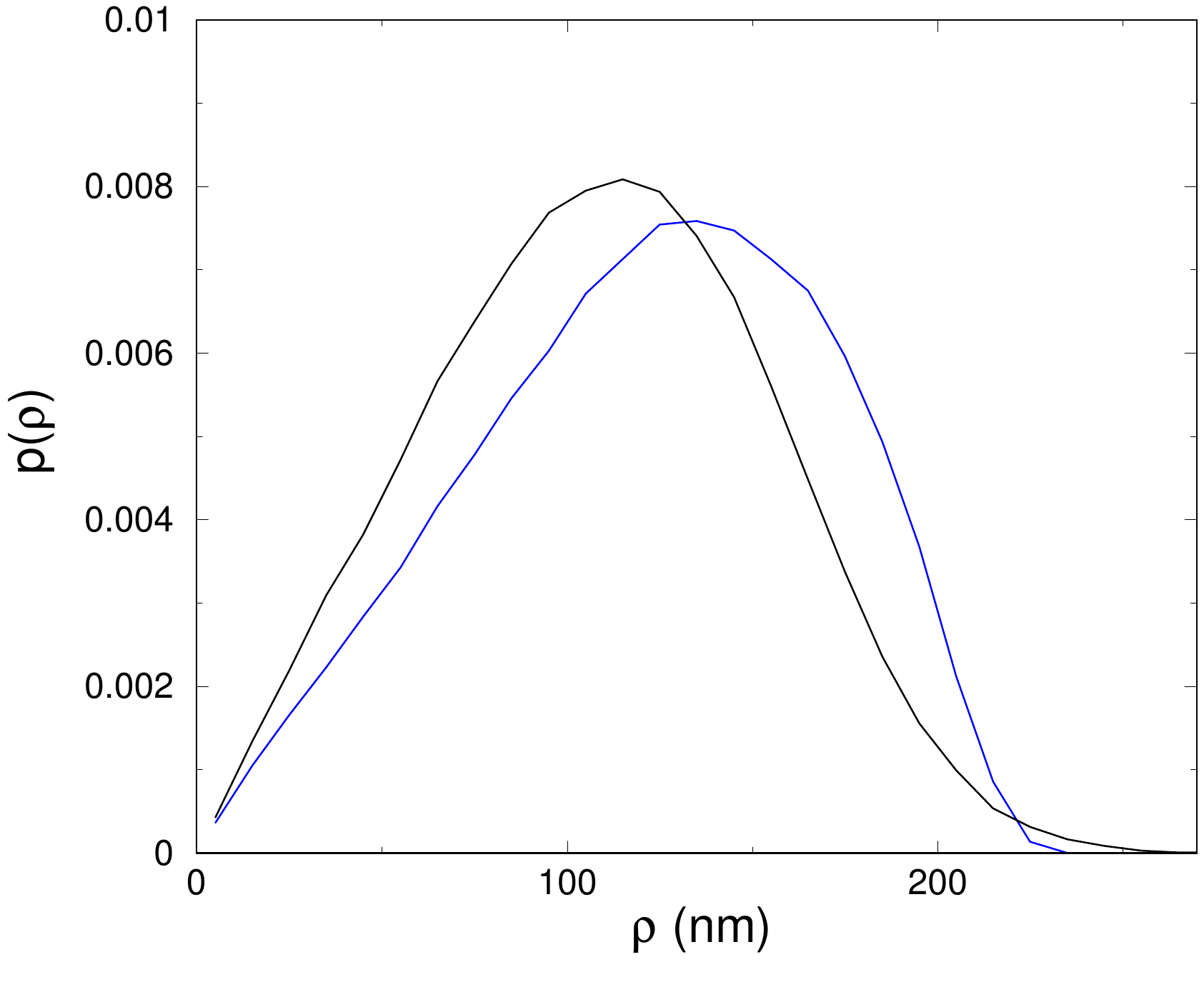}
\caption{Experimental (black) and numerical (blue) distributions $p(\rho)$ for $L=401$~bp and $R=150$~nm.\label{comp1}}
\end{figure}

\begin{figure*}[ht]
\includegraphics[width=0.3\linewidth]{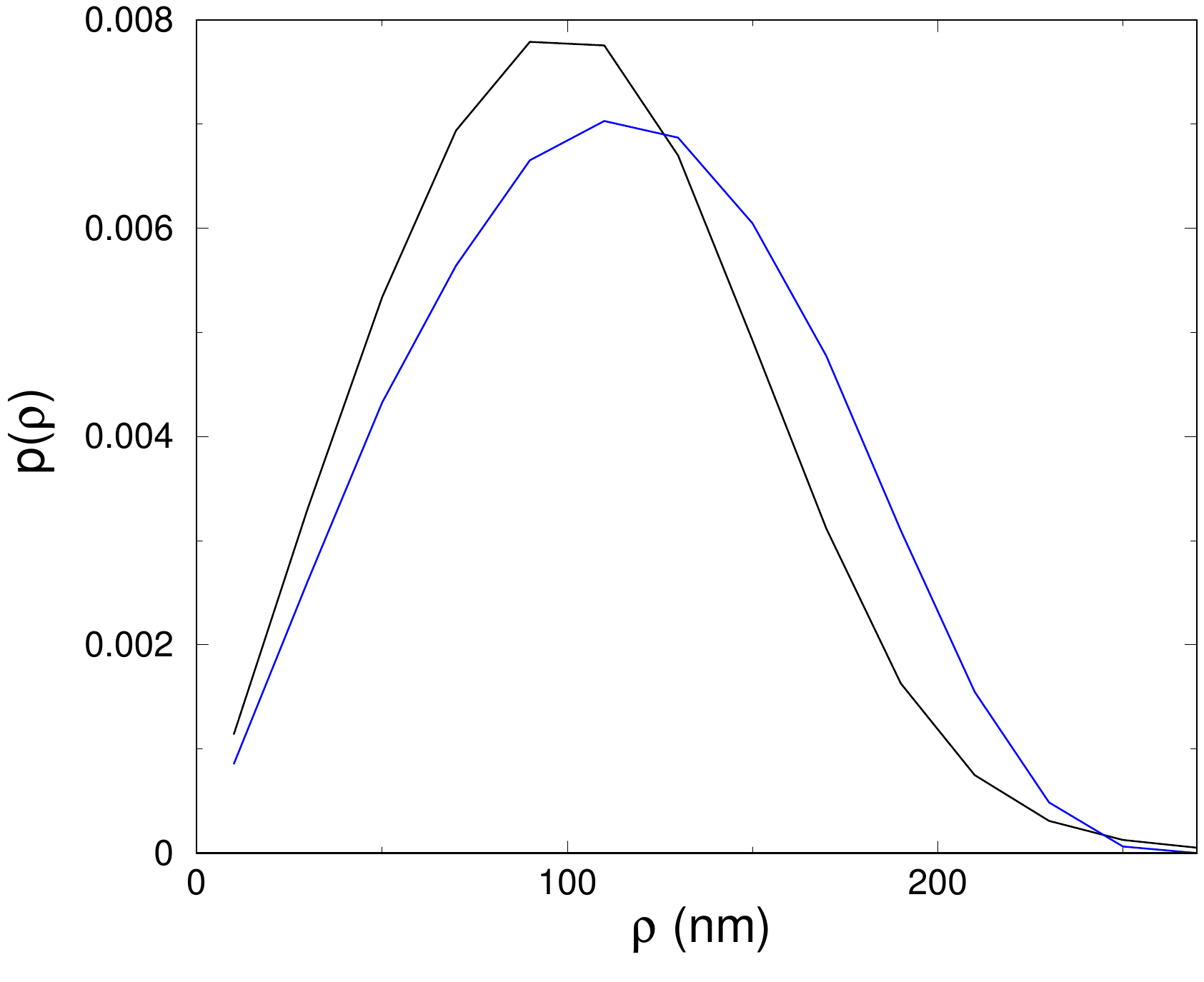}
\includegraphics[width=0.3\linewidth]{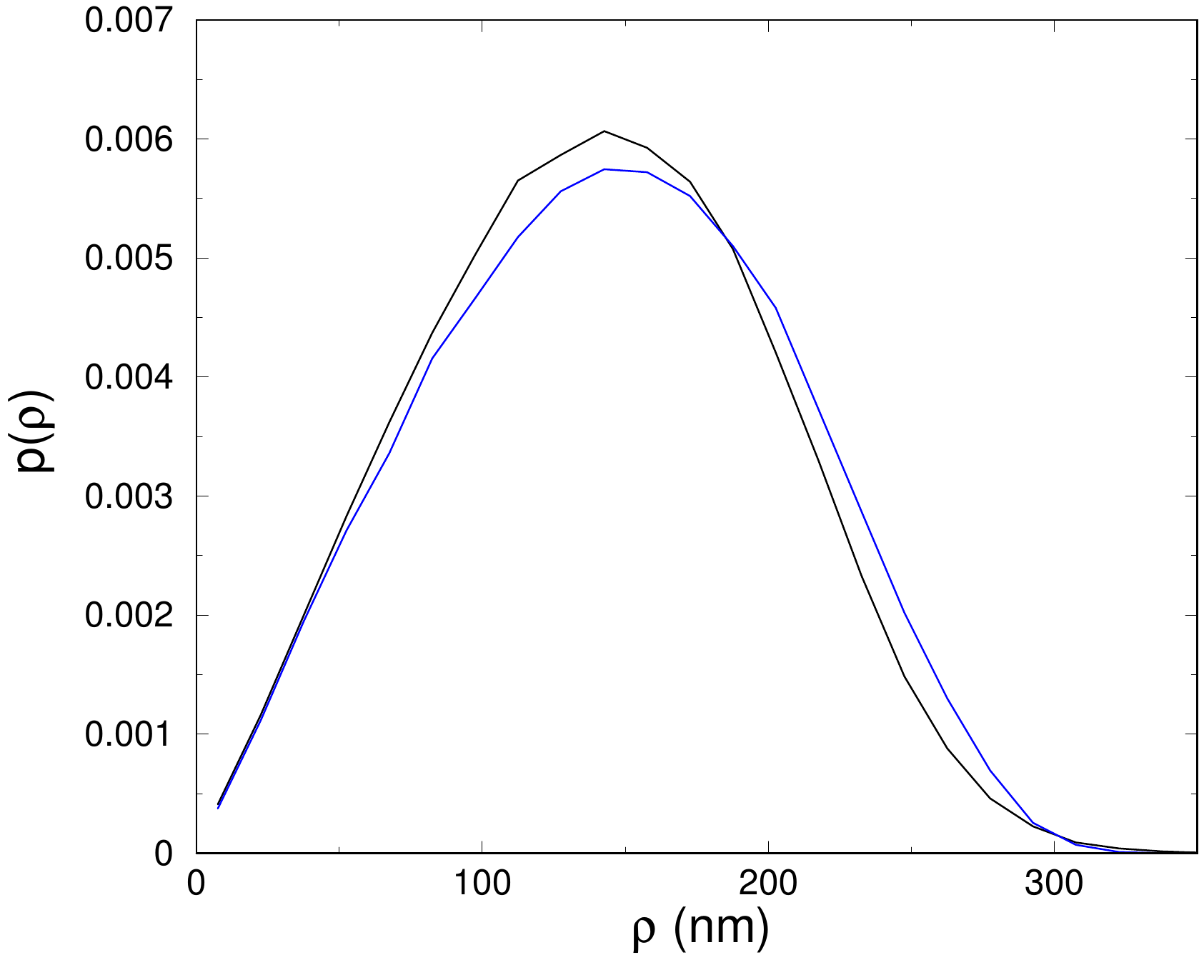}
\includegraphics[width=0.3\linewidth]{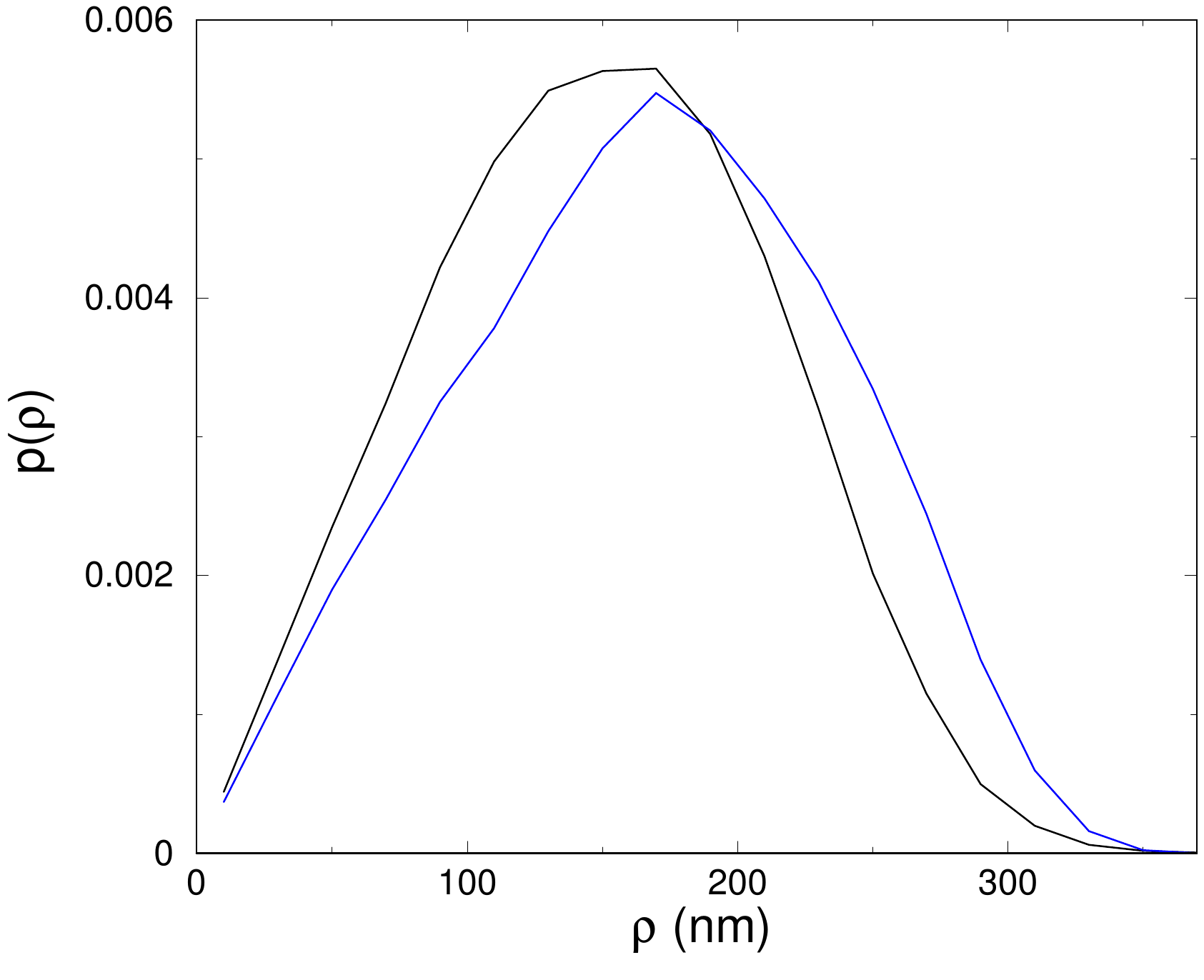}\\
\includegraphics[width=0.3\linewidth]{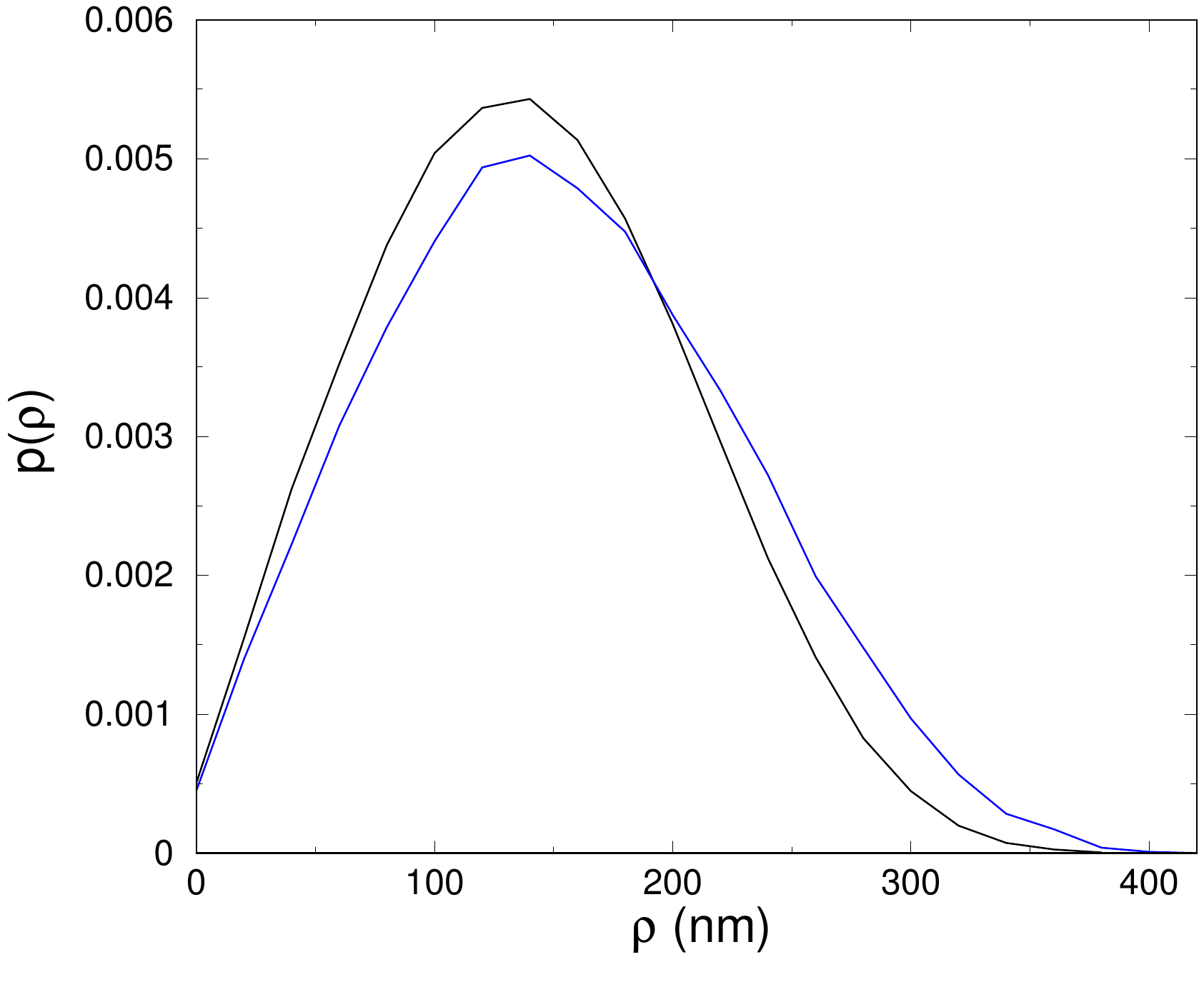}
\includegraphics[width=0.3\linewidth]{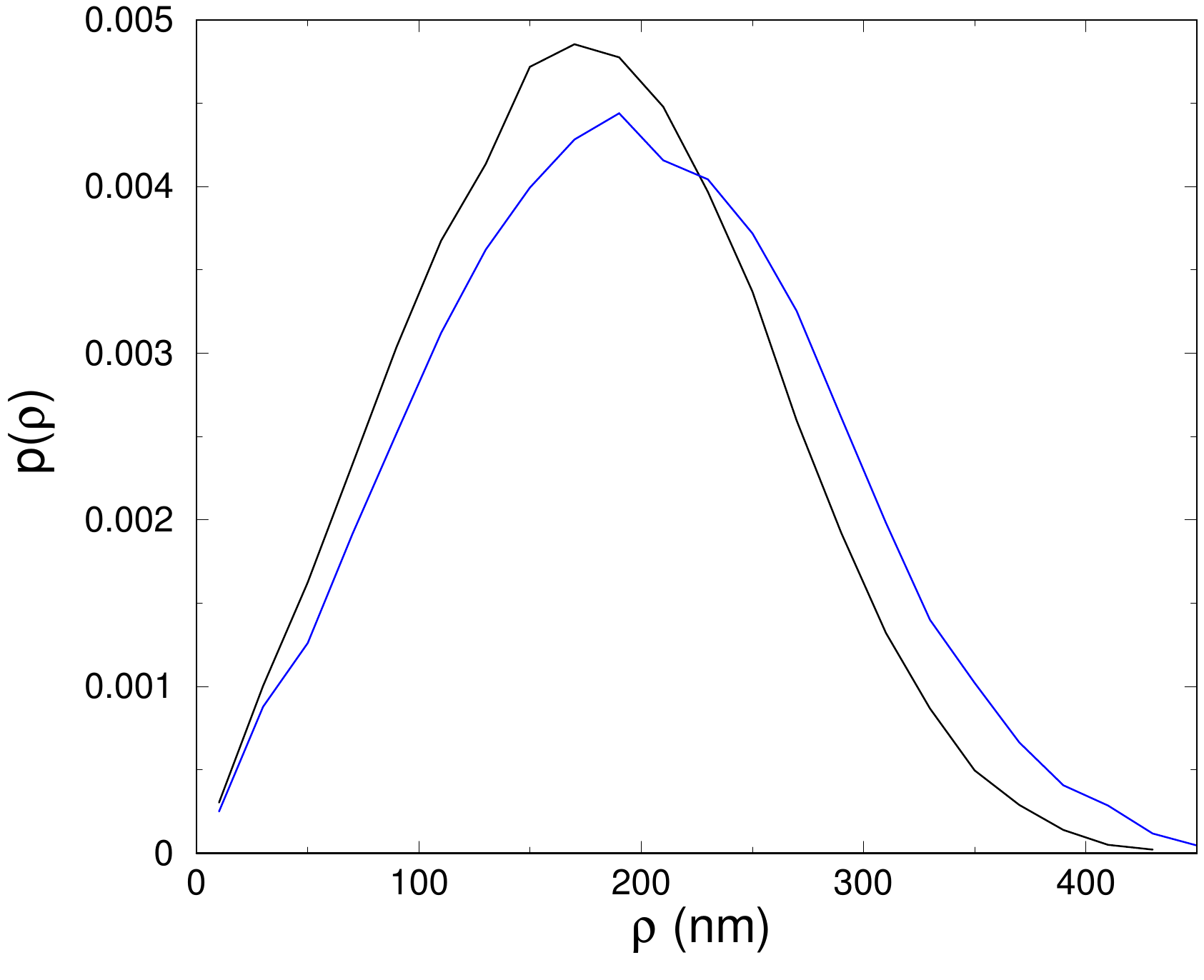}
\includegraphics[width=0.3\linewidth]{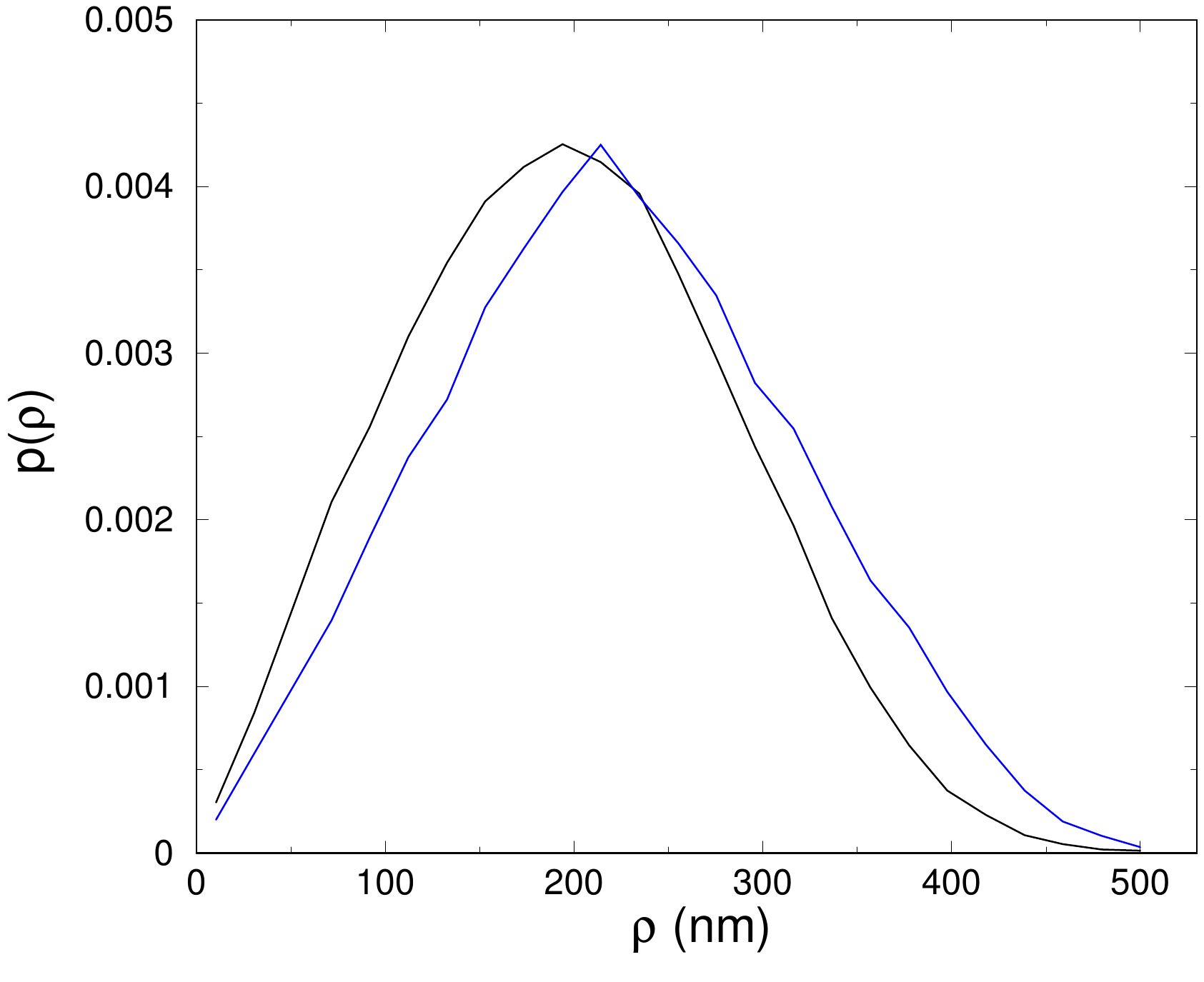}\\
\includegraphics[width=0.3\linewidth]{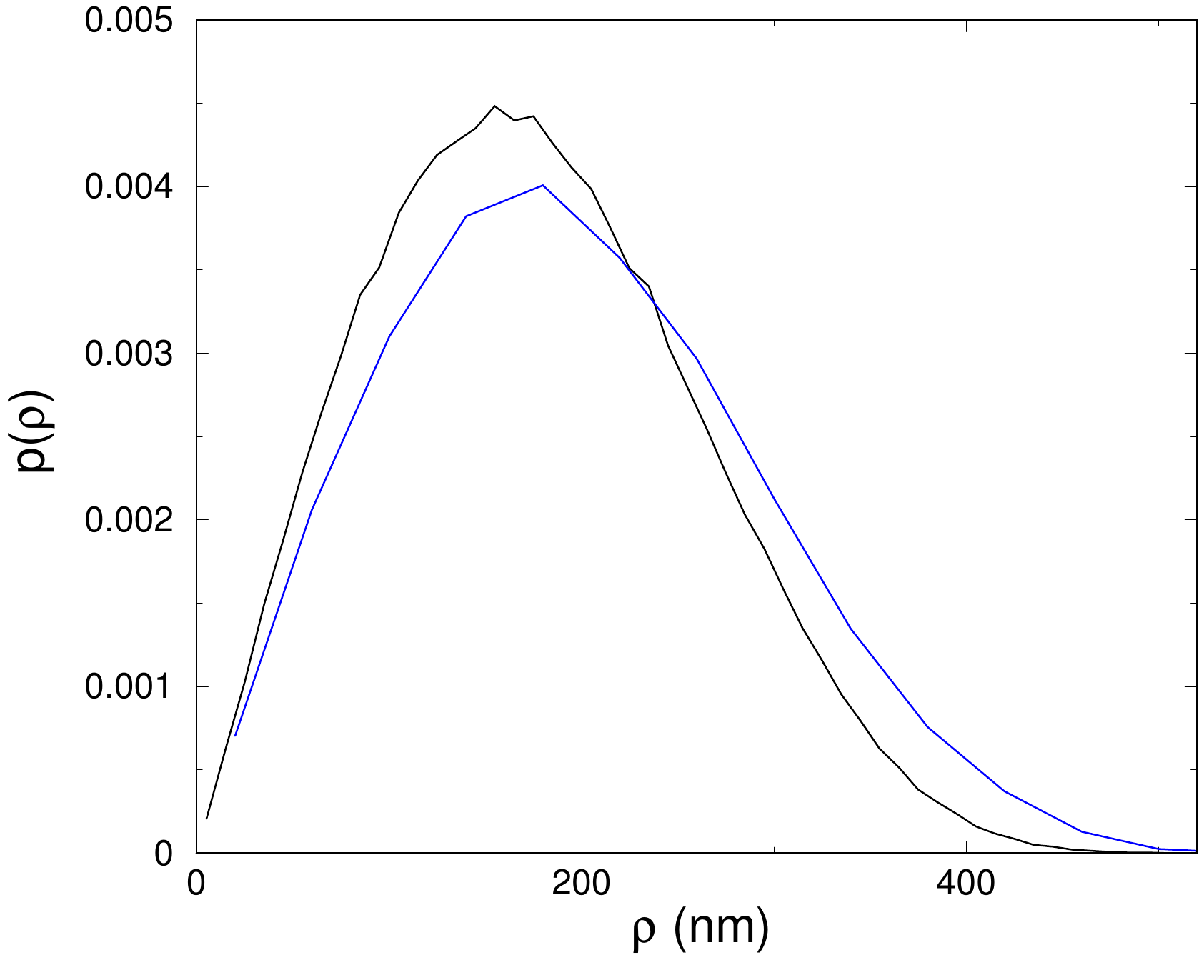}
\includegraphics[width=0.3\linewidth]{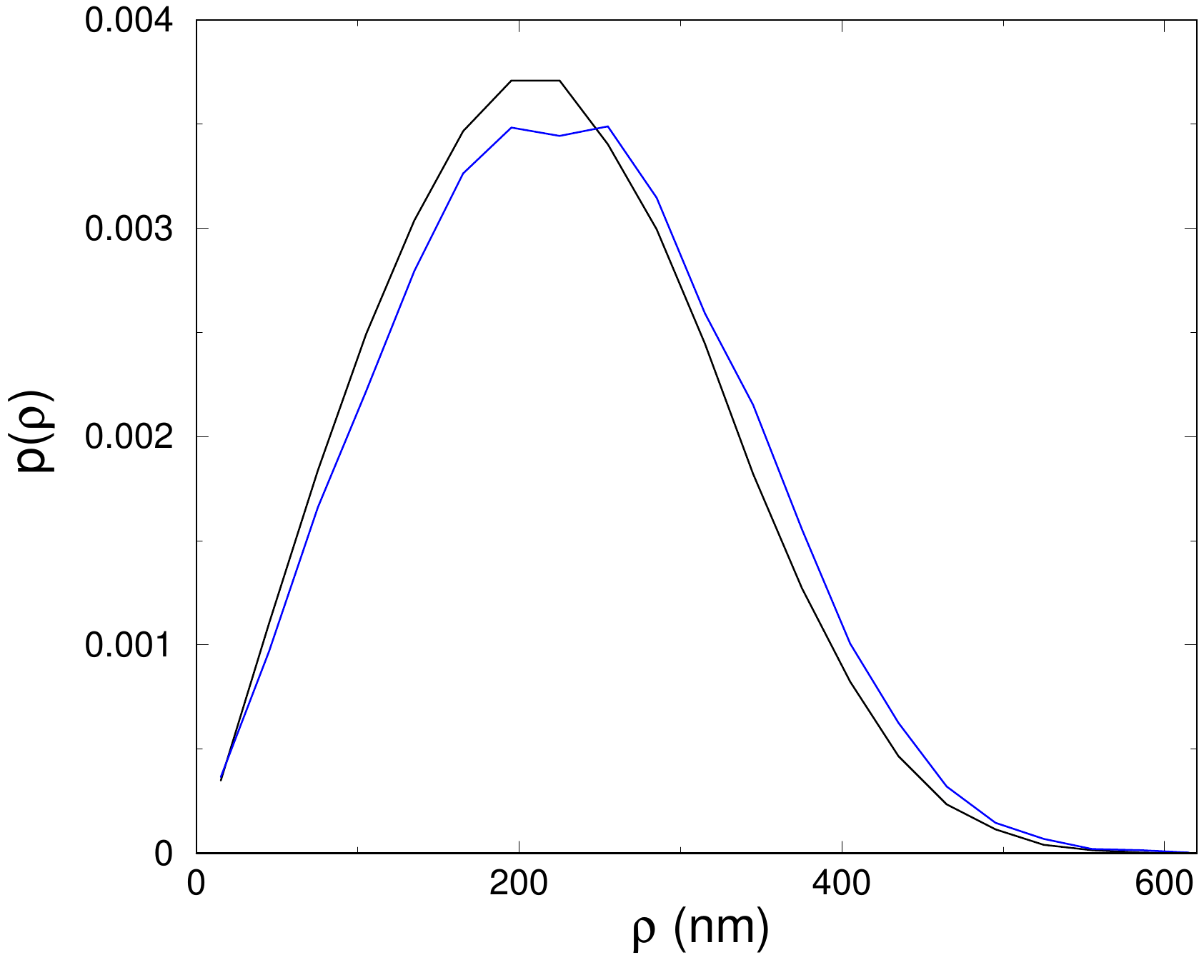}
\includegraphics[width=0.3\linewidth]{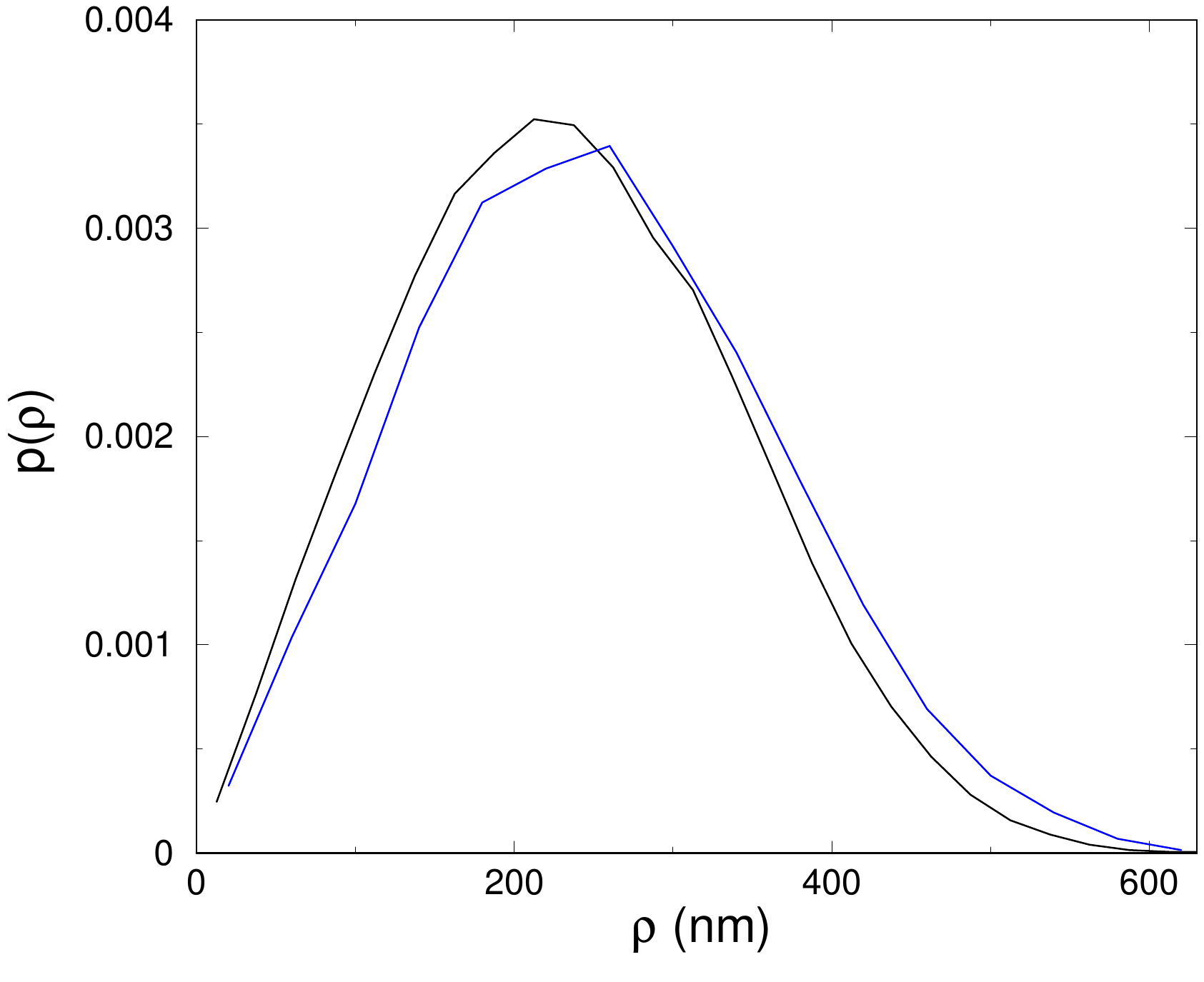}\\
\caption{Experimental (black) and numerical (blue) distributions $p(\rho)$. From top to bottom: $L=798$, 1500, 2080~bp. From left to right: $R=20$, 100, 150~nm.\label{comp2}}
\end{figure*}

Figures~\ref{comp1} and~\ref{comp2} show experimental and numerical probability 
distributions $p(\rho)$, in the semi-flexible regime, for DNA of various lengths (from 401 to 2080~bp) and various particle radii (from 20 to 150~nm). The agreement is systematically good. To quantify further the small discrepancies, the values of $\Delta \rpar$ are compared into deeper detail in the main text.

%\newpage

\subsection*{Effects of exposure times $T_{\rm ex}$  on experimental distributions}

\begin{figure}[ht]
\begin{center}
\includegraphics[height=6cm]{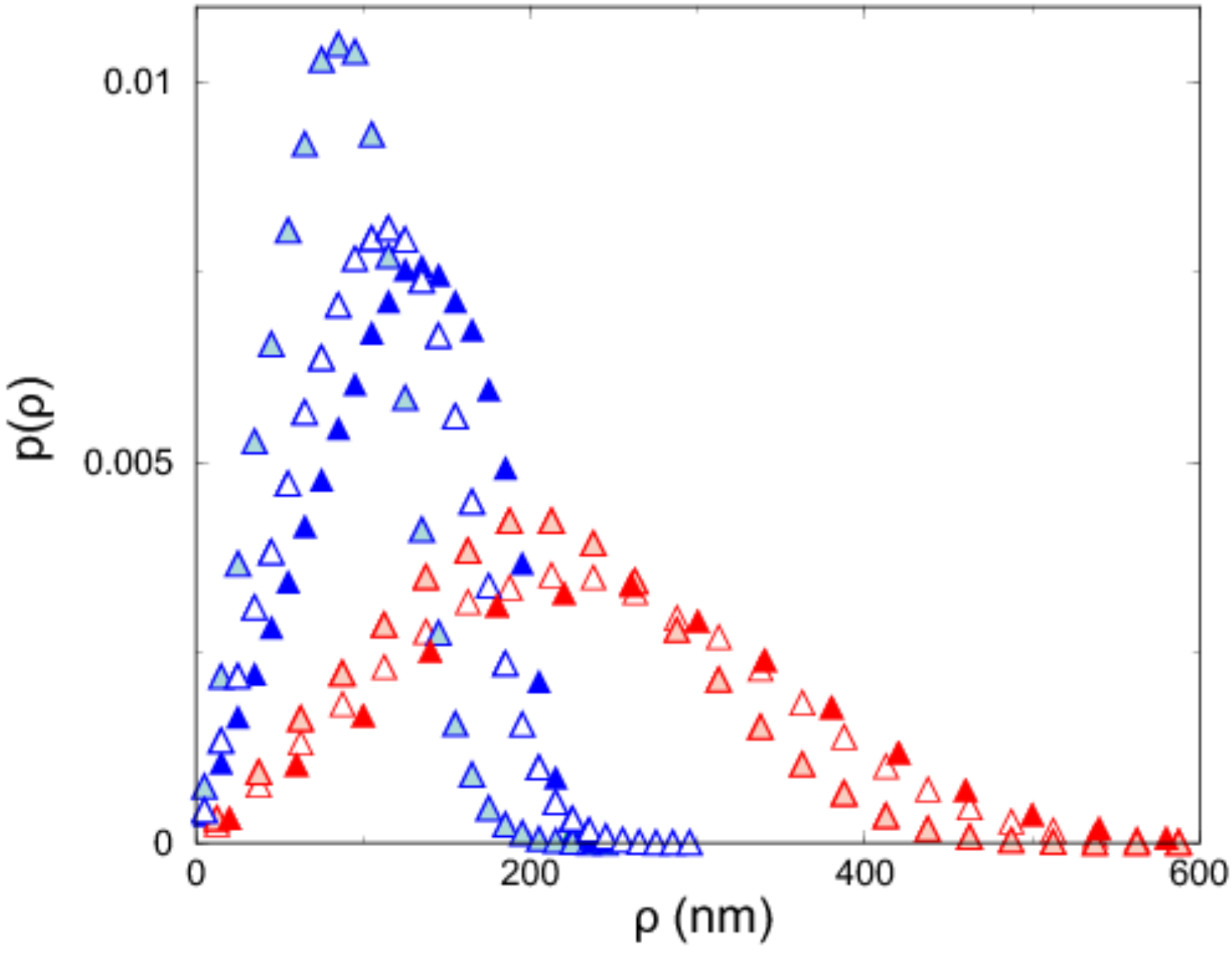}
\caption{\footnotesize Experimental probability distributions $p(\rho)$ for two exposure times, $T_{\rm ex}= 5$ (open symbols) and 40~ms (light red and red symbols). For comparison, solid symbols correspond to numerical distributions. Blue: $R=150$~nm and $L=401$~bp ($\tau_\parallel \simeq 20$~ms, see Table 1 in the main text); Red: $R=150$~nm and $L=2080$~bp ($\tau_\parallel \simeq 57$~ms).
\label{comparaison:5:40}}
\end{center}
\end{figure}

Beyond their width $\Delta \mathbf{r}_{\parallel}$, probability distributions $p(\rho)$ are also affected when $T_{\rm ex}$ becomes of the order of magnitude of $\tau_{\parallel}$, as shown in Fig.~\ref{comparaison:5:40}. In particular, because of the central-limit theorem, their Gaussian character is restored when $T_{\rm ex} \gg  \tau_{\parallel}$, even though the real distributions are far from Gaussian in the rigid or semi-flexible regime.

But recovering real distributions from their measured counterparts is a difficult task that is out of scope of the present work: it requires to de-convolute the averaging effect, which depends not only on $\rho$ but also on $z$, because diffusion time-scales are $\br$-dependent.  The examination of equations~(7,8) reveals that lowering the exposure time $T_{\rm ex}$ is sufficient to reduce averaging effects. When $T_{\rm ex} \ll \tau_{\parallel}$, they become negligible. But we have shown in this work that the {\em a priori} determination of the adapted values of $T_{\rm ex}$ is not an easy task because the quantification of blurring effects cannot rely on simple assumptions. 

\subsection*{Simple calculation of relaxation times} 

Some authors~\cite{Brinkers09SI} estimate relaxation times $\tau_\parallel$ as follows. 
If one considers the complex as a spherical particle attached to an ideal spring (the DNA molecule), ignoring hydrodynamic corrections, excluded-volume effects, and interaction between the beads and the wall, one can easily compute the relaxation time of the particle:
\begin{equation}
\tau_{\parallel} = \frac{\langle \mathbf{r}_{\parallel}^2 \rangle}{2D_{\rm part}},
\end{equation}
where $D_{\rm part}=k_BT/(6 \pi \eta R)$ and $\langle \mathbf{r}_{\parallel}^2 \rangle$ is calculated as follows. If $K$ is the spring constant of the DNA molecule, then 
\begin{equation}
K = \frac32 \frac{k_BT}{\ell_{\rm p} L},
\end{equation}
assuming a 3D phantom semi-flexible chain of persistence length $\ell_{\rm p}$
and $L \gg \ell_{\rm p}$~\cite{DeGennes}. Owing to the equipartition theorem, 
\begin{equation}
\frac12 K \langle \mathbf{r}_{\parallel}^2 \rangle = \frac12 K \langle x^2 \rangle + \frac12 K \langle y^2 \rangle
= k_B T.
\end{equation}
Thus one gets the theoretical value
\begin{equation}
\tau_{\parallel}^{\rm th,1} = \frac{4 \pi \eta \ell_{\rm p}} {k_BT} RL,
\label{BOTE}
\end{equation}
which is equal to the ratio of the friction coefficient of the particle, $6 \pi \eta R$, to the spring constant, $K$.

\begin{figure}
\begin{center}
\ \includegraphics[width=7.1cm]{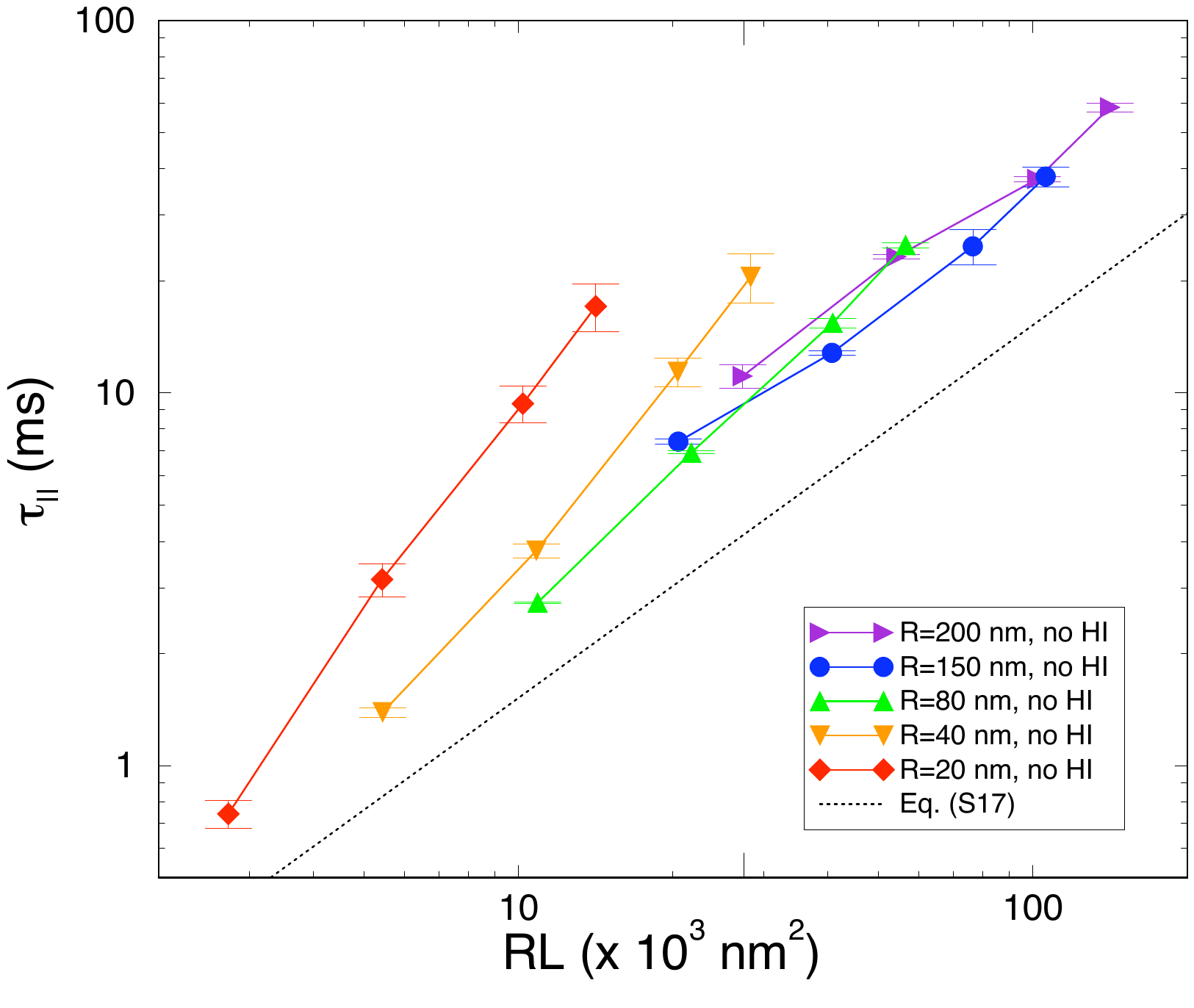} \ \ \includegraphics[width=7.1cm]{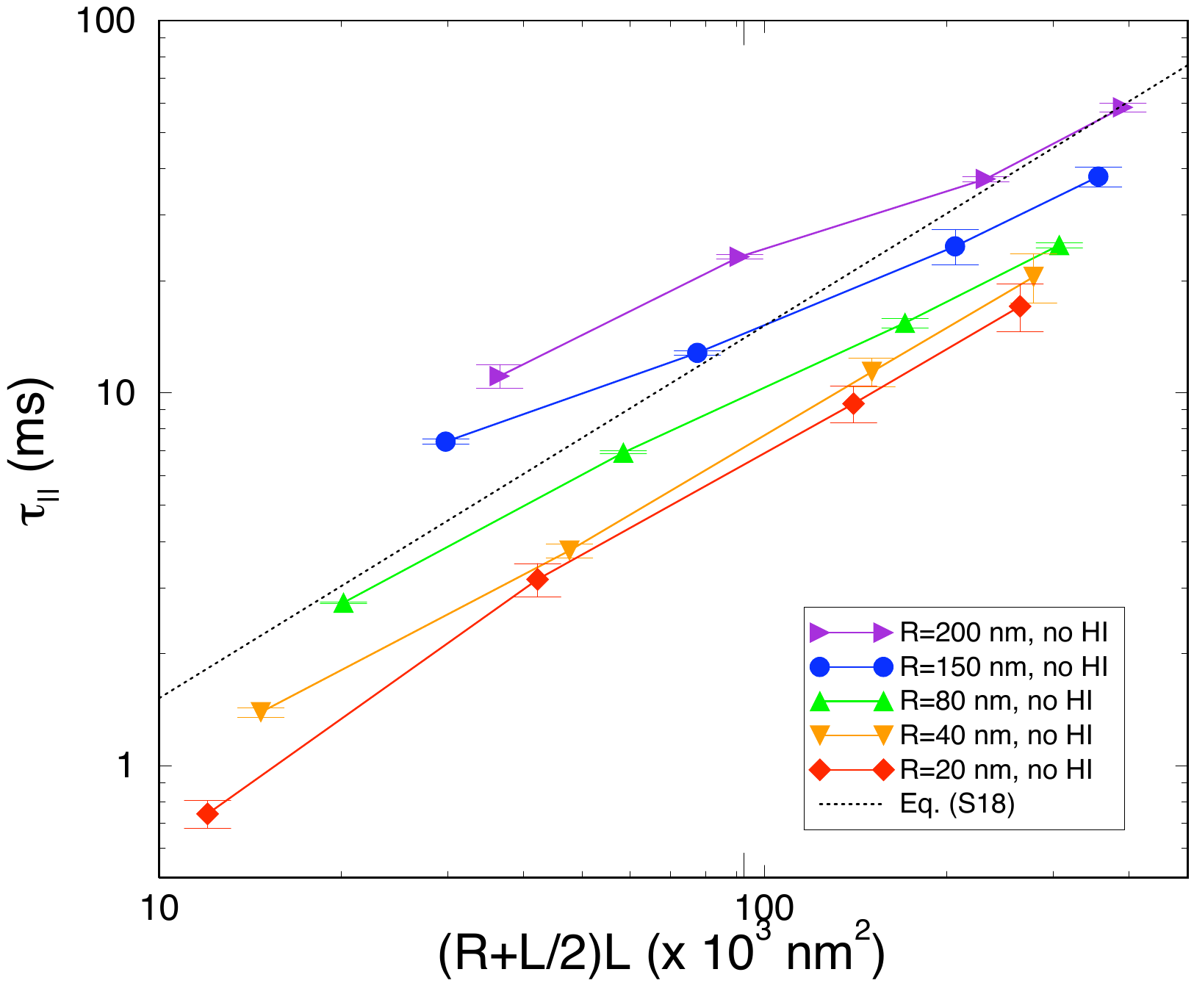} \
\end{center}
\caption{Left: Numerical values of the relaxation time $\tau_{\parallel}^{\rm sim}$ in function of $RL$ for $R>0$ (same data as in Fig. 3). The dotted line shows the values $\tau_{\parallel}^{\rm th,1}$ predicted by  equation~(\ref{BOTE}). Right: same data in function of $(R+L/2)L$. The dotted line shows the values $\tau_{\parallel}^{\rm th,2}$ predicted by  equation~(\ref{BOTE2}). Log-log coordinates.
\label{tau:RL}}
\end{figure}

Figure~\ref{tau:RL} (Left) shows our numerical relaxation times in function
of $RL$, together with the values
predicted by equation~(\ref{BOTE}). 
Note that in this case, we only compare our numerical values with no hydrodynamic corrections near the wall to this theoretical prediction, since they are not included in the simple theory. The above simple approach appears to be reasonable when
$R$ is large. More quantitatively, if $R\geq L/6$, then the numerical
relaxation time is less than three times the above theoretical
estimate. In addition, if $R$ is large, the scaling $\tau_{\parallel}^{\rm th,1} \sim L$ is correct. In the best cases however, equation~(\ref{BOTE}) still
underestimates $\tau_\parallel^{\rm sim}$ by a factor 2.

By contrast, when $R$ is small~-- the case of interest in the present work~-- the above approximations fail, by up to a factor 10. Even the scaling in equation~(\ref{BOTE}) is incorrect since $\tau_{\parallel}^{\rm sim} \propto L^2$ in this case, because the present approach neglects the own dynamics of the polymer that dominates at small $R$ ($R<L/6$). 

To improve the previous approximation, one can thus replace the diffusion coefficient of the particle by that of the particle-DNA complex, using equation (11) of the main text, which leads to
\begin{equation}
\tau_{\parallel}^{\rm th,2} = \frac{4 \pi \eta \ell_{\rm p}} {k_BT} (R+L/2)L = \left(1+\frac{L}{2R}\right)
\tau_{\parallel}^{\rm th,1}
\label{BOTE2}
\end{equation}
in the Rouse approximation. Figure~\ref{tau:RL} (Right) shows that this theoretical value is better than the previous one since it only fails to predict the numerical values by a factor (at most) 2.5. The scaling is correct at small $R$ since in this case $\tau_{\parallel}^{\rm th,2} \sim L^2$. But it is now incorrect at larger $R$. 

Thus one cannot ignore the interplay between the (finite-size) polymer, the particle and the wall when inferring accurately relaxation times.

\subsection*{Acquisition periods $T_{\rm ac}$ }

In the last section of the paper, we explain why the acquisition
period of the detector, $T_{\rm ac}$ cannot be much larger than the
measured relaxation time, $\tau_{\rm m}$. Indeed, if we focus for
example on the correlation function $C(t)$ used to measure $\tau_{\rm
 m}$, it reads $C(t)=C(0) \exp(-t/\tau_{\rm m})$, but the only
experimentally accessible values are $C(T_{\rm ac}),C(2T_{\rm
 ac}),\ldots$ Here we are looking for the largest values of $T_{\rm
 ac}$ that can be used in practice. In this case, $C(2T_{\rm ac})$ is
so close to 0 that it does bear any useful information due to
statistical noise. Thus the only relevant information, if any, comes
from $C(T_{\rm ac})=C(0)\exp(-T_{\rm ac}/\tau_{\rm m})$. If we define
$Q=C(T_{\rm ac})/C(0)$, then the fitting procedure essentially amounts
to writing
\begin{equation}
\tau_{\rm m} = -\frac{T_{\rm ac}}{\ln Q}.
\end{equation}
It follows that the relative error on the measure of $\tau_{\rm m}$ is
\begin{equation}
\frac{\Delta \tau_{\rm m}}{\tau_{\rm m}} = \frac{\tau_{\rm m}}{T_{\rm ac}}
\frac{\Delta Q}{Q} \frac1{\sqrt{n}},
\end{equation}
where $\Delta Q$ is the experimental dispersion of measures of $Q$ on
different trajectories and $n$ is the number of trajectories for given
experimental conditions. If $T_{\rm ac}$ grows, then $Q$ goes to 0 and
this relative error diverges. Thus $T_{\rm ac}$ must not exceed a
limiting value. Our less favorable experimental case if for
$L=798$~bp, $R=20$~nm, and $T_{\rm ac}=8.9$~ms, in which case
$\tau_{\rm m}=4.4$~ms and $\tau_{\rm m}/T_{\rm ac}=0.5$. We have
measured that $\Delta Q/Q=0.7$ in this case. With $n=34$, we get a
satisfactory relative error of 6\%. If using a higher $T_{\rm ac}$,
the relative error on $Q$ and thus on $\tau_{\rm m}$ would rapidly
increase (exponentially with $T_{\rm ac}$), thus requiring a much
larger number $n$ of realizations. We conclude that the condition
$\tau_{\rm m}/T_{\rm ac}\gtrsim0.5$, in other words $T_{\rm
 ac}\lesssim 2\tau_{\rm m}$, is reasonable in the present context.

\subsection*{Measuring the amplitude of movement: a comprehensive example}

At the end of the Results and Discussion section of the paper, we explain how the relative error on the measure of the amplitude of movement $\Delta \br_\parallel^2=
\langle \rpar^2 \rangle$ decreases with the averaging-interval length, $T_{\rm av}$~:
\begin{equation}
\lambda(T_{\rm av})\equiv\frac{{\rm Err}(\br_\parallel^2)}{\langle \br_\parallel^2 \rangle} \propto \sqrt{\frac1{T_{\rm av}}}.
\end{equation}

Figure~S9 displays $\lambda(T_{\rm av})$ measured for two experimental data sets and two acquisition periods, $T_{\rm ac}=13.46$ and 40.38~ms. The second acquisition period is simply mimicked  by discarding 2 points out of 3 in the data sets. The absolute error ${\rm Err}(\br_\parallel^2)$ is calculated as the root-mean-square deviation of the measure of $\langle \br_\parallel^2 \rangle$ on successive intervals of duration $T_{\rm av}$. One observes that the slope $-1/2$ at long times is consistent with the expected behavior. At short times, this behavior is potentially affected by the correlations between successive measurements of $\br_\parallel^2$. 

\begin{figure*}[ht]
\ \hfill \includegraphics[width=0.4\linewidth]{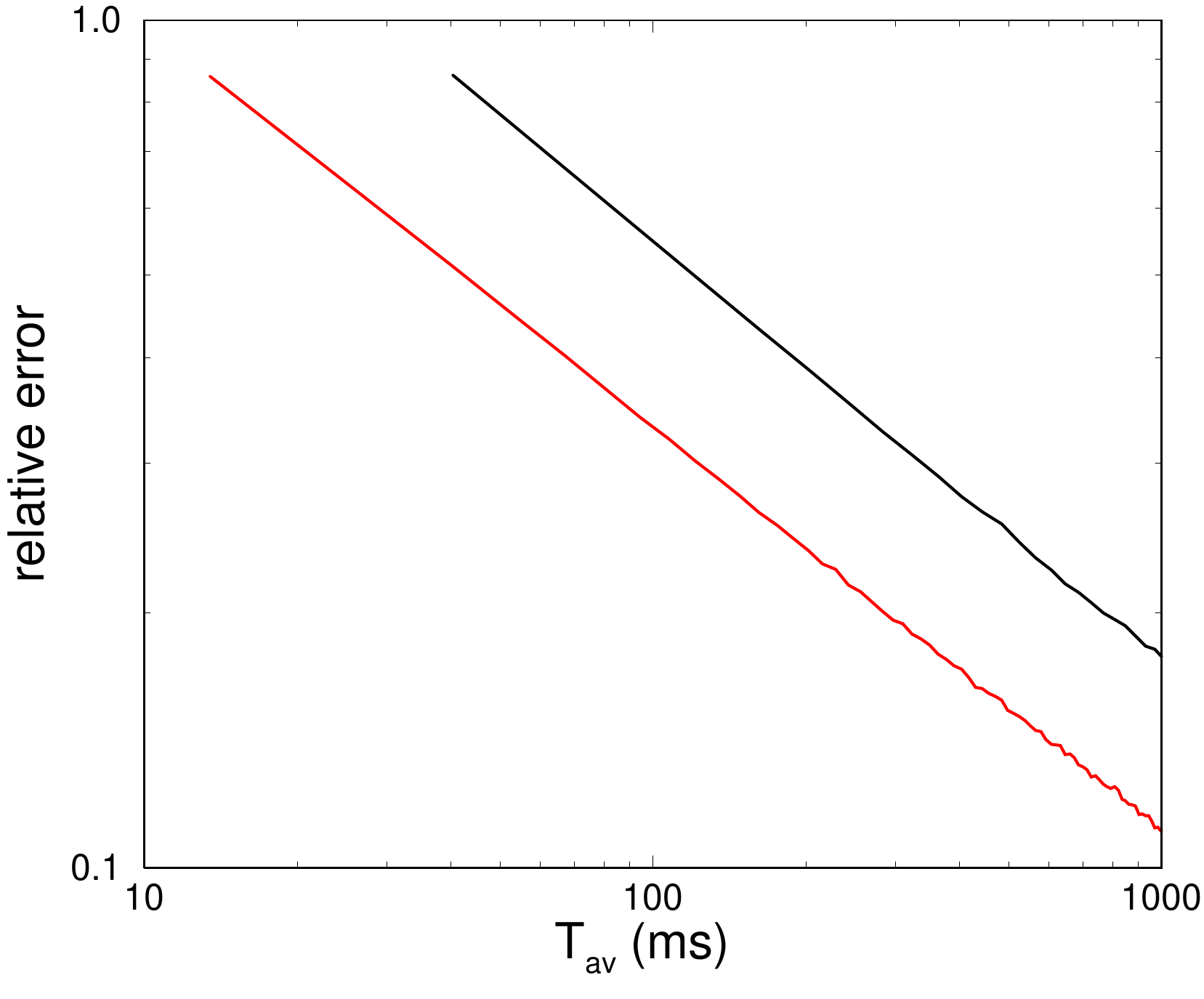}\hfill
\includegraphics[width=0.4\linewidth]{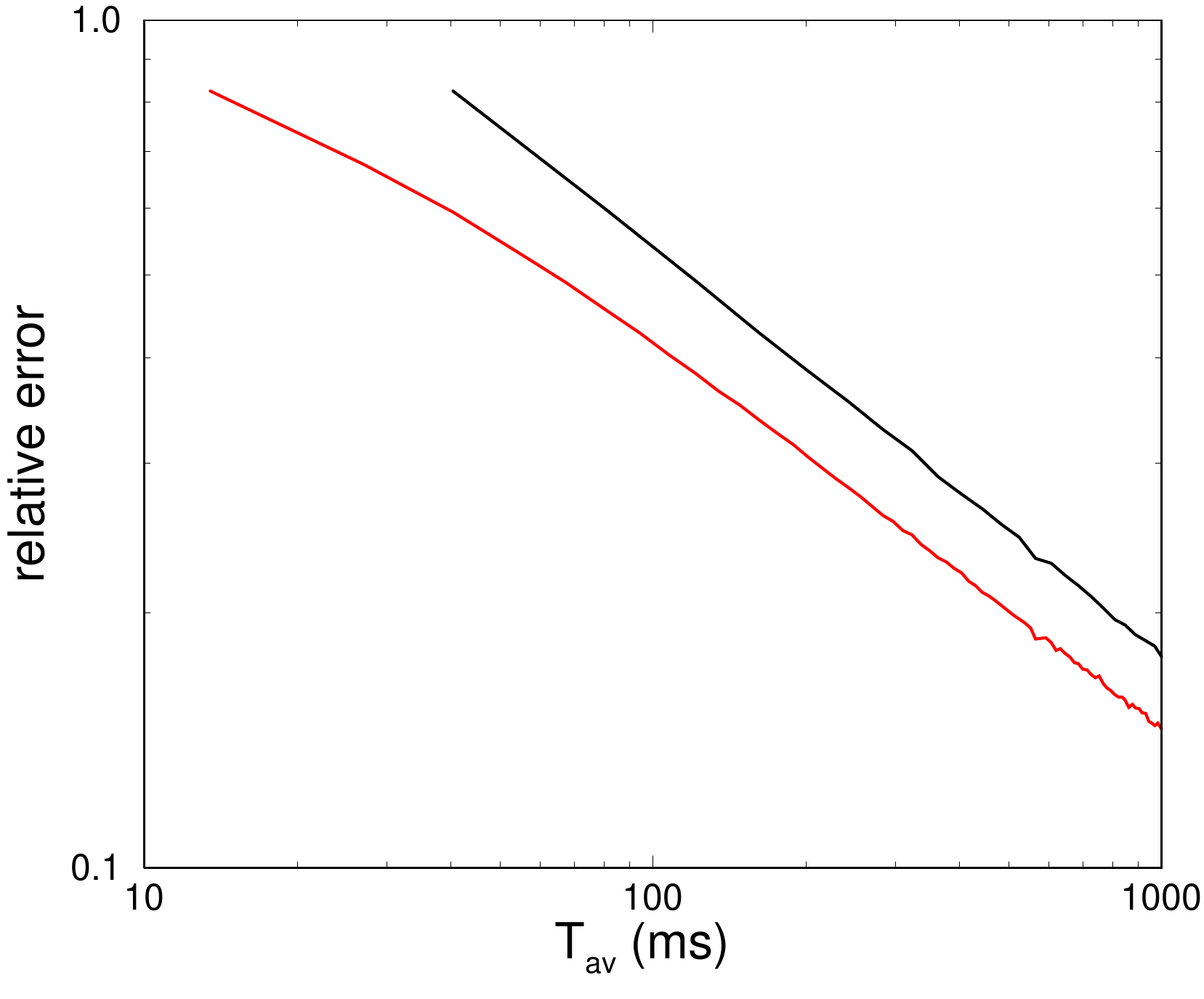} \hfill \
\caption{Relative error $\lambda(T_{\rm av})$ on the measure of the amplitude of movement $\langle \br_\parallel^2 \rangle$, extracted from experimental data for $L=2080$~bp and $R=20$~nm (left) and 100~nm (right). Upper curves: $T_{\rm ac}=40.38$~ms; Lower curves: $T_{\rm ac}=13.46$~ms.
Log-log coordinates.}
\label{rel:error}
\end{figure*}

More quantitatively, equation~(12) of the paper predicts an upper bound for the relative error. In the present case, $\tau_\parallel\simeq  8.3$ and 36.3~ms for $R=20$ and 100~nm, 
and $\tau_{\parallel,2} \approx \tau_\parallel/3$. Thus in the case  $R=100$~nm
and $T_{\rm ac}=13.46$~ms, $\tau_{\parallel,2} \gtrsim T_{\rm ac}$ and one expects
$\lambda(T_{\rm av}=1s) < \sqrt{\tau_\parallel/T_{\rm av}} = 0.19$. We obtain experimentally $\lambda(T_{\rm av}=1{\rm s})=0.15$, in agreement with this upper bound. In the three remaining cases, we are in the regime $\tau_{\parallel,2} \ll T_{\rm ac}$ and the upper bound becomes $\lambda(T_{\rm av}=1{\rm s}) <\sqrt{T_{\rm ac}/T_{\rm av}}$ because successive images are uncorrelated. At $T_{\rm ac}=40$~ms and $T_{\rm av}=1$~s, $\lambda(T_{\rm av}=1{\rm s}) < 0.20$, while we measure $\lambda(T_{\rm av}=1{\rm s})=0.18$ for both radii. Finally, for $R=20$~nm and $T_{\rm ac}=13.46$~ms, the upper bound is 0.12 and we measure 0.11. In all four cases, the upper bound matches experimental data.

\subsection*{Detecting conformational changes: protein binding/unbinding}

We consider the same particle-DNA complex as in the
example described in the Results and Discussion section ($L=798$~bp, $R=20$~nm, $N=25$ and $\tau_\parallel \simeq 3$~ms). However, instead of studying looping events, we focus on DNA kinking
induced by protein binding on a specific locus. Indeed, some
proteins are known to strongly bend the molecule when adsorbed on
it, in such a way that binding can be detected by TPM~\cite{dixitSI,Pouget06SI}. 
We suppose that the locus is situated on bead $B_{12}$, in the
middle of the chain, and that the associated bending energy is now
$\frac{\epsilon}{2a} (1 -\cos (\theta_{12}-\theta_{\rm kink}))$
instead of $\frac{\epsilon}{2a} (1 -\cos \theta_{12})$.
$\theta_{\rm kink}$ is chosen equal $\pi/2$ in the present example. We have measured that in the kinked state,
the particle amplitude of movement $\Delta \mathbf{r}_\parallel^2$ is 
reduced by about 13\%: $\Delta \mathbf{r}_{\parallel,\rm kinked}^2=
0.87 \; \Delta \mathbf{r}_{\parallel,\rm unkinked}^2$ where $\Delta \mathbf{r}_{\parallel,\rm unkinked}=124$~nm.

\begin{figure}[h]
\begin{center}
\includegraphics[width=0.9\linewidth]{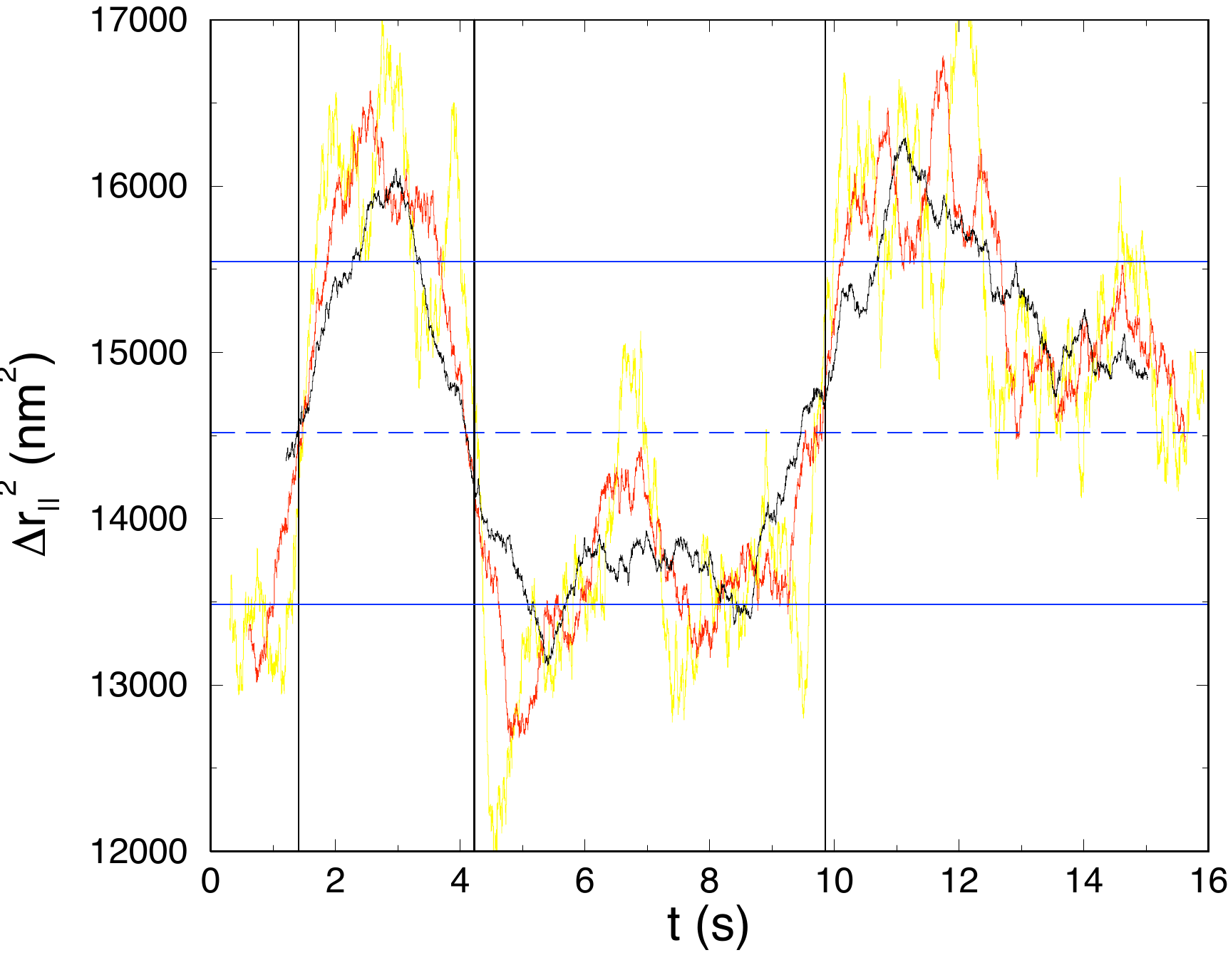}\hspace{1cm}
\includegraphics[width=0.9\linewidth]{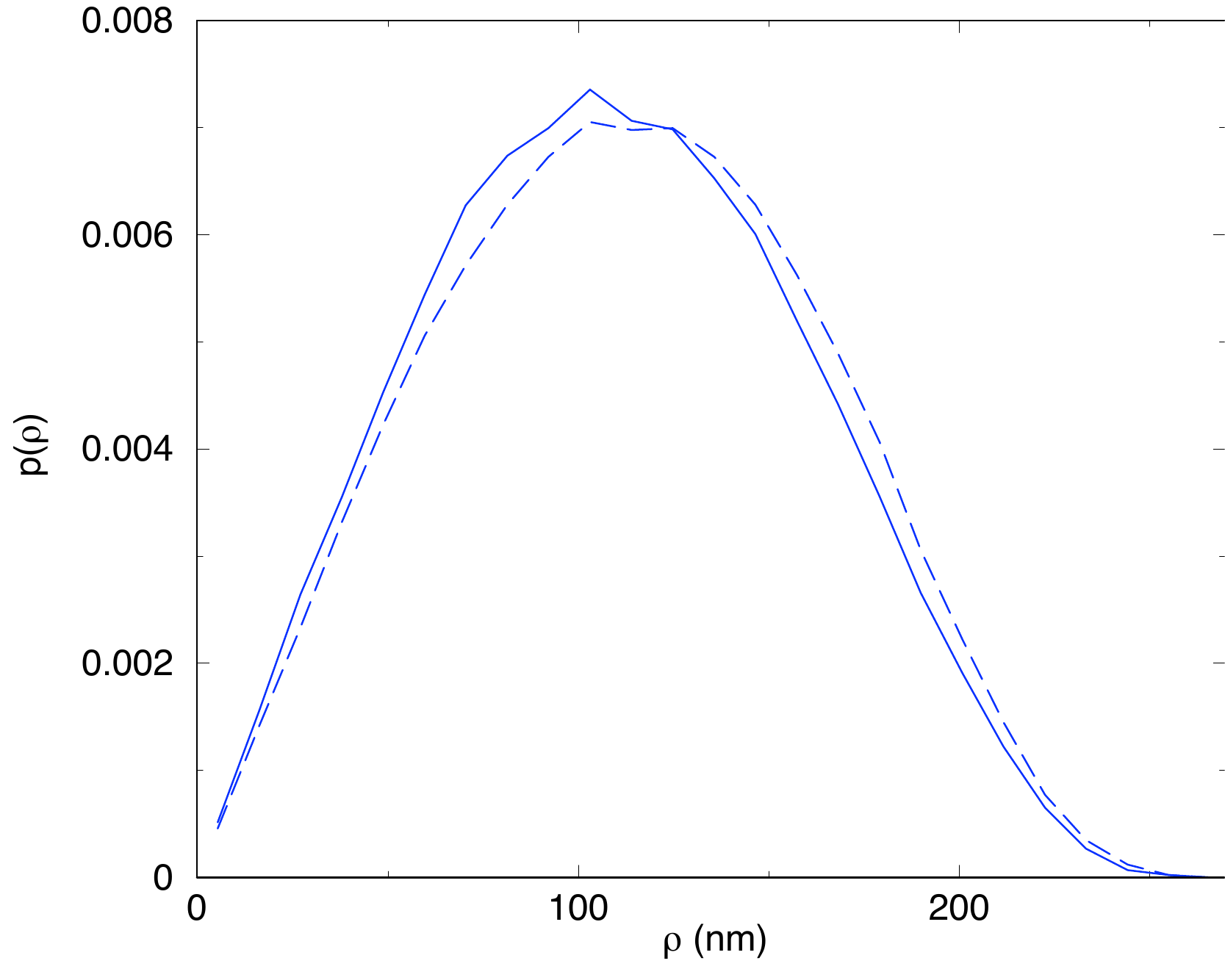}
\caption{Left: Simulated amplitude of movement $\Delta
  \mathbf{r}_\parallel^2(t)$ for a $L=798$~bp DNA and
  a $R=20$~nm particle. The different plots represent $\Delta
  \mathbf{r}_\parallel^2$ averaged on an interval of duration
  $T_{\rm av}=0.6$~s (yellow), 1.2~s (red) and 2.4~s (black). The
  vertical lines indicate the protein binding and unbinding
  events in the simulation. The horizontal solid lines show the
  expected values of $\Delta \mathbf{r}_\parallel^2$ in the kinked
  (bottom) and unkinked (top) states. The dashed line is the threshold
  separating the two states for detection purposes. Right:
  Distribution $p(\rho)$ corresponding to the same simulation (solid line),
together with $p(\rho)$ for the unkinked case (dashed line, same as Fig.~S4).
\label{kinking}} 
\end{center}
\end{figure}

The procedure is the same as in the DNA looping example described in the main text. But in the present case, the two amplitudes are very close, which makes detection
of binding/unbinding events more delicate. We set the threshold
inbetween $\Delta \mathbf{r}_{\parallel,\rm kinked}^2$ and $\Delta
\mathbf{r}_{\parallel,\rm unkinked}^2$, at $0.93 \; \Delta
\mathbf{r}_{\parallel,\rm unkinked}^2$. Thus relative amplitude of fluctuations
must not be larger than $\lambda=0.07$ in both states to detect them
unambiguously. Using again equation~(12), this sets $T_{\rm av}^* = \tau_\parallel /\lambda^2
= 0.69$~s. Even though the particle-DNA complex is the
same as in the previous example, one must use a much larger averaging
time because the two state amplitudes are much closer. Again, as
examplified in Fig.~S10 (left), if $ T_{\rm
av}>T_{\rm av}^*$, then false detections are scarce. If $ T_{\rm
av} \gg T_{\rm av}^*$, they occur with a vanishing probability. Note that $T_{\rm av}^*$ was determined on the basis of a 68\% confidence interval (see main text). An averaging time value four times as large as $T_{\rm av}^*$ ensures a 95\% confidence interval [equation~(12)].

We also emphasize that in the present case, due to the closeness
mentionned above, the two states are not distinguishable on histograms
$p(\rho)$. As illustrated in Fig.~S10 (right), $p(\rho)$ is
not bimodal and is very similar to the same distribution without
kinking. Whereas identifying
bimodal distributions is useful in the context of TPM to determine
conformational changes~\cite{Pouget06SI}, it is useless in the present context. 
By contrast, plotting $\Delta \mathbf{r}_{\parallel,\rm kinked}^2$ vs time
is efficient, provided that one knows in advance what averaging time
$ T_{\rm av}$ must be used to distinguish between both conformations 
with a good confidence.

\end{document}